\documentclass[stslayout, noinfoline, reqno]{imsart}

\RequirePackage[OT1]{fontenc}
\RequirePackage{amsthm,amsmath}
\RequirePackage{natbib}
\RequirePackage[colorlinks,citecolor=blue,urlcolor=blue]{hyperref}

\arxiv{stat.CO/00000000}

\startlocaldefs

\usepackage{lmodern}
\usepackage{amsthm}
\usepackage{amsfonts}
\usepackage{amssymb}
\usepackage{nccmath}
\usepackage{mathtools}
\usepackage[capitalise]{cleveref}
\usepackage[overload]{textcase}
\usepackage[dvipsnames]{xcolor}

\usepackage{graphicx}
\graphicspath{{images/}}

\usepackage{subcaption}
\usepackage{acro}
\acsetup{
  tooltip = true,
  long-format = \it,
  short-format = \sc
}

\DeclareAcronym{kf}{short=kf, long=Kalman filter}
\DeclareAcronym{enkf}{short=enkf, long=ensemble Kalman filter}
\DeclareAcronym{etkf}{short=etkf, long=ensemble transform Kalman filter}
\DeclareAcronym{letf}{short=letf, long=linear ensemble transform filter}
\DeclareAcronym{pf}{short=pf, long=particle filter}
\DeclareAcronym{etpf}{short=etpf, long=ensemble transform particle filter}
\DeclareAcronym{sletpf}{short=sletpf, long=smooth local ensemble transform particle filter}
\DeclareAcronym{ot}{short=ot, long=optimal transport}
\DeclareAcronym{nwp}{short=nwp, long=numerical weather prediction}
\DeclareAcronym{rmse}{short=rmse, long=root mean squared error}
\DeclareAcronym{pde}{short=pde, long=partial differential equation}
\DeclareAcronym{spde}{short=spde, long=stochastic partial differential equation}
\DeclareAcronym{sde}{short=sde, long=stochastic differential equation}
\DeclareAcronym{pou}{short=pou, long=partition of unity}
\DeclareAcronym{mcmc}{short=mcmc, long=Markov chain Monte Carlo}
\DeclareAcronym{dft}{short=dft, long=discrete Fourier transform}
\DeclareAcronym{idft}{short=idft, long=inverse discrete Fourier transform}
\DeclareAcronym{ssm}{short=ssm, long=state-space model}
\DeclareAcronym{ks}{short=ks, long=Kuramoto--Sivashinksy}
\DeclareAcronym{st}{short=st, long=stochastic turbulence}

\DeclareMathAlphabet{\mathitsf}{\encodingdefault}{\sfdefault}{m}{sl}
\SetMathAlphabet{\mathitsf}{bold}{\encodingdefault}{\sfdefault}{bx}{sl}
\DeclareSymbolFont{bbold}{U}{bbold}{m}{n}
\DeclareSymbolFontAlphabet{\mathbbold}{bbold}

\newcommand{\reals}{\mathbb{R}}

\newcommand{\naturals}{\mathbb{N}}
\newcommand{\complexs}{\mathbb{C}}

\newcommand{\lpa}{\left(}
\newcommand{\rpa}{\right)}
\newcommand{\lsb}{\left[}
\newcommand{\rsb}{\right]}
\newcommand{\set}[1]{\mathcal{#1}}
\newcommand{\metric}[1]{\mathnormal{#1}}
\newcommand{\op}[1]{{#1}}
\newcommand{\prob}{\mathbb{P}}
\newcommand{\expc}{\mathbb{E}}
\newcommand{\dr}{\mathrm{d}}
\newcommand{\gvn}{\,|\,}
\newcommand{\gau}{\mathcal{N}}

\newcommand{\tr}{^{\mkern-1.5mu\mathsf{T}}}

\newcommand{\cns}[1]{\mathtt{#1}}
\newcommand{\mtx}[1]{{#1}}
\newcommand{\vct}[1]{{#1}}
\newcommand{\idmtx}[1][]{\mtx{\mathrm{I}}_{#1}}
\newcommand{\onevct}[1][]{\vct{\mathrm{1}}_{#1}}
\newcommand{\indc}[1]{\mathbbold{1}_{#1}}
\newcommand{\rvar}[1]{\mathitsf{#1}}
\newcommand{\rvct}[1]{\vct{\rvar{#1}}}
\newcommand{\rmtx}[1]{\mtx{\rvar{#1}}}
\newcommand{\otilde}{\widetilde{\mathcal{O}}}
\newcommand{\range}[2][1]{#1\colon\mkern-4mu#2}
\newcommand{\srange}[2][1]{{#1\mkern-1mu:#2}}
\newcommand{\idxset}[4]{\lbrace #1 \rbrace_{#2\in\srange[#3]{#4}}}
\newcommand{\sumrange}[3]{\sum_{#1\in\srange[#2]{#3}}}
\newcommand{\pred}[1]{\vec{#1}}
\newcommand{\pss}[1]{^{\mkern1mu #1}}
\newcommand{\recip}[1]{\mfrac{1}{#1}}

\newsavebox{\foobox}
\newcommand{\slantbox}[2][0]{\mbox{%
        \sbox{\foobox}{#2}%
        \hskip\wd\foobox
        \pdfsave
        \pdfsetmatrix{1 0 #1 1}%
        \llap{\usebox{\foobox}}%
        \pdfrestore
}}
\newcommand{\unslant}[2][-.25]{\slantbox[#1]{$#2$}}
\newcommand{\uppi}{\mkern0.5mu\unslant\pi\mkern-2mu}

\DeclareMathOperator{\median}{median}
\DeclareMathOperator*{\argmin}{argmin}
\DeclareMathOperator{\sinc}{sinc} 

\DeclarePairedDelimiter{\ceil}{\lceil}{\rceil}
\DeclarePairedDelimiter{\floor}{\lfloor}{\rfloor}

\numberwithin{equation}{section}
\theoremstyle{plain}

\theoremstyle{plain}
\newtheorem{assumption}{Assumption}

\endlocaldefs

\begin{document}

\begin{frontmatter}
\title{%
  A scalable optimal-transport based local particle filter
}
\runtitle{%
  A scalable optimal-transport based local particle filter
}

\begin{aug}
\author{\fnms{Matthew M.} \snm{Graham}\thanksref{e1} \ead[label=e1,mark]{m.m.graham@nus.edu.sg}}
\kern-0.7em\and
\author{\fnms{Alexandre H.} \snm{Thiery}\thanksref{e2} \ead[label=e2,mark]{a.h.thiery@nus.edu.sg}}

\address{\printead{e1,e2}}
\runauthor{M. M. Graham \& A. H. Thiery}

\affiliation{%
  Department of Statistics and Applied Probability,
  National University of Singapore
}

\end{aug}

\begin{abstract}
  Filtering in spatially-extended dynamical systems is a challenging problem with significant practical applications such as numerical weather prediction. Particle filters allow asymptotically consistent inference but require infeasibly large ensemble sizes for accurate estimates in complex spatial models. Localisation approaches, which perform local state updates by exploiting low dependence between variables at distant points, have been suggested as a potential resolution to this issue. Naively applying the resampling step of the particle filter locally however produces implausible spatially discontinuous states. The ensemble transform particle filter replaces resampling with an optimal-transport map and can be localised by computing maps for every spatial mesh node. The resulting local ensemble transport particle filter is however computationally intensive for dense meshes. We propose a new optimal-transport based local particle filter which computes a fixed number of maps independent of the mesh resolution and interpolates these maps across space, reducing the computation required and allowing it to be ensured particles remain spatially smooth. We numerically illustrate that, at a reduced computational cost, we are able to achieve the same accuracy as the local ensemble transport particle filter, and retain its improved robustness to non-Gaussianity and ability to quantify uncertainty when compared to local ensemble Kalman filters.
\end{abstract}

\acresetall



\tableofcontents

\end{frontmatter}

\section{Introduction}
\label{sec:introduction}

A natural paradigm for modelling geophysical systems such as the atmosphere is as \emph{spatially-extended dynamical systems}: one or more state variables defined over a spatial domain are evolved through time according to a set of \acp{spde}. In this article we will consider the problem of inferring the distribution of the unknown state of such a system given noisy observations at a sequence of time points. As well as being an important problem in its own right, state inference is also a vital sub-component of tasks such as forecasting the future state of a system and inferring values for any free parameters in the numerical model used \citep{fearnhead2017particle}.

A key issue in performing state inference in spatially-extended systems is the typically high dimension of the state space. To allow numerical simulation of the \ac{spde} model the spatial domain is discretised in to a mesh (also known as a grid); the system state can then be represented as a finite-dimensional vector consisting of the concatenated values of the state variables at the nodes of the mesh. The resulting state dimension is therefore a multiple of the number of mesh nodes which can be very large. For example in the global atmospheric models used in current operational \ac{nwp} systems the mesh size can be of the order $10^8$ or higher \citep{bauer2015quiet}. 

For large state dimensions, even inference in linear-Gaussian models\footnote{Throughout this article we will for brevity refer to dynamical models with linear state update and observation operators and additive Gaussian noise processes as \emph{linear-Gaussian}.} using the \ac{kf} \citep{kalman1960new} is computationally infeasible due to the high processing and memory costs of operations involving the full covariance matrix of the state distribution. This motivated the development of \ac{enkf} methods \citep{evensen1994sequential,burgers1998analysis} which use an ensemble of particles to represent the state distribution rather than the full mean and covariance statistics. As the ensemble sizes used are typically much smaller than the state dimension\footnote{Current operational \ac{nwp} ensemble systems are limited to $\sim 50$ particles due to the high computational cost of numerically integrating the particles forward in time \citep{buizza2005comparison}.}  the computational savings can be considerable.

Although \ac{enkf} methods are only consistent in an infinite ensemble limit for linear-Gaussian models \citep{furrer2007estimation,legland2011large}, they have been empirically found to perform well in models with weakly non-linear state update and observation operators, even when using relatively small ensembles of size much less than the state dimension \citep{evensen2009data}; the performance of the \ac{enkf} in non-asymptotic regimes has been theoretically investigated in several recent works \citep{kelly2014well,del2018stability,bishop2018stability,tong2016nonlinear}.
A key aspect in allowing \ac{enkf} methods to be scaled to large spatially-extended geophysical models is the use of \emph{spatial localisation} \citep{houtekamer1998data,hamill2001distance}. Localisation exploits the observation that there is often low statistical dependence between state variables at distant points in spatially-extended systems. In \ac{enkf} methods this property is used to improve the noisy covariance estimates resulting from the small ensemble sizes used by removing spurious correlations between distant state variables.

\Ac{enkf} methods have been successfully applied in a variety of settings, including operational \ac{nwp} systems \citep{bonavita2008ensemble,clayton2013operational}, however the quality of the state distribution estimates is fundamentally limited by the linear-Gaussian assumptions made by the underlying \ac{kf} updates. For models with non-Gaussian noise processes or strongly non-linear state update or observation operators, \ac{enkf} methods tend to produce poor estimates of the state distribution \citep{lei2010comparison}.

\Acp{pf} \citep{gordon1993novel,del1996non} offer an alternative ensemble-based approach to sequential state inference that unlike \ac{enkf} methods provides consistent estimates for non-Gaussian distributions. The simplest variant, the bootstrap \ac{pf}, alternates propagating the ensemble members forward in time under the model dynamics, with resampling according to weights calculated from the likelihood of the particles given the observed data.

While \acp{pf} offer asymptotically consistent inference for general state space models, in practice they typically suffer from \emph{weight-degeneracy} in high-dimensional systems: after propagation only a single particle has non-negligible weight. For even simple linear-Gaussian models, \acp{pf} have been shown to require an ensemble size which scales exponentially with the number of observations to avoid degeneracy \citep{snyder2008obstacles,bengtsson2008curse,snyder2011particle}. 

Given the importance of localisation in scaling \ac{enkf} methods to large spatial systems, it is natural to consider whether \ac{pf} methods can be localised to overcome weight-degeneracy issues \citep{snyder2008obstacles,van2009particle}. \citet{rebeschini2015can} analysed a simple local \ac{pf} scheme in which the spatial domain is partitioned into disjoint blocks and independent \acp{pf} run for each block, with local particle weights computed from the observations within each block. The authors demonstrate this \emph{block \ac{pf}} algorithm can overcome the need to exponentially scale the ensemble size with dimension to prevent degeneracy. However as the variables in each block are resampled independently from those in other blocks, dependencies between blocks are ignored; this introduces a systematic bias that is difficult to control \citep{bertoli2014adaptively}.

\begin{figure}

\begin{minipage}[t]{.1911\linewidth}
  \centering
  \begin{subfigure}[b]{\linewidth}
    \includegraphics[width=\linewidth]{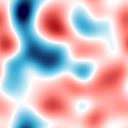}
    \caption{True state.}\label{sfig:gaussian-field-true-state}
  \end{subfigure}
  \begin{subfigure}[b]{\linewidth}
    \includegraphics[width=\linewidth]{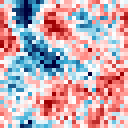}
    \caption{Observed.}\label{sfig:gaussian-field-noisy-obs}
  \end{subfigure}
\end{minipage}~
\begin{minipage}[t]{.39\linewidth}
  \centering
  \begin{subfigure}[b]{\linewidth}
    \includegraphics[width=.49\linewidth]{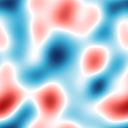}
    \includegraphics[width=.49\linewidth]{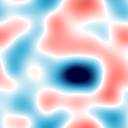}
    \caption{Prior samples.}
    \label{sfig:gaussian-field-predictive-dist-samples}
  \end{subfigure}
  \begin{subfigure}[b]{\linewidth}
    \includegraphics[width=.49\linewidth]{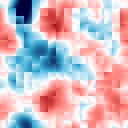}
    \includegraphics[width=.49\linewidth]{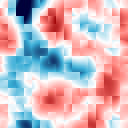}
    \caption{Posterior samples: block \ac{pf}.
    }
    \label{sfig:gaussian-field-block-pf-16x16}
  \end{subfigure}
\end{minipage}~
\begin{minipage}[t]{.39\linewidth}
  \centering
  \begin{subfigure}[b]{\linewidth}
    \includegraphics[width=.49\linewidth]{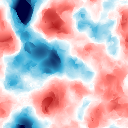}
    \includegraphics[width=.49\linewidth]{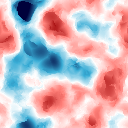}
    \caption{\centering Posterior samples: local \ac{etpf}.
    }
    \label{sfig:gaussian-field-letpf-per-gridpoint}
  \end{subfigure}
  \begin{subfigure}[b]{\linewidth}
    \includegraphics[width=.49\linewidth]{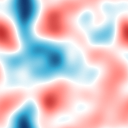}
    \includegraphics[width=.49\linewidth]{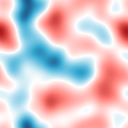}
    \caption{Posterior samples: this article.
    }
    \label{sfig:gaussian-field-sletpf-gaspari-cohn-8x8-pou}
  \end{subfigure}
\end{minipage}

\caption{Examples of local \ac{pf} assimilation updates applied to a Gaussian process model. The smooth true state field is shown in panel \subref{sfig:gaussian-field-true-state} and corresponding noisy observations in \subref{sfig:gaussian-field-noisy-obs}. Panel \subref{sfig:gaussian-field-predictive-dist-samples} shows prior samples and \subref{sfig:gaussian-field-block-pf-16x16}--\subref{sfig:gaussian-field-sletpf-gaspari-cohn-8x8-pou} approximate posterior samples after applying different local \ac{pf} assimilation updates. In each of \subref{sfig:gaussian-field-predictive-dist-samples}--\subref{sfig:gaussian-field-sletpf-gaspari-cohn-8x8-pou} 2 out of 40 samples are shown.}
\label{fig:gaussian-field-resampling-schemes-examples}

\end{figure}

This issue is illustrated for a two-dimensional Gaussian process model in \cref{fig:gaussian-field-resampling-schemes-examples}. The smooth true state field, shown in \cref{sfig:gaussian-field-true-state}, is partially and noisily observed (\cref{sfig:gaussian-field-noisy-obs}). While the samples in the prior ensemble (\cref{sfig:gaussian-field-predictive-dist-samples}) reflect the smoothness of the true state field, the posterior samples shown in \cref{sfig:gaussian-field-block-pf-16x16}, computed using a block \ac{pf} assimilation update show spatial discontinuities at the block boundaries. Such discontinuities can cause numerical instabilities in the computation of spatial derivatives when integrating the \acp{spde} model to forward propagate the particles.

The \acf{etpf} \citep{reich2013nonparametric} uses an \ac{ot} map to linearly transform an ensemble instead of resampling. The \ac{etpf} can be localised by computing \ac{ot} maps for each mesh node using local particle weights \citep{cheng2015assimilating}; updating the particles using the resulting spatially varying maps significantly reduces the introduction of spatial discontinuities compared to independent resampling. This can be seen in the samples computed using the local \ac{etpf} shown in \cref{sfig:gaussian-field-letpf-per-gridpoint}, which show greater spatial regularity than the block \ac{pf} samples in \cref{sfig:gaussian-field-block-pf-16x16}, though they remain less smooth than the true state field.

The requirement in the local \ac{etpf} to solve an \ac{ot} problem at every node can be computationally burdensome when the mesh size is large. Solving each \ac{ot} problem has complexity $\otilde(\cns{P}^3)$ where $\cns{P}$ is the ensemble size ($\otilde$ indicates limiting complexity excluding polylogarithmic factors); although solvers can be run in parallel this still represents a large computational overhead. 

In this article we propose an alternative smooth and computationally scalable local \ac{etpf} scheme. A finite set of \emph{patches} which cover the spatial domain are defined, with a non-negative \emph{bump function} supported on the patch. The set of bump functions is constrained to be a \ac{pou}: the functions sum to unity at all points in the spatial domain. A single \ac{ot} map is calculated for each spatial patch. The \ac{pou} is then used to interpolate these local per-patch maps across the spatial domain, defining maps for all nodes in the spatial mesh.

Through an appropriate choice of bump functions this scheme can maintain a prescribed level of smoothness in the transformed state fields while also significantly reducing the number of \ac{ot} problems needing to be solved. Examples posterior samples computed using the proposed scheme are shown in \cref{sfig:gaussian-field-sletpf-gaspari-cohn-8x8-pou}. Here the \ac{pou} is a set of smooth bump functions tiled in a $8\times 8$ grid. As well as giving more plausibly smooth fields than those computed using the local \ac{etpf}, in this example the number of \ac{ot} problems solved was reduced from to 16\,384 to 64. 

The remainder of the article is structured as follows. In \cref{sec:ensemble-filtering} we briefly introduce our notation and some preliminaries on the filtering problem and ensemble methods, followed by a review of \ac{spde} models and existing local filtering approaches in \cref{sec:spatial-systems}. The new method we propose is described in \cref{sec:proposed-method} and a numerical study comparing the approach to existing local ensemble filters is presented in \cref{sec:experiments}, with a concluding discussion in \cref{sec:conclusion}.

\section{Ensemble approaches to filtering}
\label{sec:ensemble-filtering}

\subsection{Notation}

Random variables are denoted by sans-serif symbols, e.g. $\rvar{x}$, and $\rvar{x} \sim \mu$ indicates $\rvar{x}$ has distribution $\mu$. The probability of an event $\rvar{x}$ taking a value in a set $\set{A}$ is $\prob(\rvar{x}\in\set{A})$ and the expected value of $\rvar{x}$ is $\expc[\rvar{x}]$. 
The conditional probability of $\rvar{x} \in \set{A}$ given $\rvar{y} = y$ is denoted $\prob(\rvar{x}\in\set{A}\gvn\rvar{y}=y)$ and likewise the conditional expectation of $\rvar{x}$ given $\rvar{y} = y$ is $\expc[\rvar{x}\gvn\rvar{y}=y]$. A Gaussian distribution with mean $\vct{m}$ and covariance $\mtx{C}$ is denoted $\gau(\vct{m},\mtx{C})$. The set of integers from $\cns{A}$ to $\cns{B}$ inclusive is $\range[\cns{A}]{\cns{B}}$ and quantities sub- or superscripted by an integer range indicate an indexed set, e.g. $\phi_{\srange{\cns{M}}} = \idxset{\phi_m}{m}{1}{\cns{M}}$. 
The $\cns{D}$ vector of ones is $\onevct[\cns{D}]$ and the $\cns{D}\times\cns{D}$ identity matrix $\idmtx[\cns{D}]$, with the subscript omitted when unambiguous. The indicator function on a set $\set{S}$ is $\indc{\set{S}}$. The set of real numbers is $\reals$, non-negative reals $\reals_{\geq 0}$ and complex numbers $\complexs$. For $z \in \complexs$, $\Re(z)$ and $\Im(z)$ indicate its real and imaginary parts.

\subsection{State-space models}

The class of models we aim to perform inference in is \acp{ssm}. Let $\set{X}$ be a vector-space representing the \emph{state-space} of the system of interest. We assume observations of the system are available at a set of $\cns{T}$ times, with the observations at each discrete \emph{time index} $t \in \range{\cns{T}}$ belonging to a common vector-space $\set{Y}$. We denote the unknown system state at each time index as a random variable $\rvar{x}_t \in \set{X}$ and the corresponding observations as a random variable $\rvct{y}_t \in \set{Y}$. The modelled state dynamics are assumed to be Markovian and specified by a set of \emph{state-update operators} $\op{F}_{\srange{\cns{T}}}$ such that
\begin{equation}
  \label{eq:generate-states}
  \rvct{x}_{1} = \op{F}_{1}(\rvct{u}_{1}),
  ~ \rvct{u}_{1} \sim \mu_{1};
  \quad
  \rvct{x}_{t} = \op{F}_{t}(\rvct{x}_{t-1}, \rvct{u}_{t}),
  ~ \rvct{u}_{t} \sim \mu_{t}
  ~~ \forall t \in \range[2]{\cns{T}},
\end{equation}
with each $\rvct{u}_t \in \set{U}$ a \emph{state noise} variable drawn from a distribution $\mu_t$, representing the stochasticity in the state initialisation and dynamics at each time step. The observations $\rvct{y}_t$ at each time index $t$ are assumed to depend only on the current state $\rvct{x}_t$ and are generated via a set of \emph{observation operators}  $\op{G}_{\srange{\cns{T}}}$,
\begin{alignat}{2}
  \label{eq:generate-observation}
  \rvct{y}_{t} &= \op{G}_t(\rvct{x}_{t}, \rvct{v}_{t}),
  &&\quad \rvct{v}_{t} \sim \nu_{t}
  \quad \forall t \in \range{\cns{T}}.
\end{alignat}
Any stochasticity in the observation process at each time index is introduced by the \emph{observation noise} variable $\rvct{v}_t \in \set{V}$ with distribution $\nu_t$. In \acp{ssm} where the operators $\op{F}_{\srange{\cns{T}}}$ and $G_{\srange{\cns{T}}}$
are all linear and the distributions $\mu_{\srange{\cns{T}}}$ and $\nu_{\srange{\cns{T}}}$ are all Gaussian -- the aforementioned linear-Gaussian case --  the joint distribution on all states $\rvar{x}_{\srange{\cns{T}}}$ and observations $\rvar{y}_{\srange{\cns{T}}}$ is Gaussian and a \ac{kf} can be used to perform exact inference. In this article we will focus on approximate inference methods for \acp{ssm} outside this class where exact inference is intractable.

We require that the conditional distributions on $\rvct{y}_t$ given $\rvct{x}_t$ have known densities $g_{\srange{\cns{T}}}$ with respect to a common dominating measure $\upsilon$ on $\set{Y}$, i.e.
\begin{equation}\label{eq:observation-density}
  \prob(\rvct{y}_t \in \dr\vct{y} \gvn \rvct{x}_t = \vct{x}_t) =
  g_t(\vct{y}\gvn \vct{x}_t) \, \upsilon(\dr\vct{y})
  \quad \forall t \in \range{\cns{T}}.
\end{equation}
For the state updates we assume only that the state-update operators $\op{F}_t$ can be computed for any set of inputs and that we can generate samples from the state noise distributions $\mu_t$; the resulting state transition distributions will not necessarily have tractable densities. 

\subsection{Filtering and predictive distributions}

Our main objects of interest from an inference perspective are the \emph{filtering distributions}: the conditional distributions on the state at time index $t \in \range{\cns{T}}$ given the observations at time indices up to and including $t$. We will denote the filtering distribution at each time index $t$ as
\begin{equation}\label{eq:filtering-distribution}
  \pi_{t}(\dr\vct{x}) =
  \prob(
    \rvct{x}_t \in \dr\vct{x}
    \gvn
    \rvct{y}_{\srange{\cns{T}}}=\vct{y}_{\srange{\cns{T}}}
  ).
\end{equation}
The \emph{filtering problem} is then the task of inferring the filtering distributions $\pi_{\srange{\cns{T}}}$ given a \ac{ssm} for the system and a sequence of observations $\vct{y}_{\srange{\cns{T}}}$.

A further concept that will be important for our discussion of inference methods is the \emph{predictive distribution} on the state at the next time index $t + 1$ given the observations up to the current time index $t$. We will denote the predictive distribution at time index $t$ as
\begin{equation}\label{eq:predictive-distribution}
  \pred{\pi}_{t+1}(\dr\vct{x}) =
  \prob(
    \rvct{x}_{t+1} \in \dr\vct{x}
    \gvn
    \rvct{y}_{\srange{\cns{T}}}=\vct{y}_{\srange{\cns{T}}}
  ).
\end{equation}

\subsection{Prediction and assimilation updates}

A key property for filtering algorithms is that the filtering distribution at any time index can be expressed recursively in terms of the distributions at the previous time indices. Generally this recursion is split into two steps, here termed the \emph{prediction} and \emph{assimilation} updates.

The prediction update transforms the filtering distribution $\pi_{t}$ to the predictive distribution $\pred{\pi}_{t+1}$. This update corresponds to propagating the state distribution forward in time according to the modelled dynamics, with no new observations introduced. Denoting the Dirac measure at a point $\vct{x} \in \set{X}$ by $\delta_{\vct{x}}$ the prediction update can be expressed as
\begin{equation}\label{eq:prediction-update}
  \pred{\pi}_{t+1}(\dr\vct{x}) =
  \int_{\set{U}} \int_{\set{X}}
    \delta_{\op{F}_{t+1}(\vct{x}', \vct{u})}(\dr\vct{x}) \,
  \pi_{t}(\dr\vct{x}') \, \mu_{t+1}(\dr\vct{u}).
\end{equation}

The assimilation update then relates the predictive distribution $\pred{\pi}_{t+1}$ to the filtering distribution at the next time step $\pi_{t+1}$. It corresponds to an application of Bayes' theorem, with the predictive distribution forming the prior and the filtering distribution at the next time index the posterior after a new observed data point has been assimilated. The observation density $g_{t+1}$ defines the likelihood term, with the assimilation update then
\begin{equation}\label{eq:assimilation-update}
  \pi_{t+1}(\dr\vct{x}) =
  \frac
    {g_{t+1}(\vct{y}_{t+1} \gvn \vct{x})}
    {\int_{\set{X}} g_{t+1}(\vct{y}_{t+1} \gvn \vct{x}') \, \pred{\pi}_{t+1}(\dr \vct{x}')}
  \pred{\pi}_{t+1}(\dr\vct{x}).
\end{equation}

The combination of prediction and assimilation updates together define a map from the filtering distribution at time index $t$ to the distribution at $t+1$:
$$ \cdots \longrightarrow \pi_{t} \xrightarrow{\textrm{prediction}} \pred{\pi}_{t+1} \xrightarrow{\textrm{assimilation}} \pi_{t+1} \longrightarrow \cdots; $$
sequentially alternating prediction and assimilation updates is in theory therefore all that is needed to compute the filtering distributions at all times indices. In practice however for most \acp{ssm} the integrals in \cref{eq:prediction-update,eq:assimilation-update} will be intractable to solve exactly, necessitating some form of approximation.

\subsection{Ensemble filtering}\label{subsec:ensemble-filtering}

A particularly common approximation is to use an ensemble of state particles to represent the filtering distribution at each time index. Specifically the filtering distribution $\pi_{t}$ at time index $t$ is represented by an empirical measure defined by placing point masses at the values of a set of $\cns{P}$ state particles $\rvct{x}_{t}\pss{\srange{\cns{P}}}$
\begin{equation}\label{eq:empirical-filtering-distribution}
  \pi_{t}(\dr\vct{x}) \approx
  \recip{\cns{P}} \sumrange{p}{1}{\cns{P}} \delta_{\rvct{x}\pss{p}_{t}}(\dr\vct{x}).
\end{equation}

A key advantage of using an ensemble representation of the filtering distribution is that a simple algorithm can be used to implement a prediction update consistent with \cref{eq:prediction-update}. Specifically if a set of $\cns{P}$ independent state noise samples $\rvct{u}_{t+1}\pss{\srange{\cns{P}}}$ are generated from $\mu_{t+1}$, then given particles $\rvct{x}_{t}\pss{\srange{\cns{P}}}$ approximating $\pi_{t}$, a new set of $\cns{P}$ particles can be computed as
\begin{equation}\label{eq:empirical-prediction-update}
  \pred{\rvct{x}}\pss{p}_{t+1} = \op{F}_{t+1}(\rvct{x}\pss{p}_{t}, \rvct{u}\pss{p}_{t+1})
  \quad \forall p \in \range{\cns{P}}.
\end{equation}
This new particle ensemble can then be used to form an empirical measure approximation to the predictive distribution $\pred{\pi}_{t+1}$
\begin{equation}\label{eq:empirical-predictive-distribution}
  \pred{\pi}_{t+1}(\dr\vct{x}) \approx
  \recip{\cns{P}}
  \sumrange{p}{1}{\cns{P}} \delta_{\pred{\rvct{x}}\pss{p}_{t+1}}(\dr\vct{x}).
\end{equation}

\subsection{Linear ensemble transform filters}\label{subsec:linear-ensemble-transform-filters}

Although \cref{eq:empirical-prediction-update} specifies an approach for performing a prediction update, a method for approximating the assimilation update in \cref{eq:assimilation-update} to account for the observed data is also required. One possibility is to require that the filtering ensemble $\rvct{x}\pss{\srange{\cns{P}}}_{t}$ is formed as a linear combination of the predictive ensemble $\pred{\rvct{x}}\pss{\srange{\cns{P}}}_{t}$ 
\begin{equation}\label{eq:letf-assimilation-update}
  \rvct{x}\pss{p}_{t} = \sum_{q\in\srange{\cns{P}}} \rvar{a}_t^{p,q} \pred{\rvct{x}}\pss{q}_{t}
\end{equation}
where $\rvar{a}_t\pss{\srange{\cns{P}},\srange{\cns{P}}} \in \reals^{\cns{P}\times\cns{P}}$ are a set of coefficients describing the transformation. In general the coefficients may depend non-linearly on both the observation $\vct{y}_t$ and predictive ensemble particles $\pred{\rvct{x}}\pss{\srange{\cns{P}}}_{t}$, however the form of the update constrains the filtering ensemble $\rvct{x}\pss{\srange{\cns{P}}}_{t}$ to lie in the linear subspace spanned by the predictive ensemble members. The class of ensemble filters using an assimilation update of the form in \cref{eq:letf-assimilation-update} was termed \acp{letf} in \citet{cheng2015assimilating}, and encompasses both ensemble Kalman and particle filtering methods, as will be discussed in the following subsections.

\subsection{Ensemble Kalman filters}\label{subsec:ensemble-kalman-filters}

In a linear-Gaussian \ac{ssm} the predictive and filtering distributions are Gaussian at all time indices: $\pi_t = \gau(\vct{m}_t, \mtx{C}_t)$ and $\pred{\pi}_t = \gau(\pred{\vct{m}}_t, \pred{\mtx{C}}_t)$ for all $t \in \range{\cns{T}}$, and so can be fully described by the mean and covariance parameters. The \acf{kf} \citep{kalman1960new} gives an efficient scheme for performing exact inference in linear-Gaussian \acp{ssm} by iteratively updating the mean and covariance parameters. For an observation operator and noise distribution
\begin{equation}\label{eq:linear-gaussian-observation}
  \op{G}_t(\vct{x}, \vct{v}) = \mtx{H}_t\vct{x} + \vct{v},
  \quad
  \rvct{v}_t \sim \gau(\vct{0}, \mtx{R}_t),
\end{equation}
the \ac{kf} assimilation update can be written
\begin{subequations}\label{eq:kf-assimilation-update}
\begin{align}
  \label{eq:kf-assimilation-update-covariance}
  \mtx{C}_t &= \pred{\mtx{C}}_t - \pred{\mtx{C}}_t\mtx{H}_t\tr(\mtx{R}_t+\mtx{H}_t\pred{\mtx{C}}_t\mtx{H}_t\tr)^{-1}\mtx{H}_t\pred{\mtx{C}}_t,\\
  \label{eq:kf-assimilation-update-mean}
  \vct{m}_t &= 
  \pred{\vct{m}}_t + \mtx{C}_t\mtx{H}_t\tr\mtx{R}_t^{-1}(\vct{y}_t - \mtx{H}_t\pred{\vct{m}}_t).
\end{align}
\end{subequations}
\Acf{enkf} methods are a class of \acp{letf} which use an assimilation update consistent with the \ac{kf} updates in \cref{eq:kf-assimilation-update} for linear-Gaussian \acp{ssm} in the limit of an infinite ensemble, in effect replacing the predictive mean $\pred{\vct{m}}_t$ and covariance $\pred{\mtx{C}}_t$ with ensemble estimates. The use of an ensemble representation rather than the full means and covariances used in the \ac{kf} both gives a significant computational gain (by avoiding the need to store and perform operations on the full covariance matrices) while also allowing application of the approach to \acp{ssm} with non-linear state updates via the prediction update in \cref{eq:empirical-prediction-update}.

The originally proposed \ac{enkf} method \citep{evensen1994sequential,burgers1998analysis} generates simulated observations from the observation model in \cref{eq:linear-gaussian-observation} for each predictive ensemble member to form a Monte Carlo estimate of the $\mtx{R}_t+\mtx{H}_t\pred{\mtx{C}}_t\mtx{H}_t\tr$ term in \cref{eq:kf-assimilation-update-covariance}. Although simple to implement, the introduction of artificial observation noise adds an additional source of variance which can be significant for small ensemble sizes. This additional variance can be eliminated by the use of \emph{square-root} \ac{enkf} variants \citep{anderson2001ensemble,bishop2001adaptive,whitaker2002ensemble} which typically giving more stable and accurate filtering for small ensemble sizes. 

Of particular interest here is the \ac{etkf} proposed by \citet{bishop2001adaptive}, with this approach particularly efficient in the regime of interest where the ensemble size $\cns{P}$ is much smaller than the state and observation dimensionalities. As we will use a localised variant of the \ac{etkf} as a baseline in the numerical experiments in \cref{sec:experiments} we outline the \ac{etkf} algorithm in \cref{app:ensemble-transform-kalman-filter} and show how it can be expressed in the form of the \ac{letf} assimilation update in \cref{eq:letf-assimilation-update}.

\subsection{Particle filters}\label{subsec:particle-filters}

Particle filtering offers an alternative \ac{letf} approach that gives consistent estimates of the filtering distributions as $\cns{P} \to \infty$ for the non-Gaussian case. The \ac{pf} assimilation update transforms the empirical approximation to the predictive distribution $\pred{\pi}_{t}$ in \cref{eq:empirical-predictive-distribution} to an empirical approximation of the filtering distribution $\pi_{t}$ by attaching importance weights to the predictive ensemble
\begin{equation}\label{eq:bootstrap-particle-weights}
  \tilde{\rvar{w}}\pss{p}_{t} = g_{t}(\vct{y}_{t}\gvn\pred{\rvct{x}}\pss{p}_{t}),
  ~~
  \rvar{w}\pss{p}_{t} = \frac
    {\tilde{\rvar{w}}\pss{p}_{t}}
    {\sumrange{q}{1}{\cns{P}} \tilde{\rvar{w}}\pss{q}_{t}}
  ~~ \forall p \in \range{\cns{P}},
  \quad
  \pi_{t}(\dr\vct{x}) \approx 
  \sumrange{p}{1}{\cns{P}} \rvar{w}\pss{p}_{t} \delta_{\pred{\rvct{x}}\pss{p}_{t}}(\dr\vct{x}).
\end{equation}
Directly iterating this importance weighting scheme, at each time index propagating the ensemble forward in time according to \cref{eq:empirical-prediction-update} and incrementally updating a set of (unnormalised) importance weights gives an algorithm termed \emph{sequential importance sampling}. While appealingly simple, sequential importance sampling requires an exponentially growing ensemble size as the number of observation times $\cns{T}$ increases. The key additional step in particle filtering is to resample the particle ensemble according to the importance weights between prediction updates. That is the filtering distribution ensemble at time index $t$ is defined in terms of the corresponding predictive distribution ensemble as
\begin{equation}\label{eq:pf-assimilation-update}
  \rvct{x}\pss{p}_{t} = \sumrange{q}{1}{\cns{P}} \rvar{r}\pss{p,q}_{t} \pred{\rvct{x}}\pss{\mkern1mu q}_{t}
  \quad \forall p \in \range{\cns{P}},
\end{equation}
where $\rvar{r}\pss{\srange{\cns{P}},\srange{\cns{P}}}_t \in \lbrace 0,1 \rbrace^{\cns{P}\times\cns{P}}$ are a set of binary random variables satisfying
\begin{equation}\label{eq:resampling-variable-conditions}
  \sumrange{q}{1}{\cns{P}}
  \rvar{r}\pss{p,q}_{t} = 1,
  ~~
  \expc\bigg[
    \sumrange{q}{1}{\cns{P}}
    \rvar{r}\pss{q,p}_{t}
    \gvn
    \rvar{w}\pss{p}_{t} = w\pss{p}_{t}
  \bigg] =
  \cns{P} w\pss{p}_{t}
  \quad \forall p \in \range{\cns{P}}.
\end{equation}
This has the effect of removing particles with low weights from the ensemble and so ensures computational effort is concentrated on the most plausible particles. There are multiple algorithms available for generating random variables $\rvar{r}\pss{\srange{\cns{P}},\srange{\cns{P}}}_t$ satisfying \cref{eq:resampling-variable-conditions} - see for example the reviews in \citep{douc2005comparison,hol2006resampling,gerber2019negative}. Distributed versions of particle filters have recently been proposed and analyzed \citep{bolic2005resampling,verge2015parallel,whiteley2016role,sen2019particle,lee2015forest}. 

The iterated application of prediction updates according to \cref{eq:empirical-prediction-update} and resampling assimilation updates according to \cref{eq:pf-assimilation-update} together defines the \emph{bootstrap} \ac{pf} algorithm. Although simple, the bootstrap \ac{pf} algorithm does not exploit all the information available at each time index -- specifically the prediction update in \cref{eq:empirical-prediction-update} does not take in to account future observations. Alternative \ac{pf} schemes can be employed which use prediction updates which take in to account future observations. Although such schemes typically express the resulting particle weights in terms of the state transition densities we describe in \cref{app:alt-particle-filter-proposals} how they can be implemented in \acp{ssm} with intractable transition densities.

While adjusting the prediction update can significantly improve performance compared to the bootstrap \ac{pf} for a fixed ensemble size, when applied to systems with high state and observation dimensionalities these \ac{pf} methods will still tend to suffer from weight degeneracy. In particular, even when using `locally optimal' updates in a simple linear-Gaussian model, the resulting \ac{pf} has been shown to still generally require an ensemble size which still grows exponentially with the dimension of the observation space to avoid weight degeneracy \citep{snyder2008obstacles,snyder2015performance}.

\subsection{Ensemble transform particle filters}\label{subsec:ensemble-transform-particle-fitlers}

Although typically the resampling variables in \ac{pf} assimilation updates are generated independently of the predictive ensemble particle values given the weights, this is not required. \citet{reich2013nonparametric} exploited this flexibility to propose an alternative particle filtering approach termed the \acf{etpf} which uses \ac{ot} methods to compute a resampling scheme which minimises the expected distances between the particles before and after resampling.

A valid resampling scheme can be parametrised by a set of resampling probabilities $\rho_{t}\pss{\srange{\cns{P}},\srange{\cns{P}}} \in [0,1]^{\cns{P}\times\cns{P}}$ with $\rho_{t}\pss{p,q} = \prob\left(\rvar{r}_{t}\pss{p,q} = 1 \gvn \rvar{w}\pss{q}_{t} = w\pss{q}_{t}\right)$ satisfying
\begin{equation}\label{eq:resampling-probability-conditions}
  \sumrange{q}{1}{\cns{P}} \rho_{t}\pss{p,q} = 1,
  \quad
  \sumrange{q}{1}{\cns{P}} \rho_{t}\pss{q,p} = \cns{P} w\pss{p}_{t}
  ~~ \forall p \in \range{\cns{P}}.
\end{equation}
A simple choice satisfying \cref{eq:resampling-probability-conditions} is $\rho\pss{p,q}_{t} = w_{t}\pss{q} ~~\forall p \in \range{\cns{P}},\, q \in \range{\cns{P}}$ with this corresponding to the probabilities used in standard \ac{pf} resampling schemes.

If we denote the set of resampling probabilities satisfying \cref{eq:resampling-probability-conditions} for a given set of weights $w_{t}\pss{\srange{\cns{P}}}$ by $\set{R}(w_{t}\pss{\srange{\cns{P}}})$ and the realisations of the predictive particles $\pred{\rvct{x}}\pss{\srange{\cns{P}}}_t$ at time index $t$ by $\pred{\vct{x}}\pss{\srange{\cns{P}}}_t$, \citet{reich2013nonparametric} instead proposed to compute the resampling probabilities as the solution to the optimal transport problem
\begin{equation}\label{eq:etpf-optimal-transport-problem}
  \rho_{t}\pss{\srange{\cns{P}},\srange{\cns{P}}}
  =
  \argmin_{\varrho\pss{\srange{\cns{P}},\srange{\cns{P}}} \in \set{R}(w_{t}\pss{\srange{\cns{P}}})}
  \sumrange{p}{1}{\cns{P}} \sumrange{q}{1}{\cns{P}} \varrho\pss{p,q} \left| \pred{\vct{x}}\pss{p}_t - \pred{\vct{x}}\pss{q}_t \right|_2^2.
\end{equation}

The optimal transport problem can be posed as a linear program and efficiently solved using the network simplex algorithm \citep{orlin1997polynomial} with a computational complexity of order $\otilde(\cns{P}^3)$. While the resulting resampling probabilities could then be used to generate binary variables $\rvar{r}\pss{\srange{\cns{P}},\srange{\cns{P}}}_t$ and the standard \ac{pf} resampling assimilation update in \cref{eq:pf-assimilation-update} applied, \citet{reich2013nonparametric} instead proposes to use the resampling probabilities to directly update the particles as follows
\begin{equation}\label{eq:etpf-assimilation-update}
  \rvct{x}\pss{p}_{t} = \sumrange{q}{1}{\cns{P}} \rho\pss{p,q}_{t} \pred{\rvct{x}}\pss{\mkern1mu q}_{t}
  \quad \forall p \in \range{\cns{P}}.
\end{equation}
For $\cns{P} \to \infty$ this assimilation update remains consistent as, due to properties of the optimal transport problem solution, the resampling probabilities tend to binary $\lbrace 0,1 \rbrace$ values \citep[Theorem 1]{reich2013nonparametric} and thus \cref{eq:etpf-assimilation-update} becomes equivalent to updating using realisations of the binary random variables.

While the \ac{etpf} does not in itself help overcome the weight degeneracy issue, the deterministic and distance minimising nature of the \ac{etpf} update naturally lends itself to spatial localisation approaches which can help overcome the poor scaling of \acp{pf} with dimensionality, as will be discussed in the following section.
\section{Spatial models and local ensemble filters}
\label{sec:spatial-systems}

Our particular focus in this article is on filtering in models of spatially-extended dynamical systems. Let $\set{S}$ be a $\cns{D}$-dimensional compact metric space equipped with distance function $\metric{d} : \set{S} \times \set{S} \to [0,\infty)$, representing the spatial domain the state of the modelled system is defined over, and $\set{Z} \subseteq \reals^{\cns{N}}$ be the space the state variables at each spatial coordinate in $\set{S}$ take values in. The state-space of the system is then a function space $\set{Z}^\set{S}$ with the state at each time index  $\rvct{z}_t : \set{S} \to \set{Z}$ a spatial field. The dynamics of the system will typically be modelled by a set of \acp{spde}, with $\rvct{z}_{\srange{\cns{T}}}$ then corresponding to a solution of these equations at $\cns{T}$ times, given an initial state sampled from some distribution.

In practice in most problems we cannot solve the \ac{spde} model exactly and instead use numerical integration schemes to generate approximate solutions. The states are assumed to be restricted to a function space with a fixed dimensional representation, with typically a state field $\rvct{z}_t: \set{S} \to \set{Z}$ represented as a linear combination of a finite set of $\cns{M}$ basis functions $\beta_{m} : \set{S} \to \reals$
\begin{equation}\label{eq:spatial-field-basis-expansion}
  \rvct{z}_t(s) = \sum_{m\in\srange{\cns{M}}}\rvct{x}_{t,m} \beta_m(s)
  \quad\forall t\in\range{\cns{T}},\,s\in\set{S},
\end{equation} 
with coefficients $\rvct{x}_{t,m} \in \set{Z} ~~\forall m \in \range{\cns{M}}$. For the purposes of inference we will therefore consider the state space to be a vector space $\set{X} = \set{Z}^{\cns{M}} \subseteq \reals^{\cns{M}\cns{N}}$ with state vectors consisting of the concatenation of the basis function coefficients.

Typically the basis functions will be defined by partitioning the spatial domain $\set{S}$ in to a \emph{mesh} of polytopic spatial elements, for example triangles or quadrilaterals for $\cns{D} = 2$. The vertices of these polytopes (and potentially additional points such as the midpoints of edges) define a collection of $\cns{M}$ \emph{nodes} with spatial locations $s_{\srange{\cns{M}}}$. Typically each node is associated with a basis function $\beta_m$ satisfying
\begin{equation}\label{eq:spatial-basis-function-conditions}
  \beta_m(s_m) = 1,
  \quad
  \beta_m(s_n) = 0 
  \quad
  \forall 
  m \in \srange{\cns{M}}, \,
  n \in \srange{\cns{M}},\, n \neq m,
\end{equation}
which combined with \cref{eq:spatial-field-basis-expansion} implies that $\rvct{z}_t(s_m) = \rvar{x}_{t,m} ~\forall m\in\range{\cns{M}}$.

We will assume that there are $\cns{L}$ observations $\rvar{y}_{t,\srange{\cns{L}}}$ at every time point, each of dimension $\cns{K}$, with the overall observation vector $\rvct{y}_t$ then a length $\cns{K}\cns{L}$ vector
\begin{equation}
  \rvct{y}_t\tr = 
  [\rvct{y}_{t,1}\tr ~ \rvct{y}_{t,2}\tr ~ \cdots ~ \rvct{y}_{t,\cns{L}}\tr]
  \quad\forall t\in\range{\cns{T}}.
\end{equation}
We also assume that $\rvar{y}_{t,l} \perp \rvar{y}_{t,m} \gvn \rvar{x}_{t} ~~\forall l\neq m$ i.e. the observations are conditionally independent given the state and that each observation $\rvar{y}_{t,l}$ depends only on the value of the state field $z_t$ at a fixed spatial location $s^{\textrm{o}}_l$. Together these two assumptions mean we can express the logarithm of the observation density as
\begin{equation}\label{eq:log-likelihood-sum}
  \log g_t(\vct{y}_t \gvn \vct{x}_t) =
  \sum_{l\in\srange{\cns{L}}} \log g_{t,l}(y_{t,l} \gvn \vct{z}_t(s^{\textrm{o}}_l))
  \quad\forall t\in\range{\cns{T}}.
\end{equation}

\subsection{Decay of spatial correlations}

The combination of high state and observation space dimensionalities, and low feasible ensemble sizes, make filtering in spatial \acp{ssm} a significant computational challenge. Fortunately \acp{ssm} of spatially extended systems often also exhibit a favourable \emph{decay of spatial correlations} property which can be exploited to make approximate filtering more tractable by performing local updates to the particles.

If we assume the spatial field $\rvct{z}_{t}$ is defined as in \cref{eq:spatial-field-basis-expansion} and $\rvct{x}_t$ is distributed according to the filtering distribution $\pi_t$ then the spatial correlation function $c_{t,f} : \set{S} \times \set{S} \to [0,1]$ of a square integrable function $f \in \set{L}^2$ is defined as
\begin{equation}
  c_{t,f}(s, s') =
  \frac
    {
      \expc[f(\rvct{z}_{t}(s))f(\rvct{z}_{t}(s'))] - 
      \expc[f(\rvct{z}_{t}(s))]\expc[f(\rvct{z}_{t}(s'))]
    }
    {
      \lpa
      \expc\left[(f(\rvct{z}_{t}(s)) - \expc[f(\rvct{z}_{t}(s))])^2\right]
      \expc\left[(f(\rvct{z}_{t}(s')) - \expc[f(\rvct{z}_{t}(s'))])^2\right]
      \rpa^{\frac{1}{2}}
    },
\end{equation}
and  the \emph{maximal spatial correlation function} as $\bar{c}_{t}(s, s') = \sup_{f\in\set{L}^2} c_{t,f}(s, s')$.

The decay of spatial correlations property can then be stated as
\begin{equation}\label{eq:decay-of-spatial-correlations}
  \bar{c}_{t}(s, s') \to 0
  \quad\textrm{as}\quad
  d(s,s') \to \infty
  \quad
  \forall s \in \set{S}, s' \in\set{S},
\end{equation}
which indicates that the dependence between state variables at distinct spatial locations decays to zero as the distance between the locations increases. 

While it will typically not be possible to analytically verify \cref{eq:decay-of-spatial-correlations} holds exactly, it has been empirically observed that models of spatially extended systems in which the underlying dynamics are governed by local interactions between the state variables exhibit an approximate decay of correlations property. In particular weak long-range spatial correlations are a defining feature of \emph{spatio-temporal chaos} \citep{hunt2007efficient} with many spatial models of interest, such as the atmospheric models used in \ac{nwp}, exhibiting such behaviour.

\subsection{Local linear ensemble transform filters}

For \acp{ssm} exhibiting a decay of spatial correlations property, localising the \ac{letf} assimilation update in \cref{eq:letf-assimilation-update}, as proposed by \citet{cheng2015assimilating}, can offer significant performance gains compared to algorithms employing global updates. Rather than using a single set of transform coefficients $\rvar{a}_t\pss{\srange{\cns{P}},\srange{\cns{P}}}$ for the assimilation update, $\cns{M}$ sets of coefficients $\rvar{a}_{t,\srange{\cns{M}}}\pss{\srange{\cns{P}},\srange{\cns{P}}}$ are defined, one for each spatial mesh node location $s_{\srange{\cns{M}}}$ with the assimilation update then
\begin{equation}\label{eq:local-letf-assimilation-update}
  \rvct{x}\pss{p}_{t,m} = \sum_{q\in\srange{\cns{P}}} \rvar{a}_{t,m}\pss{p,q} \pred{\rvct{x}}\pss{q}_{t,m}
  \quad
  \forall p\in\range{\cns{P}},\,\forall m\in\range{\cns{M}}.
\end{equation}
As previously mentioned, the global \ac{letf} update in \cref{eq:letf-assimilation-update} restricts the filtering ensemble members $\rvct{x}\pss{\srange{\cns{P}}}_t$ to lie in the $\cns{P}$ dimensional linear subspace of $\set{X}$ spanned by the predictive ensemble $\pred{\rvct{x}}\pss{\srange{\cns{P}}}_t$. When $\set{X}$ is high-dimensional, as is generally the case in spatially extended models, this can be highly restrictive. 

The local \ac{letf} update in \cref{eq:local-letf-assimilation-update} overcomes this restriction of the global \ac{letf} update, with the filtering ensemble members now formed from local linear combinations of the predictive ensemble members and thus no longer constrained to a $\cns{P}$ dimensional linear subspace. In particular for models exhibiting a decay of correlations property, the state variables at each mesh node can be updated using coefficients computed using only the subset of observations which are within some localisation radius of the mesh node while still retaining accuracy. 

Local variants of the \ac{enkf} \citep{houtekamer1998data,hamill2001distance} are the prototypical examples of local \acp{letf}, and have been successfully used to perform filtering in large complex spatio-temporal models including operational ensemble \ac{nwp} systems \citep{bowler2009local}. In \cref{app:ensemble-transform-kalman-filter} we briefly introduce a local variant of the \ac{etkf} \citep{hunt2007efficient} which we use as a baseline in the numerical experiments.

\subsection{Local particle filters}

It has been speculated that spatial localisation may be key to achieving useful results from \acp{pf} in large spatio-temporal models \citep{morzfeld2017collapse} based on its importance to the success of \ac{enkf} methods in such models. In \citet{farchi2018comparison} the authors systematically compare a wide range of localised \ac{pf} and related algorithms which have been proposed in the literature including localised variants of the \ac{etpf} which we will discuss in the following subsection. Below we briefly introduce concepts from a local \ac{pf} algorithm proposed by \citet{penny2015local} which are relevant to this article, however we refer readers to \citet{farchi2018comparison} for a much more extensive review.

For the standard \ac{pf}, the logarithms of the unnormalised particle weights are
\begin{equation}\label{eq:global-particle-log-weights}
  \log\tilde{\rvar{w}}\pss{p}_{t} = 
  \sum_{l\in\srange{\cns{L}}} 
    \log g_{t,l}\left(
      \vct{y}_{t,l}\gvn\pred{\rvct{z}}\pss{p}_{t}(s_l^{\textrm{o}})
    \right)
  \quad \forall p \in \range{\cns{P}}.
\end{equation}
i.e. a summation of contributions due to the observations at all locations $s_{\srange{\cns{L}}}^{\textrm{o}}$. 

For a model exhibiting a decay of spatial correlations property we would expect that only a local subset of observations should have a strong influence on the distribution of the state variables at each mesh node. We can formalise this intuition into a concrete approach for computing local particle weights via the use of a \emph{localisation function} $\ell_r : [0, \infty) \to [0, 1]$ and localisation radius $r$ satisfying
\begin{equation}\label{eq:localisation-function-conditions}
  \ell_r(0) = 1, \quad \ell_r(d) = 0 \quad\forall d > r > 0.
\end{equation}
Local unnormalised weights for each mesh node can then be defined
\begin{equation}\label{eq:local-particle-log-weights}
  \log\tilde{\rvar{w}}\pss{p}_{t,m} =
  \sum_{l\in\srange{\cns{L}}} 
    \log g_{t,l}\left(
      \vct{y}_{t,l}\gvn\pred{\rvct{z}}\pss{p}_{t}(s^{\textrm{o}}_l)
    \right)
    \ell_r(d(s_m, s_l^{\textrm{o}}))
  \quad \forall p \in \range{\cns{P}},\, m\in\range{\cns{M}},
\end{equation}
and corresponding local normalised weights
\begin{equation}\label{eq:local-particle-normalised-weights}
  \rvar{w}\pss{p}_{t,m} = 
  \frac{\tilde{\rvar{w}}\pss{p}_{t,m}}{ 
    \sum_{q\in\srange{\cns{P}}} \tilde{\rvar{w}}\pss{q}_{t,m}}
  \quad \forall p \in \range{\cns{P}},\, m\in\range{\cns{M}}.
\end{equation}
This formulation for the local particle weights has the desired property of using only a local subset of observations to update the state variables at each mesh node (with the terms in the sum zero when $d(s_m, s_l^{\textrm{o}}) > r$).

Typical choices for the localisation function include the \emph{uniform} or top-hat function $\ell_r(d) = \indc{[0,r]}(d)$ and the \emph{triangular} function $\ell_r(d) = (1 - \frac{d}{r})\indc{[0,r]}(d)$. In this article we exclusively use the smooth and compactly supported 5\textsuperscript{th} order piecewise rational function proposed by \citet{gaspari1999construction} and defined as
\begin{equation}\label{eq:gaspari-and-cohn-localisation}
  \ell_r(d) = \begin{cases}
    -8\frac{d^5}{r^5} + 8\frac{d^4}{r^4} + 5\frac{d^3}{r^3} - \frac{20}{3}\frac{d^2}{r^2} + 1 & 0 \leq d < \frac{r}{2} \\
    \frac{8}{3}\frac{d^5}{r^5} - 8\frac{d^4}{r^4} + 5\frac{d^3}{r^3} + \frac{20}{3}\frac{d^2}{r^2} - 10\frac{d}{r} + 4 - \frac{1}{3}\frac{r}{d} & \frac{r}{2} \leq d < r \\
  \end{cases}.
\end{equation}

\begin{figure}[!t]

  \centering
  \begin{subfigure}[t]{.48\linewidth}
  \includegraphics[width=.49\linewidth]{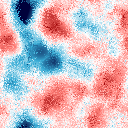}
  \includegraphics[width=.49\linewidth]{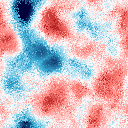}
  \caption{Independent resampling at each node.}
  \label{sfig:gaussian-field-lpf-per-gridpoint}
  \end{subfigure}
  ~
  \begin{subfigure}[t]{.48\linewidth}
  \includegraphics[width=.49\linewidth]{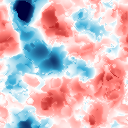}
  \includegraphics[width=.49\linewidth]{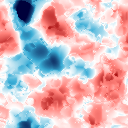}
  \caption{Coupled resampling and smoothing.}
  \label{sfig:gaussian-field-lpf-poterjoy-coupled-smoothed}
  \end{subfigure}
  
  \caption{Examples of local \ac{pf} assimilation updates applied to the same spatial Gaussian process model as Figure \ref{fig:gaussian-field-resampling-schemes-examples}.}
  \label{fig:gaussian-field-resampling-schemes-poterjoy}
  
\end{figure}

\citet{penny2015local} propose a local \ac{pf} algorithm which uses local particle weights defined as in \cref{eq:local-particle-normalised-weights} for the specific case of a Gaussian observation density and uniform localisation function $\ell_r(d) = \indc{[0,r]}(d)$. The local weights are used to generate binary resampling variables $\rvar{r}\pss{\srange{\cns{P}},\srange{\cns{P}}}_{t,\srange{\cns{M}}}$ for each mesh node satisfying
\begin{equation}\label{eq:local-resampling-variable-conditions}
  \sumrange{q}{1}{\cns{P}}
  \rvar{r}\pss{p,q}_{t,m} = 1,
  ~~
  \expc\bigg[
    \sumrange{q}{1}{\cns{P}}
    \rvar{r}\pss{q,p}_{t,m}
    \gvn
    \rvar{w}\pss{p}_{t,m} = w\pss{p}_{t,m}
  \bigg] =
  \cns{P} w\pss{p}_{t,m}
  \quad \forall p \in \range{\cns{P}},\,m\in\range{\cns{M}}.
\end{equation}
Generating the resampling variables for each mesh node independently means the state variables at adjacent mesh nodes for a post-resampling particle will typically originate from different prior particles, tending to lead to highly discontinuous and noisy spatial fields. An example of this is shown in \cref{sfig:gaussian-field-lpf-per-gridpoint} which show examples of the posterior state field samples generated using independent resampling at each mesh node with local weights for the smooth spatial Gaussian process example encountered previously in Figure \ref{fig:gaussian-field-resampling-schemes-examples}.

To ameliorate the issues associated within using independent resampling variables, it is proposed in \citet{penny2015local} to use a variant of the \emph{systematic resampling} scheme \citep{douc2005comparison} often used as variance reduction method in standard \ac{pf} algorithms. A single random standard uniform variable is used to generate the resample variables for all mesh nodes, resulting in per-node sets of resampling variables which each satisfy the marginal requirements in \cref{eq:local-resampling-variable-conditions} while also being strongly correlated to the resampling variables for other nodes. 
The correlation introduced between the resampling variables when using this `coupled resampling' scheme significantly reduces but does not eliminate the introduction of discontinuities into the resampled fields. 

Rather than directly use these resampling variables in a local equivalent to the \ac{pf} assimilation update in \cref{eq:pf-assimilation-update}, \citet{penny2015local} instead propose to use a `smoothed' update which uses a weighted average of the resampling variables at the current mesh node and all neighbouring nodes to update the particles values at each node. 
\cref{sfig:gaussian-field-lpf-poterjoy-coupled-smoothed} shows examples of posterior state fields samples computed using this smoothed assimilation update with the resampling variables generated using the coupled scheme. The previously observed discontinuities are now removed, however the samples still remain significantly less smooth than the true state used to generate the observations (\cref{sfig:gaussian-field-true-state}) and prior samples (\cref{sfig:gaussian-field-predictive-dist-samples}). 

\subsection{Local ensemble transform particle filter}

While techniques such as the smoothed and coupled resampling update used in \citet{penny2015local} can help reduce the introduction of spatial discontinuities, the resampling variables $\rvar{r}_{t,\srange{\cns{M}}}\pss{\srange{\cns{P}},\srange{\cns{P}}}$ are still calculated without taking into account the values of the predictive particles $\pred{\rvct{x}}\pss{\srange{\cns{P}}}_t$ values other than via the local particle weights. The \ac{etpf} assimilation update discussed in \cref{subsec:ensemble-transform-particle-fitlers} explicitly tries to minimise a distance between the values of the transformed and pre-update particles and does not require introducing any randomness and so is a natural candidate for a local \acp{pf} with improved spatial smoothness properties.

\citet{cheng2015assimilating} proposed a localised variant of the \ac{etpf} as a particular instance of their \ac{letf} framework. Local particle weights are calculated as in \cref{eq:local-particle-log-weights,eq:local-particle-normalised-weights} for each mesh node, and a set of \ac{ot} problems solved
\begin{equation}\label{eq:local-etpf-optimal-transport-problem}
  \rho_{t,m}\pss{\srange{\cns{P}},\srange{\cns{P}}}
  =
  \argmin_{\varrho\pss{\srange{\cns{P}},\srange{\cns{P}}} \in \set{R}(\rvar{w}_{t,m}\pss{\srange{\cns{P}}})}
  \sumrange{p}{1}{\cns{P}} 
  \sumrange{q}{1}{\cns{P}} \varrho\pss{p,q} \rvar{c}\pss{p,q}_{t,m}
  \quad
  \forall m \in\range{\cns{M}}.
\end{equation}
Here the transport cost terms $\rvar{c}\pss{\srange{\cns{P}},\srange{\cns{P}}}_{t,\srange{\cns{M}}}$ are analogous to the inter-particle Euclidean distances used in \cref{eq:etpf-optimal-transport-problem}. Rather than compute global transport costs based on distances  between the state variables values at points across the full spatial domain, \citet{cheng2015assimilating} proposed to compute localised transports costs for each mesh node index $m \in \range{\cns{M}}$ by integrating a distance between the state variables values against a localisation function centred at the mesh node location $s_m$ and with support on points $s \in \set{S} : d(s, s_m) < r'$
\begin{equation}\label{eq:local-transport-costs-fields}
  \rvar{c}\pss{p,q}_{t,m} = 
  \int_{\set{S}} 
    \left| \pred{\rvct{z}}\pss{p}_t(s) - \pred{\rvct{z}}\pss{q}_t(s) \right|_2^2
    \ell'_{r'}(d(s_m, s))
  \,\dr s
  \quad
  \forall m\in\range{\cns{M}},
  p\in\range{\cns{P}},
  q\in\range{\cns{P}}.
\end{equation}
The localisation function $\ell'_{r'}$ and localisation radius $r'$ are denoted with primes here to emphasise they may be different from those used for the local weights computation. 
A more pragmatic definition of the localised transport costs is
\begin{equation}\label{eq:local-transport-costs-coefficients}
  \rvar{c}\pss{p,q}_{t,m} = 
  \sumrange{n}{1}{\cns{M}}
    \left| \pred{\rvct{x}}\pss{p}_{t,n} - \pred{\rvct{x}}\pss{q}_{t,n}\right|_2^2
    \ell'_{r'}(d(s_m, s_n))
    \quad
    \forall m\in\range{\cns{M}},
    p\in\range{\cns{P}},
    q\in\range{\cns{P}}.
\end{equation}
In the common case of a rectilinear mesh with equal spacing between the nodes across the domain, the summation in \cref{eq:local-transport-costs-coefficients} can be seen, as a quadrature approximation to the integral in \cref{eq:local-transport-costs-fields} up to a constant multiplier which does not affect the \ac{ot} solutions.

If the localisation functions $\ell_r$ and $\ell'_{r'}$ are smooth, then both the local weights $\rvar{w}\pss{\srange{\cns{P}}}_{t,\srange{\cns{M}}}$ and local transport costs $\rvar{c}\pss{\srange{\cns{P}},\srange{\cns{P}}}_{t,\srange{\cns{M}}}$ will vary smoothly as functions of the mesh node locations $s_{\srange{\cns{M}}}$. However, the solutions $\rho\pss{\srange{\cns{P}},\srange{\cns{P}}}_{t,\srange{\cns{M}}}$ to the linear programs defined by the local optimal transport problems in \cref{eq:local-etpf-optimal-transport-problem} will not vary smoothly with the mesh node locations $s_{\srange{\cns{M}}}$ even if the local weights and transport costs do. This can be seen in the spatial Gaussian process example in \cref{fig:gaussian-field-resampling-schemes-examples}, with the local \ac{etpf} scheme used to compute the posterior samples illustrated in \cref{sfig:gaussian-field-letpf-per-gridpoint}. Although less apparent than the discontinuities in \cref{sfig:gaussian-field-block-pf-16x16}, the fields in \cref{sfig:gaussian-field-letpf-per-gridpoint} still show spatial artefacts due to the non-smooth variation of the \ac{ot} solutions.

\begin{figure}[!t]

  \centering
  \begin{subfigure}[b]{.48\linewidth}
  \includegraphics[width=.49\linewidth]{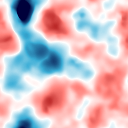}
  \includegraphics[width=.49\linewidth]{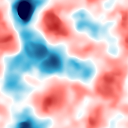}
  \caption{Regularisation coefficient $\lambda = 10^{-3}$.}
  \label{sfig:gaussian-field-letpf-entropic-1e-3}
  \end{subfigure}
  ~
  \begin{subfigure}[b]{.48\linewidth}
  \includegraphics[width=.49\linewidth]{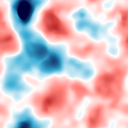}
  \includegraphics[width=.49\linewidth]{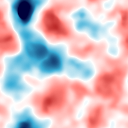}
  \caption{Regularisation coefficient $\lambda = 10^{-2}$.}
  \label{sfig:gaussian-field-letpf-entropic-1e-2}
  \end{subfigure}
  
  \caption{Examples of the \citet{cheng2015assimilating} local \ac{etpf} assimilation update applied to the same spatial Gaussian process model as Figure \ref{fig:gaussian-field-resampling-schemes-examples} using entropically regularised \ac{ot} maps for different values of the regularisation coefficient $\lambda$.}
  \label{fig:gaussian-field-letpf-entropic}
  
\end{figure}

One option to increase the smoothness of the update is to regularise the \ac{ot} problems. In particular the \emph{entropically regularised} \ac{ot} problems defined by
\begin{equation}\label{eq:local-etpf-regularised-optimal-transport-problem}
  \rho_{t,m}\pss{\srange{\cns{P}},\srange{\cns{P}}}
  =
  \argmin_{\varrho_{t,m}\pss{\srange{\cns{P}},\srange{\cns{P}}} \in \set{R}(\rvar{w}_{t,m}\pss{\srange{\cns{P}}})}
  \sumrange{p}{1}{\cns{P}} 
  \sumrange{q}{1}{\cns{P}} \left(
      \varrho\pss{p,q}_{t,m} \rvar{c}\pss{p,q}_{t,m}
      + \lambda \varrho\pss{p,q}_{t,m} (\log \varrho\pss{p,q}_{t,m} - 1)
      \right),
\end{equation}
for some positive regularisation coefficient $\lambda$ have a unique optimal solution which smoothly varies as a function of the local weights $\rvar{w}_{t,m}\pss{\srange{\cns{P}}}$ and transport costs $\rvar{c}\pss{\srange{\cns{P}},\srange{\cns{P}}}_{t,\srange{\cns{M}}}$ \citep{peyre2018computational} and tends to the solution of the non-regularised problem with the highest entropy as $\lambda \to 0$. Further the entropically regularised problems can be efficiently iteratively solved using Sinkhorn--Knopp iteration \citep{sinkhorn1967concerning,cuturi2013sinkhorn} with complexity $\otilde(\cns{P}^2)$ per problem \citep{altschuler2017near}.

\cref{sfig:gaussian-field-letpf-entropic-1e-3,sfig:gaussian-field-letpf-entropic-1e-2} show examples of posterior fields samples computed using entropically regularised local \ac{etpf} updates for two regularisation coefficients $\lambda$. It can be seen that introducing entropic regularisation increases the smoothness of the updated fields compared to the unregularised samples shown in \cref{sfig:gaussian-field-letpf-per-gridpoint} and that the level of smoothness increases with the regularisation coefficient $\lambda$. 

However the increase in smoothness comes at the cost of a decreased diversity in the post-update particles as $\lambda$ increases - in particular for the $\lambda = 10^{-2}$ case shown in \cref{sfig:gaussian-field-letpf-entropic-1e-2}, the four samples shown appear almost identical. This is a consequence of the assimilation updates in the local \ac{etpf} linearly transforming by the \ac{ot} maps as in \cref{eq:etpf-assimilation-update} as opposed to resampling using binary random variables generated according to the resampling probabilities encoded by the \ac{ot} maps. 
For the regularised \ac{ot} problems in \cref{eq:local-etpf-regularised-optimal-transport-problem}, as the regularisation coefficient $\lambda \to \infty$ we have that $\rho\pss{p,q}_{t,m} \to w_{t,m}\pss{q} ~~\forall p \in \range{\cns{P}},\, q \in \range{\cns{P}},\,m\in\range{\cns{M}}$. In this case applying the local \ac{etpf} assimilation update will tend to assigning the weighted mean of the state variables at each mesh-node to the post-update particles, and thus a lack of diversity or \emph{under-dispersion} in the post-update particles.

\citet{acevedo2017second} proposed a variant of the \ac{etpf} which overcomes this under-dispersion issue when using entropically regularised \ac{ot} maps. For each \ac{ot} map a correction terms is computed which ensures the empirical covariance of the updated particles matches the values that would be obtained using the standard \ac{pf} update. Although this \emph{second-order accurate} \ac{etpf} scheme overcomes the under-dispersion issues when using entropically regularised \ac{ot} maps, in localised variants the correction factors must be computed separately for the \ac{ot} map associated with each mesh node, with the computation of each correction factor having a $\mathcal{O}(\cns{P}^3)$ complexity, potentially negating any gains from using a cheaper Sinkhorn solver for the regularised \ac{ot} problems.

\begin{figure}[!t]

  \centering
  \begin{subfigure}[b]{.48\linewidth}
  \includegraphics[width=.49\linewidth]{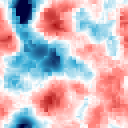}
  \includegraphics[width=.49\linewidth]{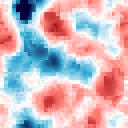}
  \caption{Block size $4 \times 4$.}
  \label{sfig:gaussian-field-letpf-block-32x32-pou}
  \end{subfigure}
  ~
  \begin{subfigure}[b]{.48\linewidth}
  \includegraphics[width=.49\linewidth]{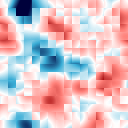}
  \includegraphics[width=.49\linewidth]{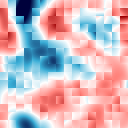}
  \caption{Block size $8 \times 8$.}
  \label{sfig:gaussian-field-letpf-block-16x16-pou}
  \end{subfigure}

  \caption{Examples of the \citet{farchi2018comparison} local block \ac{etpf} assimilation update applied to the same Gaussian process model as Figure \ref{fig:gaussian-field-resampling-schemes-examples}.}
  \label{fig:gaussian-field-letpf-block}
  
\end{figure}

In the review article of \citet{farchi2018comparison} a local \ac{etpf} variant is proposed which computes \ac{ot} maps for \emph{blocks} of state variables rather than for each mesh node individually. 
Computing \ac{ot} maps per-block rather than per-node potentially can give significant computational savings in higher spatial dimensions --- for instance for three dimensional domains, even using cubic blocks which cover just two mesh nodes in each dimension would lead to a reduction in the number of \ac{ot} problems needing to be solved by eight. In the numerical experiments in \citet{farchi2018comparison} it was found however that the accuracy of the local block \ac{etpf} method was highest when using blocks containing just one mesh-node, i.e. corresponding to the local \ac{etpf} scheme of \citet{cheng2015assimilating}. As the state variables in each block are updated independently given the computed per-block \ac{ot} maps, the poorer performance with larger blocks may be at least in part due to the spatially inhomogeneous error introduced at the block boundaries. \cref{fig:gaussian-field-letpf-block} shows examples of posterior state field samples computed using this block \ac{etpf} scheme for the earlier spatial Gaussian process example from \cref{fig:gaussian-field-resampling-schemes-examples} for two different block size; in both the boundaries of the blocks are clearly visible due to the discontinuities introduced in to the fields.
\section{Smooth and scalable local particle filtering}
\label{sec:proposed-method}

Grouping mesh nodes into spatially contiguous blocks and computing \ac{ot} maps per-block rather than per-node as proposed in \citet{farchi2018comparison} is a natural way to reduce the computational cost of local \ac{etpf} assimilation update. However this approach further decreases the smoothness of the updated fields. Here we propose an alternative approach. Rather than computing \ac{ot} maps for disjoint blocks defining a partition of the spatial domain $\set{S}$ we instead `softly' partition $\set{S}$ into patches with overlapping support, computing an \ac{ot} map for each patch and smoothly interpolating between the \ac{ot} maps associated with different patches in the overlaps. The construct we will use to both define the soft partitioning of the domain and interpolation across it is a \emph{partition of unity}.

\subsection{Partitions of unity}\label{subsec:partition-of-unity}

\begin{figure}
  \includegraphics[width=\textwidth]{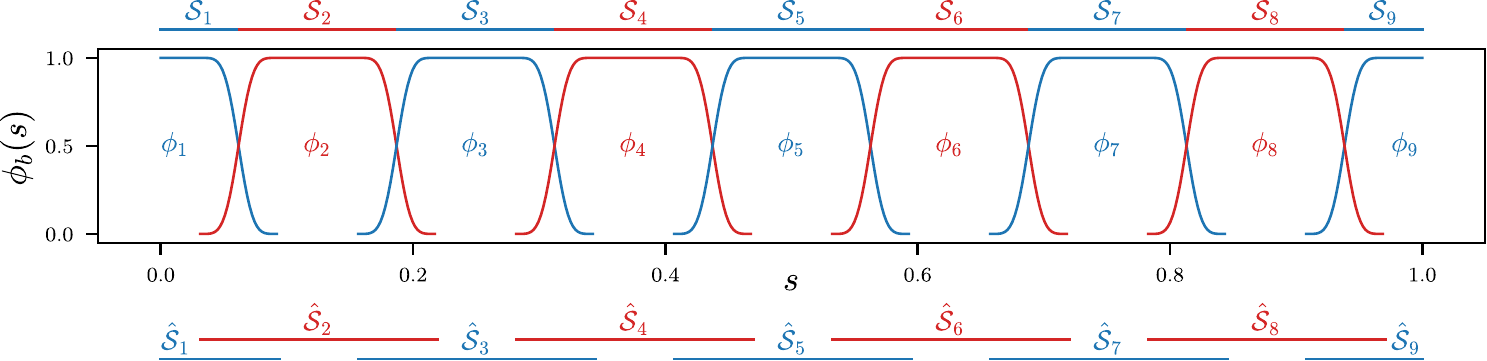}
  \caption{Example smooth partition of unity of a one-dimensional spatial domain $\set{S} = [0,1]$ with nine bump functions $\phi_{\srange{9}}$. The patches $\hat{\set{S}}_{\srange{9}}$ covering $\set{S}$ and which the bump functions have support on are visualised below the plot axes.} 
  \label{fig:pou-example}
\end{figure}

Let $\hat{\set{S}}_{\srange{\cns{B}}}$ be a cover of the spatial domain $\set{S}$ such that $\bigcup_{b=1}^{\cns{B}} \hat{\set{S}}_b = \set{S}$ with each $\hat{\set{S}}_b$ termed a \emph{patch}. We associate a \emph{bump function} $\phi_b : \set{S} \to [0,1] ~~\forall b \in \range{\cns{B}}$ with each patch with $\phi_b(s) = 0 ~~\forall s \notin \hat{\set{S}}_b, b \in \range{\cns{B}}$ and require that
\begin{equation}\label{eq:pou-sum-condition}
  \sumrange{b}{1}{\cns{B}} \phi_b(s) = 1 \quad \forall s \in\set{S}.
\end{equation}
The set of bump functions $\phi_{\srange{\cns{B}}}$ is then termed a \acf{pou} of $\set{S}$. \Acp{pou} are typically used to allow local constructions to be extended globally across a space, for instance an atlas of local charts of a manifold. Generally in such applications the bump functions will be required to be infinitely differentiable. Here we will generally not require such stringent differentiability requirements, however we will informally refer to a \emph{smooth} \ac{pou} for the case where each bump function is of at least class $\set{C}^1$ with continuous derivatives, and to a \emph{hard} \ac{pou} for the case where the cover $\hat{\set{S}}_{\srange{\cns{B}}}$ is exact, i.e. the patches are pairwise disjoint, and so the bump functions are indicators on the patches $\phi_b(s) = \indc{\hat{\set{S}}_b}(s)~\forall b \in\range{\cns{B}}$.

A useful method for constructing a \ac{pou} with specified smoothness properties on an arbitrary spatial domain is via convolution. Specifically, if $\set{S}_{\srange{\cns{B}}}$ is a partition of $\set{S}$ and $\varphi : [0,\infty) \to [0,\infty)$ is a non-negative \emph{mollifier} function satisfying
\begin{equation}
  \int_{\set{S}} \varphi \circ d(s, s') \,\dr s' = 1 \quad \forall s \in\set{S}
\end{equation}
then we can define a \ac{pou} $\phi_{\srange{\cns{B}}}$ on $\set{S}$ by convolving $\varphi$ with the indicators on $\set{S}_{\srange{\cns{B}}}$
\begin{equation}\label{eq:mollified-partition-pou}
  \phi_b(s) = \int_{\set{S}} \indc{\set{S}_b}(s) \varphi \circ d(s, s')\,\dr s'
  \quad \forall b \in \range{\cns{B}}, \, s \in \set{S}.
\end{equation}
The bump functions will then inherit any smoothness properties of the mollifier. Figure \ref{fig:pou-example} shows an example of a smooth \ac{pou} constructed in this manner. 

\subsection{Constructing smooth local linear ensemble transform filters}

We can use a \ac{pou} to define a local \ac{letf} that uses transform coefficients computed for each patch rather than mesh node. We define the per-node transform coefficients $\rvar{a}\pss{\srange{\cns{P}},\srange{\cns{P}}}_{t,m}$ in \cref{eq:local-letf-assimilation-update} in terms of a set of per-patch coefficients $\hat{\rvar{a}}^{\srange{\cns{P}},\srange{\cns{P}}}_{t,\srange{\cns{B}}}$ by
\begin{equation}
  \rvar{a}\pss{p,q}_{t,m} = 
  \sum_{b\in\srange{\cns{B}}} \hat{\rvar{a}}\pss{p,q}_{t,b} \,\phi_b(s_m)
  \quad \forall p\in\range{\cns{P}}, q\in\range{\cns{P}}, m\in\range{\cns{M}}.
\end{equation}
If the set of coefficients $\hat{\rvar{a}}^{\srange{\cns{P}},\srange{\cns{P}}}_{t,b}$ for each patch index $b \in \range{\cns{B}}$ correspond to the elements of a left stochastic matrix such that
\begin{equation}\label{eq:sletf-stochastic-matrix-conditions}
  \hat{\rvar{a}}\pss{p,q}_{t,b} \in [0,1] 
  ~\forall p\in\range{\cns{P}},q\in\range{\cns{P}} 
  \quad\textrm{and}\quad
  \sum_{q\in \srange{\cns{P}}} \hat{\rvar{a}}\pss{p,q}_{t,b} = 1 
  ~\forall p\in\range{\cns{P}},
\end{equation}
then due to the non-negativity and sum to unity properties of the \ac{pou} we have that $\rvar{a}\pss{p,q}_{t,m} \in [0,1] ~\forall p\in\range{\cns{P}},q\in\range{\cns{P}},m \in \range{\cns{M}}$ and
\begin{equation}
  \sum_{q\in \srange{\cns{P}}} \rvar{a}\pss{p,q}_{t,m} = 
  \sum_{b\in\srange{\cns{B}}}  \sum_{q\in \srange{\cns{P}}}
    \hat{\rvar{a}}\pss{p,q}_{t,b}  \,\phi_b(s_m) =  
  \sum_{b\in\srange{\cns{B}}}  \,\phi_b(s_m) =
  1
  ~~ \forall p\in\range{\cns{P}}, \, m\in\range{\cns{M}},
\end{equation}
and so that $\rvar{a}\pss{\srange{\cns{P}},\srange{\cns{P}}}_{t,\srange{\cns{M}}}$ also correspond to the elements of left stochastic matrices. 

The resulting assimilation update in terms of the values of the predictive and filtering distribution state field particles, $\pred{\rvct{z}}\pss{\srange{\cns{P}}}_{t}$ and $\rvct{z}\pss{\srange{\cns{P}}}_{t}$, at the mesh nodes $s_{\srange{\cns{M}}}$ is
\begin{equation}\label{eq:sletf-assimilation-update}
  \rvct{z}\pss{p}_{t}(s_m) =
  \sum_{q\in\srange{\cns{P}}} \sum_{b\in\srange{\cns{B}}} 
    \hat{\rvar{a}}\pss{p,q}_{t,b} \,\phi_b(s_m) \pred{\rvct{z}}\pss{\,q}_{t}(s_m)
  \quad \forall p\in\range{\cns{P}}, m\in\range{\cns{M}}.
\end{equation}
For a smooth \ac{pou} $\phi_{\srange{\cns{B}}}$ the $\cns{P}\times\cns{B}$ spatial fields defined by the pointwise products $\phi_b(s) \pred{\rvct{z}}\pss{\,q}_{t}(s)~\forall b\in\range{\cns{B}},q\in\range{\cns{P}}$ will be smooth functions of the spatial coordinate $s\in\set{S}$ if the predictive distribution state field particles $\pred{\rvct{z}}\pss{\srange{\cns{P}}}_{t}$ are themselves smooth. Each filtering distribution state field particle $\rvct{z}\pss{\srange{\cns{P}}}_{t}$ is then formed as a convex combination of these pairwise product fields, and so will also be smooth if the \ac{pou} and predictive distribution state fields are. This is illustrated for a one-dimensional example in \cref{fig:smooth-letf-1d-example} in \cref{app:smooth-local-letf-assimilation-vis}. 

\subsection{Smooth local ensemble transform particle filtering}

We now consider the specific application of the smooth local \ac{letf} scheme to define a smooth localisation of the \ac{etpf}, with in this case the coefficients $\hat{\rvar{a}}^{\srange{\cns{P}},\srange{\cns{P}}}_{t,\srange{\cns{B}}}$ corresponding to \ac{ot} maps computed for each patch. 
We first define the following notation for the distance between a subset of the spatial domain and a point. 
\begin{equation}\label{eq:distance-set-to-point}
  \underline{d}(\set{S}', s) = \inf_{s' \in \set{S}'} d(s', s)
  \quad \forall s \in \set{S}, \set{S}' \subseteq \set{S}.
\end{equation}
Analogously to the per-node case in \cref{eq:local-particle-log-weights}, the logarithms of the per-patch (unnormalised) particle weights can then be defined by
\begin{equation}\label{eq:per-patch-particle-log-weights}
  \log\tilde{\rvar{w}}\pss{p}_{t,b} =
  \sum_{l\in\srange{\cns{L}}} 
    \log g_{t,l}\left(
      \vct{y}_{t,l}\gvn\pred{\rvct{z}}\pss{p}_{t}(s^{\textrm{o}}_l)
    \right)
    \ell_r(\underline{d}(\hat{\set{S}}_b, s_l^{\textrm{o}}))
  \quad \forall b\in\range{\cns{B}},\,p \in \range{\cns{P}},
\end{equation}
As $\underline{d}(\hat{\set{S}}_b, s_l^{\textrm{o}}) = 0$ if $s_l^{\textrm{o}} \in \hat{\set{S}}_b$ and $\ell_r(0) = 1$ the weighted summation of log observation density terms in \cref{eq:per-patch-particle-log-weights} gives weight one to all the terms corresponding to observations located within a patch. Observations outside a patch but within a distance of less than $r$ are given weights between zero and one, and all observations more than a distance of $r$ from a patch are given zero weight.

Taking inspiration from the per-node case in \cref{eq:local-transport-costs-fields} we could define per-patch transport costs directly in terms of the predictive state fields $\pred{\rvct{z}}_{t}\pss{\srange{\cns{P}}}$
\begin{equation}\label{eq:per-patch-transport-costs-fields}
  \rvar{c}\pss{p,q}_{t,b} = 
  \int_{\set{S}} 
    \left| \pred{\rvct{z}}\pss{p}_t(s) - \pred{\rvct{z}}\pss{q}_t(s) \right|_2^2\,
    \ell'_{r'}(\underline{d}(\hat{\set{S}}_b, s_m))
  \,\dr s
  \quad
  \forall b\in\range{\cns{B}},
  p\in\range{\cns{P}},
  q\in\range{\cns{P}}.
\end{equation}
Although this is defined independently of the spatial discretisation used, evaluating the integrals exactly will often be intractable. Assuming the common case of equally spaced mesh nodes, we propose to define per-patch transport costs as
\begin{equation}\label{eq:per-patch-transport-costs-coefficients}
  \rvar{c}\pss{p,q}_{t,b} = 
  \sum_{m\in\set{M}}
    \left| \pred{\rvct{x}}\pss{p}_{t,m} - \pred{\rvct{x}}\pss{q}_{t,m}\right|_2^2\,
    \indc{\hat{\set{S}}_b}(s_m)
    \quad
    \forall b\in\range{\cns{B}},\,
    p\in\range{\cns{P}},\,
    q\in\range{\cns{P}},
\end{equation}
where $\set{M} \subseteq \range{\cns{M}}$ corresponds to a spatial subsampling of the mesh nodes, e.g. corresponding to every $\cns{K}$\textsuperscript{th} node in each spatial dimension, such that $|\set{M}| \approx \frac{\cns{M}}{\cns{K}^\cns{D}}$. This spatial subsampling is motivated by the observation that if the state fields are spatially smooth then the values at immediately adjacent mesh nodes will typically be very similar and there is therefore minimal loss of information in computing pointwise differences over a subset of, rather than all, mesh nodes. In addition to spatial subsampling we also define the transport costs in \cref{eq:per-patch-transport-costs-coefficients} with the fixed choice of a uniform localisation function $\ell'_{r'}$ with $r' = 0$. Empirically we found varying the choice of $\ell'_{r'}$ and $r'$ for the transport costs had little discernable effect on filtering performance. 

Given per-patch weights $\rvar{w}_{t,\srange{\cns{B}}}\pss{\srange{\cns{P}}}$ and transport costs $\rvar{c}\pss{\srange{\cns{P}},\srange{\cns{P}}}_{t,\srange{\cns{B}}}$ computed as described above, the per-patch linear transform coefficients $\hat{\rvar{a}}_{t,\srange{\cns{B}}}\pss{\srange{\cns{P}},\srange{\cns{P}}}$ are then computed as solutions to the $\cns{B}$ corresponding \ac{ot} problems 
\begin{equation}\label{eq:er-patch-optimal-transport-problems}
  \hat{\rvar{a}}_{t,b}\pss{\srange{\cns{P}},\srange{\cns{P}}}
  =
  \argmin_{\varrho\pss{\srange{\cns{P}},\srange{\cns{P}}} \in \set{R}(\rvar{w}_{t,b}\pss{\srange{\cns{P}}})}
  \sumrange{p}{1}{\cns{P}} 
  \sumrange{q}{1}{\cns{P}} \varrho\pss{p,q} \rvar{c}\pss{p,q}_{t,b}
  \quad
  \forall b \in\range{\cns{B}}.
\end{equation}

We will subsequently refer to instances of this framework as \acp{sletpf}. To define a \ac{sletpf} method for a given spatial \ac{ssm}, we need to specify: a localisation function and radius $\ell_r$ and $r$ to compute the local weights; the set of mesh nodes $\set{M}$ to use in computing the local transport costs; a \ac{pou} of the spatial domain. 

For the \ac{sletpf} local weight calculation in \cref{eq:per-patch-particle-log-weights}, the number of non-zero log observation density terms in the sum is dependent on both the localisation function and the size of the patches $\hat{\set{S}}_{\srange{\cns{B}}}$ used to define the \ac{pou}. We can define an effective number of observations considered per patch as
\begin{equation}\label{eq:per-patch-effective-observations}
  n_b =
  \sum_{l\in\srange{\cns{L}}}  \ell_r(\underline{d}(\hat{\set{S}}_b, s_l^{\textrm{o}}))
  \quad \forall b\in\range{\cns{B}}.
\end{equation}
To avoid weight degeneracy we will typically need to control the $n_{\srange{B}}$ values through the choice of \ac{pou} and localisation radius $r$, with the results of \citet{rebeschini2015can} suggesting $n_{\srange{B}}$ should roughly scale with $\log\cns{P}$. To approximately minimise $\max(n_{\srange{B}})$ for a given number of patches $\cns{B}$, as a heuristic we suggest the patches should be chosen such that each contains a roughly equal number of observations. We discuss approaches for defining a partition of the spatial domain based on the observation locations to achieve this in \cref{app:partitioning-the-spatial-domain}.

The choice of the number of patches $\cns{B}$ to use will typically be based on a tradeoff between several factors. Reducing computational cost favours using fewer patches, while the need to control $\max(n_{\srange{B}})$ and so the tendency for weight degeneracy favours using a greater number of smaller patches. More complex is the dependency of the approximation error introduced by localisation. Using larger patches and a greater number of observations to update the state variables within each patch should reduce the approximation error for the updates within each patch. However for a fixed $r$ using larger patches will also lead to great disparities in the local weights calculated for each patch using \cref{eq:per-patch-particle-log-weights} and so the transform coefficients for adjacent patches. If using a hard \ac{pou} this will typically lead to spatial discontinuities in the state particles across patch boundaries after applying the assimilation update, with the downstream effect of such discontinuities potentially negating any reduction in the approximation error within the patches. 

If using a smooth \ac{pou} the mesh nodes in the overlaps between patches will be updated using a interpolation of the transform coefficients for each of the patches, allowing smaller numbers of patches $\cns{B}$ to be used while still retaining smoothness. In the numerical experiments in \cref{sec:experiments} we show that using a smooth \ac{pou} allows use of a number of patches $\cns{B}$ less than the number of mesh nodes $\cns{M}$ while still retaining accurate filtering distribution estimates.

\subsection{Computational cost}

The computational cost of the per-node local \ac{etpf} assimilation updates proposed in \citet{cheng2015assimilating} is dominated by solving the $\cns{M}$ \ac{ot} problems leading to an overall $\otilde(\cns{M}\cns{P}^3)$ scaling for the computational cost. For the \ac{sletpf}, the number of \ac{ot} problems is determined by the number of patches $\cns{B}$ and so the cost of solving the \ac{ot} problems is $\otilde(\cns{B}\cns{P}^3)$. When $\cns{B} \ll \cns{M}$ the relative cost of the other computations in the overall assimilation update can become significant however. To derive a relationship for the overall scaling of the computational cost of the proposed \ac{sletpf} we make the following assumptions.
\begin{assumption}\label{ass:total-nodes-per-patch}
The maximum number of patches covering any mesh node is independent of and much smaller than $\cns{B}$ and so the sum across all patches of the number of mesh nodes within each patch scales independently of $\cns{B}$, i.e.
\begin{equation}
  \label{eq:total-nodes-per-patch}
  \sumrange{b}{1}{\cns{B}} \sumrange{m}{1}{\cns{M}} \indc{\hat{\set{S}}_b}(s_m) 
  = \mathcal{O}(\cns{M}).
\end{equation}
\end{assumption}
For \acp{pou} in which each patch overlaps with only a fixed number of `neighbour' patches this will hold. If a uniform subsampling scheme is used to define the set of mesh node indices $\set{M}$ used in computing the transport costs, then as a corollary we will also have that the total number of subsampled mesh nodes contained within all patches scales independently of $\cns{B}$, i.e.
\begin{equation}
  \label{eq:total-subsampled-nodes-per-patch}
  \sumrange{b}{1}{\cns{B}} \sum_{m\in\set{M}} \indc{\hat{\set{S}}_b}(s_m)
  = \mathcal{O}(|\set{M}|).
\end{equation}

\begin{assumption}\label{ass:total-observations-per-patch}
The sum across all patches of the number observations within a distance $r$ from a patch is less than the number of mesh nodes $\cns{M} > \cns{L}$, i.e.
\begin{equation}
  \label{eq:total-observations-per-patch}
  \sumrange{b}{1}{\cns{B}} \sumrange{l}{1}{\cns{L}} \indc{[0,r]}(\underline{d}(\hat{\set{S}}_b, s^{\mathrm{o}}_l)) 
  < \cns{M}.
\end{equation}
\end{assumption}
We will typically have that the number of observations locations $\cns{L}$ is small compared to the number of mesh nodes $\cns{M}$ and the localisation radius $r$ will be set to limit the number of observations considered per patch to a small subset of all observations so this will usually hold.

Under \cref{ass:total-nodes-per-patch} the cost of calculating the $\cns{B}\cns{P}^2$ transport costs using \cref{eq:per-patch-transport-costs-coefficients} is $\mathcal{O}(|\set{M}|\cns{P}^2)$ as we need to evaluate the distance between the $\cns{P}(\cns{P}-1)$ pairs of particles at $|\set{M}|$ mesh nodes and from \cref{eq:total-subsampled-nodes-per-patch} only $\mathcal{O}(|\set{M}|)$ terms in the summations for each of the particle pairs need to be evaluated.

The update to the particles in \cref{eq:sletf-assimilation-update} for a general set of per-patch linear transform coefficients $\hat{\rvar{a}}_{t,\srange{\cns{B}}}\pss{\srange{\cns{P}},\srange{\cns{P}}}$ will have a cost of $\mathcal{O}(\cns{M}\cns{P}^2)$ under \cref{ass:total-nodes-per-patch}. However for transform coefficients computed as the solution to discrete \ac{ot} problems, at most $2\cns{P}-1$ of of the $\cns{P}^2$ coefficients for each patch are non-zero \citep{reich2013nonparametric}. In this case the assimilation update in \cref{eq:sletf-assimilation-update} therefore has a $\mathcal{O}(\cns{M}\cns{P})$ cost. 

Under \cref{ass:total-observations-per-patch}, the computation using \cref{eq:per-patch-particle-log-weights} of the $\cns{B}\cns{P}$ per-patch weights will cost less than $\mathcal{O}(\cns{M}\cns{P})$ as we need to evaluate $\cns{L}\cns{P} < \cns{M}\cns{P}$ log observation density factors, and from \cref{eq:total-observations-per-patch} less than $\cns{M}$ terms in the summations for each of the $\cns{P}$ particles will be non-zero and so need to be evaluated.

Under these assumptions, the overall computational cost of each \ac{sletpf} assimilation step therefore scales as $\otilde\left(\cns{B}\cns{P}^3 + |\set{M}|\cns{P}^2 + \cns{M}\cns{P}\right)$.

\section{Numerical experiments}
\label{sec:experiments}

To evaluate the performance of the proposed approach, we perform filtering in two \ac{spde} test models, comparing our proposed scheme to the local \ac{etpf} \citep{cheng2015assimilating} and local \ac{etkf} \citep{hunt2007efficient}. Rather than measure performance in terms of the distance between the estimated mean of the filtering distribution and the true state used to generate the observations, as is common in similar work e.g. \citet{farchi2018comparison}, here we measure the errors in the ensemble estimates of expectations with respect to the true filtering distributions. This gives more directly interpretable results as a filter which exactly computes the expectations would give a zero error, unlike the difference between the mean and true state which will in general be non-zero even if the mean is computed exactly. We are also to able to assess the accuracy of a broader range of features of the filtering distribution estimates, for example their quantification of uncertainty via measures of dispersion.

To allow such comparisons, we require models for which ground truth values for expectations with respect to the filtering distributions can be computed. To this end our first model is based on a linear-Gaussian \ac{spde} model for which the true filtering distribution can be exactly computed using a Kalman filter. For the second model, we use a more challenging \ac{spde} model with non-linear state dynamics. Here our `ground-truth' for the filtering distributions is based on long runs of a \ac{mcmc} method.

\subsection{Evaluating the accuracy of filtering estimates}
\label{subsec:metrics}

For both models we consider several metrics for evaluating the accuracy of the different local ensemble filters' estimates of the filtering distributions.

The first two metrics we consider are the time- and space-averaged \acp{rmse} of the ensemble estimates of the filtering distributions means and standard deviations, to reflect respectively the filters' accuracy in estimating the central tendencies and dispersions of the filtering distributions. Denote $\mu_{\srange{\cns{T}}}$ and $\sigma_{\srange{\cns{T}}}$ as the true means and standard deviations under $\pi_{\srange{\cns{T}}}$
\begin{equation}
  \label{eq:filtering-dist-mean-and-std-true}
  \vct{\mu}_{t} = \int_{\set{X}} \vct{x} \,\pi_{t}(\dr\vct{x})
  \quad
  \textrm{and}
  \quad
  \vct{\sigma}^2_{t} = \int_{\set{X}} (\vct{x} - \vct{\mu}_t)\odot(\vct{x} - \vct{\mu}_t) \,\pi_{t}(\dr\vct{x})
  \quad
  \forall t\in\range{\cns{T}},
\end{equation}
and $\hat{\mu}_{\srange{\cns{T}}}$ and $\hat{\sigma}_{\srange{\cns{T}}}$ as the corresponding means and standard deviations under the empirical ensemble estimates to the filtering distributions $\hat{\pi}_t(\dr\vct{x})=\sum_{p=1}^{\cns{P}} \delta_{\vct{x}_t\pss{p}}(\dr\vct{x})$, 
\begin{equation}
  \label{eq:filtering-dist-mean-and-std-est}
  \hat{\vct{\mu}}_{t} = \int_{\set{X}} \vct{x} \,\hat{\pi}_{t}(\dr\vct{x})
  \quad
  \textrm{and}
  \quad
  \hat{\vct{\sigma}}^2_{t} = \int_{\set{X}} (\vct{x} - \hat{\vct{\mu}}_t)\odot(\vct{x} - \hat{\vct{\mu}}_t) \,\hat{\pi}_{t}(\dr\vct{x})
  \quad
  \forall t\in\range{\cns{T}}.
\end{equation}
We then define the time- and space-averaged \acp{rmse} of the estimates as
\begin{align}\label{eq:mean-est-rmse}
  \textsc{rmse}(\hat{\vct{\mu}}_{\srange{\cns{T}}},\vct{\mu}_{\srange{\cns{T}}}) &= 
  \sqrt{\recip{\cns{T}\cns{M}}
  \sumrange{t}{1}{\cns{T}}\sumrange{m}{1}{\cns{M}}
    \left(\hat{\mu}_{t,m} - \mu_{t,m}\right)^2},
  \\
  \label{eq:std-est-rmse}
  \textrm{and}\quad
  \textsc{rmse}(\hat{\vct{\sigma}}_{\srange{\cns{T}}}, \vct{\sigma}_{\srange{\cns{T}}}) &= 
  \sqrt{\recip{\cns{T}\cns{M}}\sumrange{t}{1}{\cns{T}}\sumrange{m}{1}{\cns{M}}
  \left(\hat{\sigma}_{t,m} - \sigma_{t,m}\right)^2}.
\end{align}
In both cases lower values of these metrics are better, with a value of zero indicating the mean or standard deviation estimates exactly match the true values. 

The two metrics discussed so far concentrate on the accuracy of estimates of local properties of the states, but do not reflect more global properties such as whether the ensemble filters correctly estimate the smoothness of the state fields. As a proxy measure for smoothness we use the expectation under the true filtering distributions of a finite-difference approximation of the integral across space of the magnitude of the spatial gradients of the state fields:
\begin{equation}
  \label{eq:filtering-dist-smoothness-coefficient-true}
  \gamma_{t} = \int_{\set{X}} \sumrange{m}{1}{\cns{M}} \left| x_{t,m} - x_{t,m\oplus 1} \right| \,\pi_{t}(\dr\vct{x}) \approx
  \expc\left[\int_{\set{S}} \left| \partial_s\rvar{z}_t(s)\right| \,\dr s\right]
  \quad
  \forall t\in\range{\cns{T}},
\end{equation}
with $m \oplus 1$ here indicating $m + 1 \mod \cns{M}$, with one-dimensional periodic spatial domains being used in both models considered. Defining the estimates $\hat{\gamma}_{\srange{\cns{T}}}$ of these \emph{smoothness coefficients} under the ensemble filtering distributions equivalently as
\begin{equation}
  \label{eq:filtering-dist-smoothness-coefficient-est}
  \hat{\gamma}_{t} = \int_{\set{X}} \sumrange{m}{1}{\cns{M}} \left| x_{t,m} - x_{t,m\oplus 1} \right| \,\hat{\pi}_{t}(\dr\vct{x})
  \quad
  \forall t\in\range{\cns{T}},
\end{equation}
we then define an overall measure of the accuracy of the ensemble estimates' spatial smoothness as the following time-averaged \ac{rmse}
\begin{equation}\label{eq:smoothness-est-rmse}
  \textsc{rmse}(\hat{\gamma}_{\srange{\cns{T}}},\gamma_{\srange{\cns{T}}}) = 
  \sqrt{\recip{\cns{T}}\sumrange{t}{1}{\cns{T}}
    \left(\hat{\gamma}_{t} - \gamma_{t}\right)^2}.
\end{equation}

\subsection{Stochastic turbulence model}

As our first example we use a linear-Gaussian \ac{ssm} derived from a \ac{spde} model for turbulent signals by \citet[Ch. 5]{majda2012filtering}. The governing \ac{spde} is
\begin{equation}\label{eq:spde-stochastic-turbulence}
  \dr\rvar{\zeta}(s,\tau) = 
  \left(
    \theta_1\partial^2_{s} + \theta_2\partial_s - \theta_3
  \right)\rvar{\zeta}(s,\tau) \dr\tau+ 
  (\kappa \circledast_s \dr\rvar{\eta})(s,\tau),
\end{equation}
where $\zeta : \set{S} \times \reals_{\geq 0} \to \reals$ is a real-valued space-time varying process, $\theta_1 \in \reals_{\geq 0}$ is a non-negative parameter controlling dissipation due to diffusion, $\theta_2 \in \reals$ is a parameter governing the direction and magnitude of the constant advection, $\theta_3 \in \reals_{\geq 0}$ is a non-negative parameter controlling dissipation due to damping, $\kappa : \set{S} \to \reals_{\geq 0}$ is a spatial kernel function which governs the spatial smoothness of the additive noise in the dynamics and $\rvar{\eta} : \set{S} \times \reals_{\geq 0} \to \reals$ is a space-time varying noise process. The spatial domain is a one-dimensional interval $\set{S} = [0, 1)$ with periodic boundary conditions and a distance function $d(s, s') = \min(|s - s'|, 1 - |s - s'|)$, and $\circledast_s$ represents circular convolution in space.

We use a spectral approach to define basis function expansions of the processes $\rvar{\zeta}$ and $\rvar{\eta}$ and kernel $\kappa$ using $\cns{M}=512$ mesh nodes. This results in a linear system of \acp{sde} for which the the Gaussian state transition and stationary distributions can be solved for exactly. We assume a linear-Gaussian observation model with the state noisily observed at $\cns{L} = 64$ locations and $\cns{T}=200$ time points. Full details of the model are given in \cref{app:stochastic-turbulence-model-details}.

The resulting \ac{st} \ac{ssm} is linear-Gaussian. We consider two cases in our experiments: inference in the original linear-Gaussian \ac{ssm}, and inference in a \emph{transformed} \ac{ssm} using this linear-Gaussian model as the base \ac{ssm}. The specific definition we use for a transformed \ac{ssm} is given in \cref{app:transformed-ssms} however in brief, by applying a non-linear transformation to the state of a linear-Gaussian \ac{ssm} we can construct a \ac{ssm} with non-Gaussian filtering distributions for which we can tractably estimate expectations with respect to the true filtering distributions with artbirary accuracy. Here the nonlinear transformation is chosen as $T(x) = \sinh^{-1}(\theta_4\vct{x})$ (with $\sinh^{-1}$ evaluated elementwise on vector arguments). As $|\sinh^{-1}(\theta_4 x)| \approx \log(2\theta_4|x|)$ for $|\theta_4 x| \gg 1$ this non-linearity has the effect of compressing the variation in large magnitude values, while expanding small magnitude values, and so for an appropriate choice of scaling factor $\theta_4$ tends to induce bimodality in the marginals of the transformed filtering distributions.

For both the transformed and linear-Gaussian cases we use the model parameter settings give in \cref{tab:st-model-params} and use simulated noisy observations $\vct{y}_{\srange{\cns{T}}}$ generated from the models using a shared set of Gaussian state and observation noise variable samples generated using a pseudo-random number generator. The resulting observation sequence $\vct{y}_{\srange{\cns{T}}}$ (which is the same for both models) is shown in \cref{fig:st-simulated-state-and-obs} along with the corresponding true state sequences $\vct{z}_{\srange{\cns{T}}}$ and $\vct{z}'_{\srange{\cns{T}}}$ used to generate the observations under the linear-Gaussian and transformed \acp{ssm} respectively.

We compare the performance of the local \ac{etkf}, local \ac{etpf} and our proposed \ac{sletpf} algorithm in estimating the filtering distributions for both the linear-Gaussian and transformed \acp{ssm}. The mesh size $\cns{M} = 512$ and number of observations $\cns{L}=64$ are sufficiently large that non-local \ac{pf} methods suffer from weight degeneracy even with large ensembles of up to $\cns{P} = 10^4$ particles for both the linear-Gaussian and transformed \acp{ssm}. While non-local variants of the \ac{enkf} do not suffer from weight degeneracy and can give relatively accurate filtering distribution estimates for an ensemble size of $\cns{P} \geq 10^3$, this is still much larger than the ensemble sizes typically used in for example \ac{nwp} ensemble filter systems. 
For an ensemble size $\cns{P} = 10^2$ we found the local \ac{etkf} significantly outperformed the non-local \ac{etkf} on all the metrics we consider in both the linear-Gaussian and transformed \acp{ssm}. We used $\cns{P}=10^2$ for all methods in the experiments here.

For the local \ac{etkf} we use the smooth compact Gaspari and Cohn localisation function $\ell_r$ defined in \cref{eq:gaspari-and-cohn-localisation}. We conducted a grid search over localisation radii $r \in \lbrace 0.010, 0.012, \dots 0.160\rbrace$, for each $r$ performing five independent runs of the local \ac{etkf} and recording the performance on the three metrics described in \cref{subsec:metrics}. The results for the linear-Gaussian \ac{st} model are summarised in \cref{tab:st-linear-gaussian-local-etkf-performance} and for the transformed \ac{st} model in \cref{tab:st-transformed-local-etkf-performance}. For each metric the minimum, median and maximum value recorded across the five runs is shown, for the value of $r$ which gave the minimum median value of that particular metric. The results for all $r$ values are shown in the Appendix in \cref{fig:full-letkf-grid-search-results}.

\begin{table}
  \begin{tabular}{r|lll}
     & 
    $\textsc{rmse}(\hat{\vct{\mu}}_{\srange{\cns{T}}},\vct{\mu}_{\srange{\cns{T}}})$ & 
    $\textsc{rmse}(\hat{\vct{\sigma}}_{\srange{\cns{T}}},\vct{\sigma}_{\srange{\cns{T}}})$ & 
    $\textsc{rmse}(\hat{\gamma}_{\srange{\cns{T}}},\gamma_{\srange{\cns{T}}})$\\
    \hline
    Minimum & $4.34 \times 10^{-2}$ & $1.37 \times 10^{-2}$ & $7.40 \times 10^{-4}$ \\
    Median  & $4.38 \times 10^{-2}$ & $1.38 \times 10^{-2}$ & $8.18 \times 10^{-4}$ \\ 
    Maximum & $4.43 \times 10^{-2}$ & $1.40 \times 10^{-2}$ & $9.13 \times 10^{-4}$ \\
    \hline
    Localisation radius $r$ & 0.030 & 0.034 & 0.024
  \end{tabular}
  \caption{Values of metrics at optimal localisation radii for local \ac{etkf} on linear-Gaussian \ac{st} model.}
  \label{tab:st-linear-gaussian-local-etkf-performance}
\end{table}

\begin{table}
  \begin{tabular}{r|lll}
    & 
   $\textsc{rmse}(\hat{\vct{\mu}}_{\srange{\cns{T}}},\vct{\mu}_{\srange{\cns{T}}})$ & 
   $\textsc{rmse}(\hat{\vct{\sigma}}_{\srange{\cns{T}}},\vct{\sigma}_{\srange{\cns{T}}})$ & 
   $\textsc{rmse}(\hat{\gamma}_{\srange{\cns{T}}},\gamma_{\srange{\cns{T}}})$\\
   \hline
   Minimum & $1.71 \times 10^{-1}$ & $1.93 \times 10^{-1}$ & $1.04 \times 10^{-2}$ \\
   Median  & $1.72 \times 10^{-1}$ & $1.94 \times 10^{-1}$ & $1.04 \times 10^{-2}$ \\ 
   Maximum & $1.74 \times 10^{-1}$ & $1.95 \times 10^{-1}$ & $1.05 \times 10^{-2}$ \\
   \hline
   Localisation radius $r$ & 0.030 & 0.152 & 0.160
 \end{tabular}
  \caption{Values of metrics at optimal localisation radii for local \ac{etkf} on transformed \ac{st} model.}
  \label{tab:st-transformed-local-etkf-performance}
\end{table}

The performance on all metrics for both models was relatively stable across the multiple runs. Unsuprisingly the local \ac{etkf} performs significantly better on the linear-Gaussian \ac{st} model than the transformed \ac{st} model. While for the linear-Gaussian \ac{st} model the optimal $r$ for each metric are relatively similar, for the transformed \ac{st} model the optimal $r$ differs significantly across the metrics meaning any choice of $r$ will incur a performance penalty on some metrics.

For our proposed \ac{sletpf} framework we need to choose a \ac{pou}. Here we construct the \acp{pou} by (discretely) convolving a mollifier function with the indicator functions on a partition of the spatial domain. As the observations are located on a regular grid, we partition the domain into  $\cns{B}$ equally sized intervals $\set{S}_b = [\frac{b-1}{\cns{B}}, \frac{b}{\cns{B}}) ~\forall b \in \range{\cns{B}}$. For the mollifier function $\varphi$ we use a normalised variant of the compactly supported Gaspari and Cohn localisation function $\ell_r$ in \cref{eq:gaspari-and-cohn-localisation}, the bump functions then defined as
\begin{equation}\label{eq:pou-bump-function-discrete-conv}
  \phi_b(s_n) = \sumrange{m}{1}{\cns{M}} \indc{\set{S}_b}(s_m) \, \frac{\ell_{w}\circ d(s_n, s_m)}{\sumrange{m'}{1}{\cns{M}} \ell_{w}\circ d(s_n, s_{m'})}
  \quad \forall n \in\range{\cns{M}}, b\in\range{\cns{B}}
\end{equation}
with $w$ a \emph{kernel width} parameter determining how many mesh nodes the effective smoothing kernel being discretely convolved with the indicators has support on. For $w = \cns{M}^{-1}$ the kernel is only non-zero at one mesh node, and no smoothing is applied, corresponding to a hard partition of the space. For $w > \cns{M}^{-1}$, the amount of smoothing and overlap between the patches increases with $w$.

For the experiments with the \ac{st} models we performed runs with \acp{sletpf} with \acp{pou} with five different numbers of patches $\cns{B} \in \lbrace 2^5, 2^6, 2^7, 2^8, 2^9 \rbrace$ and four different kernel widths $w \in \lbrace 512^{-1}, 256^{-1}, 128^{-1}, 64^{-1} \rbrace$. We used a Gaspari and Cohn localisation function for the local weight calculation, for each $(\cns{B}, w)$ pair performing five independent runs for all localisation radii $r \in \lbrace 0.001, 0.002, \dots 0.030 \rbrace$ where $\median(n_{\srange{\cns{B}}})$ was in the range $[1, 5]$. As noted previously the local \ac{etpf} of \citet{cheng2015assimilating} can be considered a particular instance of the \ac{sletpf} framework, here corresponding to the runs with a \ac{pou} with $\cns{B} = 512$ patches and $w = 512^{-1}$. The set of mesh nodes $\set{M}$ used
to calculate the per-patch transport costs as in \cref{eq:per-patch-transport-costs-coefficients} was constructed by subsampling $\range{\cns{M}}$ by a factor $\min(4, p_n)$ with $p_n = \cns{M}(\cns{B}^{-1} + 2w)-1$ the number of mesh nodes in each patch, ensuring that at least one node per patch was used to compute the transport costs.

\begin{figure}[t!]
  \centering
  \includegraphics[width=\textwidth]{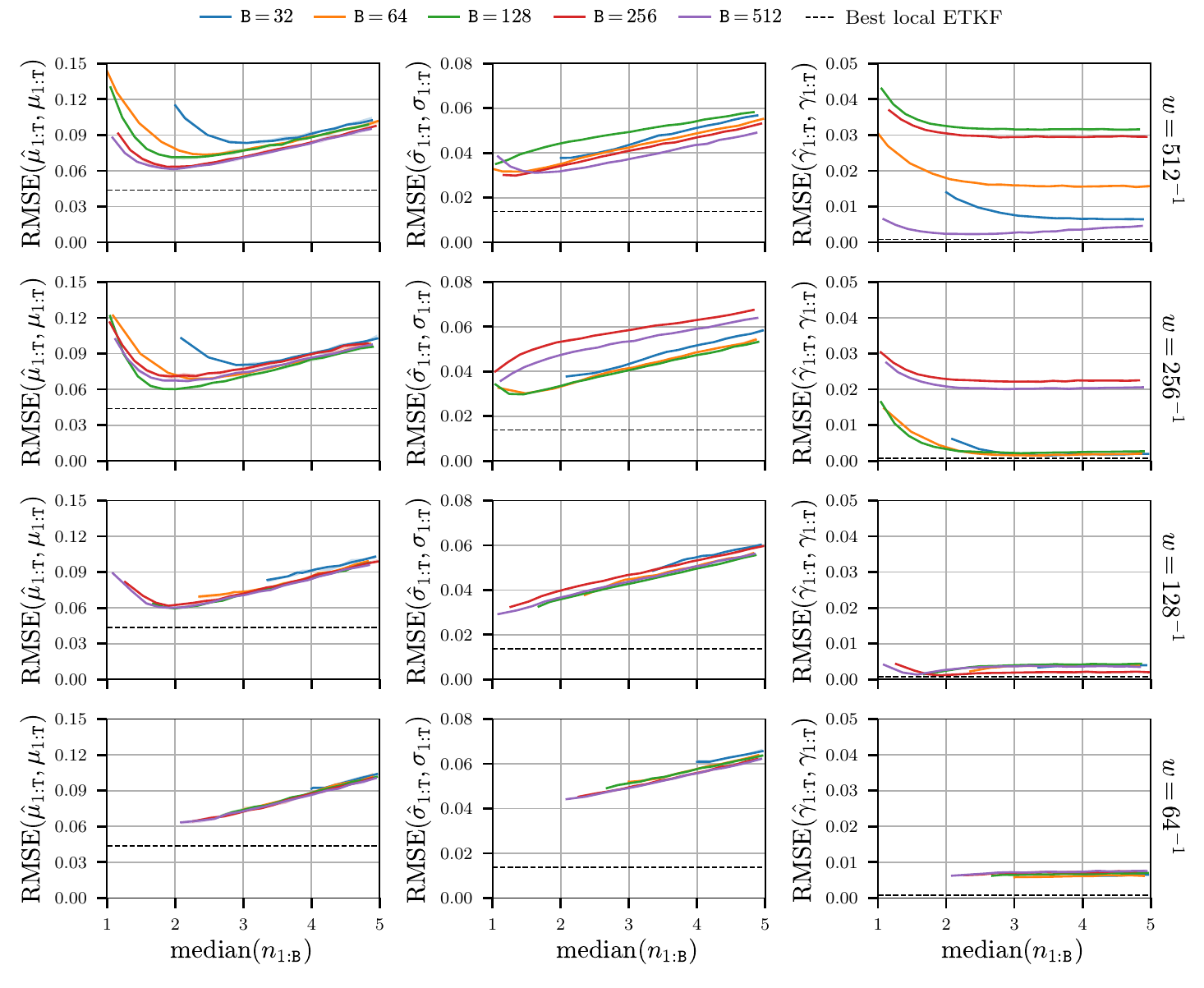}
  \caption{Comparison of accuracy of \ac{sletpf} estimates on linear-Gaussian \ac{st} \ac{ssm}.
  }
  \label{fig:st-linear-sletpf-eff-obs-vs-rmse-mean-std-smoothness}
\end{figure}

\begin{figure}[t!]
  \centering
  \includegraphics[width=\textwidth]{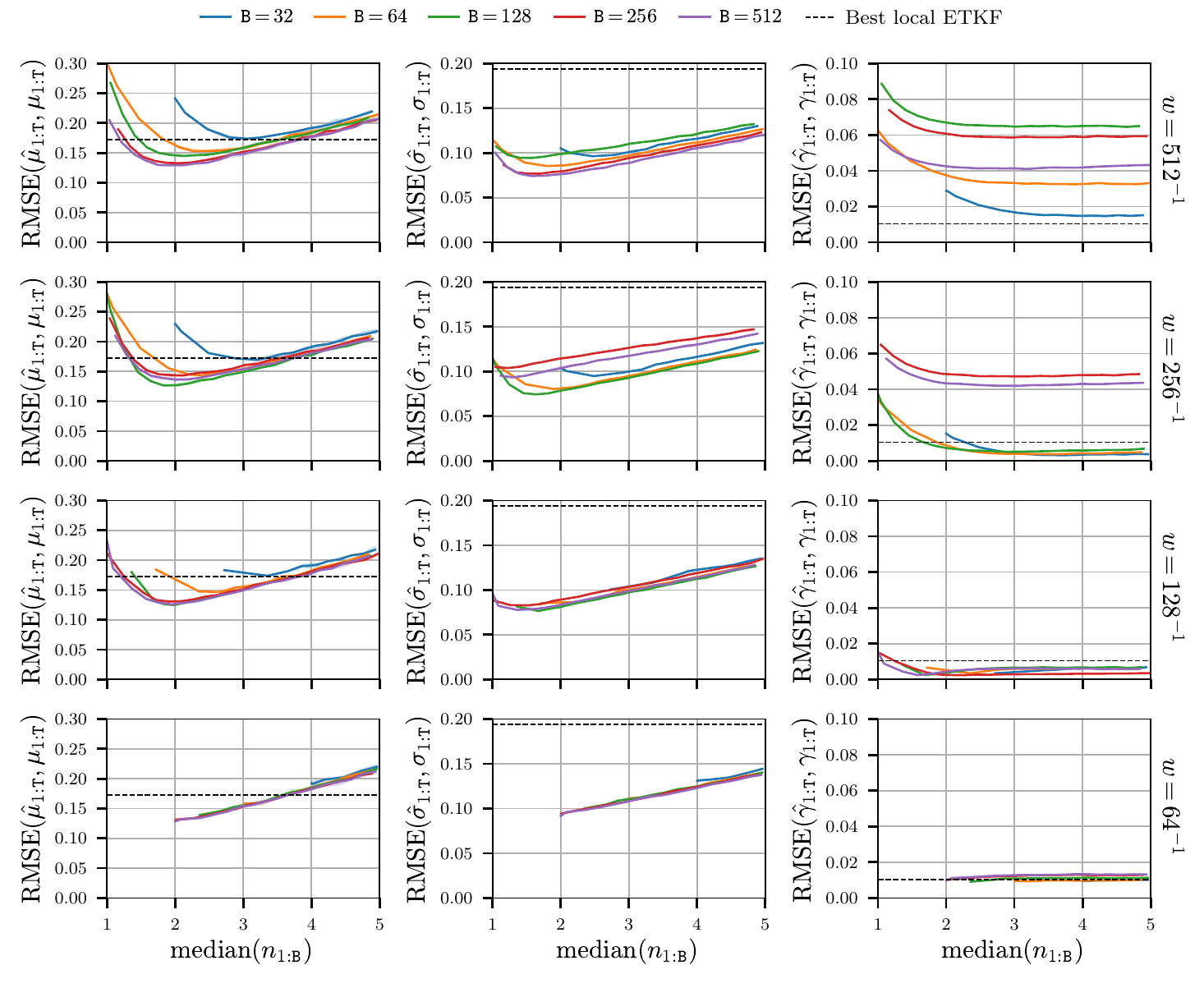}
  \caption{Comparison of accuracy of \ac{sletpf} estimates on transformed \ac{st} \ac{ssm}.
  } 
  \label{fig:st-nonlinear-sletpf-eff-obs-vs-rmse-mean-std-smoothness}
\end{figure}

The values of the three metrics recorded across all \ac{sletpf} runs for each of the $(\cns{B}, w, r)$ parameter combinations are shown for the linear-Gaussian \ac{st} model in \cref{fig:st-linear-sletpf-eff-obs-vs-rmse-mean-std-smoothness} and for the transformed \ac{st} model in \cref{fig:st-nonlinear-sletpf-eff-obs-vs-rmse-mean-std-smoothness}. In each figure, the rows of plots correspond to different kernel widths $w$ and the three columns to different metrics. On each plot the value of the relevant metric on the vertical axis is plotted against the median number of effective observations per patch on the horizontal axis (we plot against $\median(n_{\srange{\cns{B}}})$ rather than $r$ as it is more directly comparable across different values of $\cns{B}$ and $w$). The median values across the five independent runs for each of the numbers of patches $\cns{B}$ are shown by the coloured curves (see colour key at top of figures) and the surrounding lighter coloured regions indicated minimum to maximum range of values recorded across the runs (in many cases the across-run variation is too small to be visible). For each metric the best value achieved by the local \ac{etkf} (as given in \cref{tab:st-linear-gaussian-local-etkf-performance,tab:st-transformed-local-etkf-performance}) for the metric is indicated by the black horizontal dashed line.

Considering first the linear-Gaussian \ac{st} model results, we see that across all parameter combinations and metrics the local \ac{etpf} methods are outperformed by the best local \ac{etkf} results. This is as expected as the linear-Gaussian assumptions made by the \ac{etkf} are correct in this case, and by better exploiting this model structure we expect the local \ac{etkf} to outperform the more generic local \ac{etpf}. 

Concentrating on the results for filters with hard \acp{pou} without smoothing in the first row ($w = 512^{-1}$), we see that the filters with $\cns{B} = \cns{M} = 512$ patches in the \ac{pou}, corresponding to the \citet{cheng2015assimilating} scheme, outpeform filters using \acp{pou} with smaller numbers of patches across virtually all $\median(n_{\srange{\cns{B}}})$ values and metrics. This tallies with the findings of \citet{farchi2018comparison} who found that for an equivalent `block'-based local \ac{etpf} scheme the best performance was always achieved with blocks of size one.  Considering specifically the $\textsc{rmse}(\hat{\vct{\mu}}_{\srange{\cns{T}}},\vct{\mu}_{\srange{\cns{T}}})$ metric we see that as the number of patches $\cns{B}$ decreases the value of the metric across all values of  $\median(n_{\srange{\cns{B}}})$ monotonically increases (corresponding to poorer performance). The behaviours for the $\textsc{rmse}(\hat{\vct{\sigma}}_{\srange{\cns{T}}},\vct{\sigma}_{\srange{\cns{T}}})$ and $\textsc{rmse}(\hat{\vct{\gamma}}_{\srange{\cns{T}}},\vct{\gamma}_{\srange{\cns{T}}})$ metrics are more complex. For the smoothness coefficient we see that accuracy of the filter estimates initially decreases as the number of patches is increased from $\cns{B} = 512$ to $\cns{B}= 256$ and $\cns{B} = 128$. The accuracy of the smoothness estimates however then increases on decreasing the number of patches further to $\cns{B} = 64$ and again the accuracy increases on decreasing the number of patches to $\cns{B} = 32$. We believe this non-monotonic relationship between the accuracy of the smoothness estimates and the number of patches in the \ac{pou} may be explained by the spatial averaging in the computation of the smoothness coefficient: while using fewer larger patches in the \ac{pou} would be expected to introduce stronger discontinuities at the patch boundaries due to larger differences in the local weights assigned to each patch, there is a competing effect that as fewer patches are used there are fewer boundaries and so the spatially averaged error becomes lower despite the individual discontinuities at each block boundary being larger. 

Now comparing the results as the kernel width $w$ and so smoothness of the \ac{pou} is increased, there are two main trends apparent. Most prominently the variation in performance across different numbers of patches $\cns{B}$ decreases as the smoothness of the \ac{pou} increases, with many of the curves overlapping over much of their ranges for $w = 128^{-1}$ and $w = 64^{-1}$, while the optimal performance on each metric remains similar. This suggests using smooth \ac{pou} allows fewer number of patches to be used (and thus a lower computational cost of the assimilation update) while maintaining performance, contrary to what was observed for the hard \ac{pou} case where using fewer patches always decreased performance. 

A second less obvious effect is that as the kernel width $w$ is increased the lower limit for $\median(n_{\srange{\cns{B}}})$ is increased (similarly using fewer larger patches also increases the lower limit for $\median(n_{\srange{\cns{B}}})$). This is the reason for the curves starting at higher $\median(n_{\srange{\cns{B}}})$ as the kernel width increases, corresponding to the values achieved with the smallest $r$ tested ($r=0.001$). In the case of the largest kernel width tested $w = 64^{-1}$ we see that all the curves start to the right of the point at which the optimal performance is reached for the other smaller $w$. This suggests there is a drawback to making $w$ too large as it limits how far the number of observations per patch and so tendency to local weight degeneracy can be controlled; in this case it seems the best tradeoff is reached for either $w = 256^{-1}$ or $w = 128^{-1}$. Interestingly we also see that the accuracy of the smoothness and standard deviation estimates are poorer for $w = 64^{-1}$ compared to $w = 128^{-1}$ even when comparing at the same $\median(n_{\srange{\cns{B}}})$. This could be due to the greater overlap between the patches in this case, with the averaging of the particle values at the overlaps potentially acting to artificially oversmooth and reduce variation in the particles, again suggesting that the appropriate level of smoothing is a tradeoff between several factors.

The results on the transformed \ac{st} model shown in \cref{fig:st-nonlinear-sletpf-eff-obs-vs-rmse-mean-std-smoothness} show for the most part very similar trends as for the linear-Gaussian \ac{st} model. The most significant difference is the relative performance of the local \ac{etkf} and local \ac{etpf} methods, with in this case the local \ac{etpf} approaches outperforming the best local \ac{etkf} results across all parameter values for the $\textsc{rmse}(\hat{\vct{\sigma}}_{\srange{\cns{T}}},\vct{\sigma}_{\srange{\cns{T}}})$ and across a majority of the parameter values tested for the $\textsc{rmse}(\hat{\vct{\mu}}_{\srange{\cns{T}}},\vct{\mu}_{\srange{\cns{T}}})$ metric. As the only difference between these two models is the non-Gaussianity in the filtering distributions introduced by the transformation, these results support the earlier claims that \ac{pf}-based methods such as the local \ac{etpf} and \ac{sletpf} proposed in this article, are more robust to non-Gaussianity than than \ac{enkf} methods such as the local \ac{etkf}. Interestingly the relative performance loss in the local \ac{etkf} on introducing non-Gaussianity seems to be most severe in the $\textsc{rmse}(\hat{\vct{\sigma}}_{\srange{\cns{T}}},\vct{\sigma}_{\srange{\cns{T}}})$ metric, suggesting that uncertainty estimates provided by local \ac{etkf} methods on non-linear-Gaussian models should be particuarly treated with caution.
\begin{figure}
  \centering
  \includegraphics[width=\textwidth]{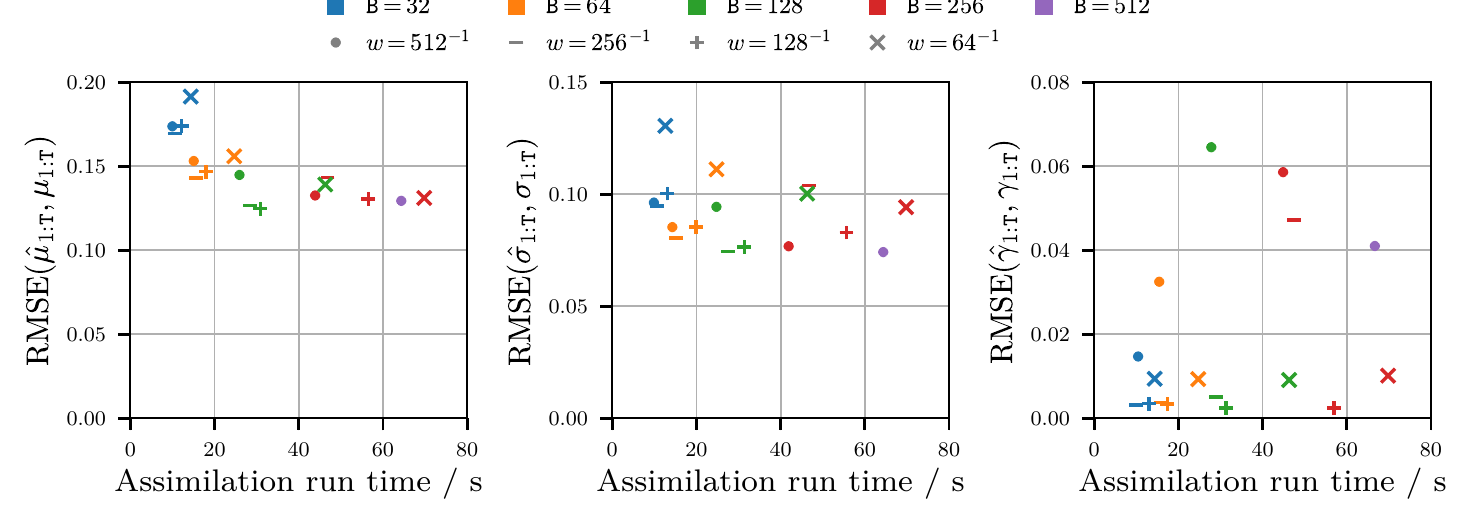}
  \caption{Accuracy versus run time for \ac{sletpf} in transformed \ac{st} \ac{ssm}.
  } 
  \label{fig:st-nonlinear-sletpf-runtime-vs-rmse-mean-std-smoothness}
\end{figure}

In addition to the accuracy of the filter estimates, we are also interested in the relative computational cost of the different methods. \cref{fig:st-nonlinear-sletpf-runtime-vs-rmse-mean-std-smoothness} shows the values of the performance metrics achieved by the different \ac{sletpf} configurations tested, against the corresponding assimilation time (i.e. total filtering time minus the time taken to integrate the model dynamics in the prediction updates) for the transformed \ac{st} \ac{ssm}. Each of the three plots corresponds to one of the performance metrics, the vertical coordinate of each marker indicates the minimum value of the metric achieved across all localisation radii $r$ for a particular $(\cns{B}, w)$ combination, with the marker colour indicating the number of patches $\cns{B}$, and the marker symbol the kernel width $w$. The horizontal coordinate of each marker indicates the median assimilation time across the five independent runs for the corresponding $(\cns{B}, w, r)$ values. For the \acp{pou} with $\cns{B} = 512$ patches, only the case without smoothing ($w = 512^{-1}$), corresponding to the \citet{cheng2015assimilating} local \ac{etpf}, is shown, with the smoother \acp{pou} in this case substantially increasing the assimilation times without any gain in accuracy.

As would be expected due to the lower number of \ac{ot} problems that need to be solved, in general the assimilation time decreases as the number of patches $\cns{B}$ in the \ac{pou} is decreased \emph{for a fixed smoothing kernel width $w$}. Note however that the assimilation time increases with the smoothing kernel width $w$ (primarily due to the increased number of non-zero terms in the summation in \cref{eq:sletf-assimilation-update}), which results for example in the assimilation time for the scheme with $\cns{B} = 256$ and $w = 64^{-1}$ ($\color{red}\times$) being slightly larger than for the runs under the \citet{cheng2015assimilating} settings of $\cns{B} = 512$ and $\cns{w} = 512^{-1}$ ($\color{Orchid}\bullet$). Although there is therefore a tradeoff in assimilation time between decreasing the number of patches $\cns{B}$ and increasing the kernel width $w$, we still find that there are combinations of $(\cns{B}, w)$ values which maintain the accuracy of the \citet{cheng2015assimilating} scheme while giving substantial reductions in assimilation time. In particular the runs with $\cns{B} = 128$ and $w = 256^{-1}$ ($\color{Green}-$) and $\cns{B} = 128$ and $w = 128^{-1}$ ($\color{Green}+$) achieve nearly identical accuracies on the mean and standard deviation \ac{rmse} metrics as $\cns{B} = 512$ and $\cns{w} = 512^{-1}$ (and a substantially improved smoothness coefficient \ac{rmse}) while reducing the assimilation time by slightly more than a factor of two. At the cost of around a 10\% increase in the mean and standard deviation \acp{rmse}, a more substantial reduction in the assimilation time by a factor of four can be achieved by using a \ac{pou} with $\cns{B} = 64$ patches and $w \in \lbrace 512^{-1}, 256^{-1}, 128^{-1} \rbrace$.

Although the absolute values of the assimilation times in \cref{fig:st-nonlinear-sletpf-runtime-vs-rmse-mean-std-smoothness} are dependent on the computational environment used to run the experiments, the relative timings should still be informative as the same \ac{sletpf} implementation was used to run all the experiments. We purposefully did not include the local \ac{etkf} runs on the plots as any differences in the assimilation times for the local \ac{etkf} versus \ac{sletpf} approaches are likely to be as much due to the particulars of the software implementations and hardware used as any fundamental differences in performance. In particular  more time was spent optimising the implementation of the \ac{sletpf} algorithm than our local \ac{etkf} implementation so the relative timings are likely to unfairly favour the \ac{sletpf} runs. The computational complexity for the local \ac{etkf} however is $\mathcal{O}(\cns{M}\cns{P}^3)$ which is the same as for the local \ac{etpf} scheme of \citet{cheng2015assimilating}, so it would be expected that there are regimes in which the \ac{sletpf} assimilation updates (with complexity $\otilde(\cns{B}\cns{P}^3 + |\set{M}|\cns{P}^2 + \cns{M}\cns{P})$) will have a computational advantage over the local \ac{etkf} updates.

\subsection{Damped stochastic Kuramoto-Sivashinsky model}

As our second test model we consider a stochastic variant of a fourth-order nonlinear \ac{pde}, often termed the \ac{ks} equation, which has been independently derived as a model of various physical phenomena \citep{kuramoto1976persistent,sivashinsky1977nonlinear} and studied as an example of a relatively simple \ac{pde} system exhibiting spatio-temporal chaos \citep{hyman1986kuramoto}. On a spatial domain $\set{S} = [0,1)$ with a distance function $d(s, s') = \min(|s - s'|, 1 - |s - s'|)$ and periodic boundary conditions, the deterministic dynamics of the \ac{ks} \ac{pde} model can be described by
\begin{equation}
  \partial_{\tau}\rvar{\zeta}(s,\tau) = 
    -\left( \frac{\partial^2_s}{\theta_1^2} + \frac{\partial^4_s}{\theta_1^4} \right) 
    \rvar{\zeta}(s,\tau) -
    \frac{\partial_s}{2\theta_1}  \left( \rvar{\zeta}^2 \right)
\end{equation}
where $\theta_1$ is a length-scale parameter, with the system dynamics becoming chaotic for large values of $\theta_1$ \citep{hyman1986kuramoto}.

As our focus in on filtering in models with stochastic dynamics, we use a related \ac{spde} model on the same spatial domain, described by
\begin{equation}
  \dr\rvar{\zeta}(s,\tau) = 
  \left(
    -\left( \frac{\partial^2_s}{\theta_1^2} + \frac{\partial^4_s}{\theta_1^4} + \theta_2 \right) 
    \rvar{\zeta}(s,\tau) -
    \frac{\partial_s}{2\theta_1}  \left( \rvar{\zeta}^2 \right)
  \right) \dr \tau +
  (\kappa \circledast_s \dr\rvar{\eta})(s,\tau)
\end{equation}
where $\zeta : \set{S} \times \set{T} \to \reals$ is a real-valued space-time varying process, $\theta_1 \in \reals_{\geq 0}$ is the non-negative length-scale parameter, $\theta_2 \in \reals_{\geq 0}$ is a non-negative parameter controlling dissipation due to damping, $\kappa : \set{S} \to \reals_{\geq 0}$ is a spatial kernel function and $\rvar{\eta} : \set{S} \times \set{T} \to \reals$ is a space-time varying noise process. In addition to the introduction of the additive noise process, we also introduce a linear damping component controlled in magnitude by $\theta_2$. This is motivated by our empirical observation in simulations that the stochastic system can become unstable when numerically integrating over long time periods without additional dampening.

We use a similar spectral approach to define the spatial basis function expansions of the state and noise processes $\zeta$ and $\eta$ and kernel $\kappa$ as for the \ac{st} model, again using $\cns{M} = 512$ mesh nodes. Full details of the discretisation used are given in \cref{app:kuramoto-sivashinsky-model-details} and the values of all the parameters used in \cref{tab:ks-model-params}. This results in a coupled non-linear system of \acp{sde} which governs the evolution of the state Fourier coefficients; unlike the linear-Gaussian dynamics of the \ac{st} model these \acp{sde} do not have an analytic solution and so need to be numerically integrated. We assume the state is observed at $\cns{T} = 200$ time points, with $\cns{S} = 10$ integrator steps performed between each observation time; the resulting state transition operators $\op{F}_{\srange{\cns{T}}}$ are non-linear and do not admit closed form transition densities.

We consider \acp{ssm} in which these \ac{ks} state dynamics are noisily observed via both linear and non-linear observation operators. In both cases the state is assumed to be observed at $\cns{L}= 64$ equispaced points in the spatial domain, with direct observations of the state values at these points in the linear case and via a hyperbolic tangent ($\tanh$) function in the non-linear case. The simulated state and observation sequences used in the experiments for both the linearly and non-linearly observed \ac{ks} \acp{ssm} are shown in \cref{fig:ks-simulated-state-and-obs} (with the same simulated state sequence being used in both cases, with only the generated observations differing). Compared to \ac{st} model, the \ac{ks} model exhibits more complex and unpredicatable state dynamics and thus can be seen as more challenging test case for the local ensemble filtering methods.

Both the linearly and non-linearly observed \ac{ks} \acp{ssm} have non-Gaussian filtering distributions which cannot be exactly inferred unlike the linear-Gaussian \ac{st} model. We therefore used a \ac{mcmc} method to generate proxy ground-truths for the filtering distributions, constructing Markov chains which left invariant the joint distribution across the $\cns{M}=512$ dimensional state vectors at all $\cns{T}=200$ time points given the observed sequence, i.e. $\prob(\rvct{x}_{\srange{\cns{T}}} \in \dr\vct{x} \gvn \rvct{y}_{\srange{\cns{T}}} = \vct{y}_{\srange{\cns{T}}})$, with the filtering distributions corresponding to marginals of this joint \emph{smoothing distribution}. Due to the large overall state dimension $\cns{M}\cns{T} \approx 10^5$ we use a gradient-based Hamiltonian Monte Carlo algorithm \citep{duane1987hybrid} to generate the chains. For each of the linear and non-linearly observed cases we ran five parallel chains of 200 samples each, with each chain using an independently seeded pseudo-random number generator. Details of the set up used for the \ac{mcmc} runs are given in \cref{app:ks-mcmc-details}. The `ground-truth' values for the filtering distributions means $\mu_{\srange{\cns{T}}}$, standard deviations $\sigma_{\srange{\cns{T}}}$ and smoothness coefficients $\gamma_{\srange{\cns{T}}}$ were estimated using the combination of the final 100 $\rvct{x}_{\srange{\cns{T}}}$ samples of each of the five chains for each \ac{ssm}, i.e. a total of 500 samples per \ac{ssm}.

\begin{table}
  \begin{tabular}{r|lll}
     & 
    $\textsc{rmse}(\hat{\vct{\mu}}_{\srange{\cns{T}}},\vct{\mu}_{\srange{\cns{T}}})$ & 
    $\textsc{rmse}(\hat{\vct{\sigma}}_{\srange{\cns{T}}},\vct{\sigma}_{\srange{\cns{T}}})$ & 
    $\textsc{rmse}(\hat{\gamma}_{\srange{\cns{T}}},\gamma_{\srange{\cns{T}}})$\\
    \hline
    Minimum & $1.41 \times 10^{-1}$ & $3.39 \times 10^{-2}$ & $2.34 \times 10^{-3}$ \\
    Median  & $1.41 \times 10^{-1}$ & $3.40 \times 10^{-2}$ & $2.36 \times 10^{-3}$ \\ 
    Maximum & $1.42 \times 10^{-1}$ & $3.40 \times 10^{-2}$ & $2.48 \times 10^{-3}$ \\
    \hline
    Localisation radius $r$ & 0.068 & 0.160 & 0.092
  \end{tabular}
  \caption{Values of metrics at optimal localisation radii for local \ac{etkf} on linearly observed \ac{ks} model.}
  \label{tab:ks-linear-local-etkf-performance}
\end{table}

\begin{table}
  \begin{tabular}{r|lll}
    & 
   $\textsc{rmse}(\hat{\vct{\mu}}_{\srange{\cns{T}}},\vct{\mu}_{\srange{\cns{T}}})$ & 
   $\textsc{rmse}(\hat{\vct{\sigma}}_{\srange{\cns{T}}},\vct{\sigma}_{\srange{\cns{T}}})$ & 
   $\textsc{rmse}(\hat{\gamma}_{\srange{\cns{T}}},\gamma_{\srange{\cns{T}}})$\\
   \hline
   Minimum & $2.87 \times 10^{-1}$ & $1.04 \times 10^{-1}$ & $4.44 \times 10^{-3}$ \\
   Median  & $2.88 \times 10^{-1}$ & $1.04 \times 10^{-1}$ & $4.49 \times 10^{-3}$ \\ 
   Maximum & $2.91 \times 10^{-1}$ & $1.05 \times 10^{-1}$ & $4.64 \times 10^{-3}$ \\
   \hline
   Localisation radius $r$ & 0.064 & 0.156 & 0.020
 \end{tabular}
  \caption{Values of metrics at optimal localisation radii for local \ac{etkf} on non-linearly observed \ac{ks} model.}
  \label{tab:ks-non-linear-local-etkf-performance}
\end{table}

As for the \ac{st} \acp{ssm}, we used $\cns{P} = 10^2$ particles for all the local ensemble filters runs on the \ac{ks} \acp{ssm}. For the local \ac{etkf} we performed an equivalent grid search as for the \ac{st} models, performing five independent runs for each localisation radius $r \in \lbrace 0.010, 0.012, \dots 0.160\rbrace$ for both the linearly and non-linearly observed \ac{ks} \acp{ssm}. The results are summarised in \cref{tab:ks-linear-local-etkf-performance,tab:ks-non-linear-local-etkf-performance}, with plots of the full grid search results shown in \cref{fig:full-letkf-grid-search-results} in \cref{app:full-letkf-grid-search-results}.

Although the absolute values of the \ac{rmse} metrics in \cref{tab:ks-linear-local-etkf-performance} are higher than for the local \ac{etkf} runs on the linear-Gaussian \ac{st} model, given the non-linear state dynamics in the \ac{ks} model mean the filtering distributions are no longer constrained to remain Gaussian, the local \ac{etkf} performs remarkably well on the linearly-observed \ac{ks} \ac{ssm}, recovering relatively accurate estimates of the filtering distribution means, standard deviations and smoothness coefficients. This is concordant with the widespread empirical success of local \ac{enkf} approaches even when applied to models with non-linear state dynamics \citep{evensen2009data}, but also suggests that the filtering distributions in this case may have remained close to Gaussian despite the non-linear dynamics.

Swapping the linear observations for a non-linear observation operator however can be seen to have a detrimental effect on the accuracy of the local \ac{etkf} estimates of the filtering distributions. The optimal values achieved for each of the three \ac{rmse} metrics shown for the non-linearly observed case in \cref{tab:ks-non-linear-local-etkf-performance} show significant increases over the corresponding figures for the linearly observed case in \cref{tab:ks-linear-local-etkf-performance}, with the errors in the standard deviation estimates showing the largest increase. This highlights that although local \ac{enkf} methods are robust to some degree of non-linearity in the dynamics or observation model, performance is still sensitive to strong departures from Gaussianity. 

\begin{figure}[t!]
  \centering
  \begin{subfigure}[b]{0.45\linewidth}
    \centering
  \includegraphics[width=\linewidth]{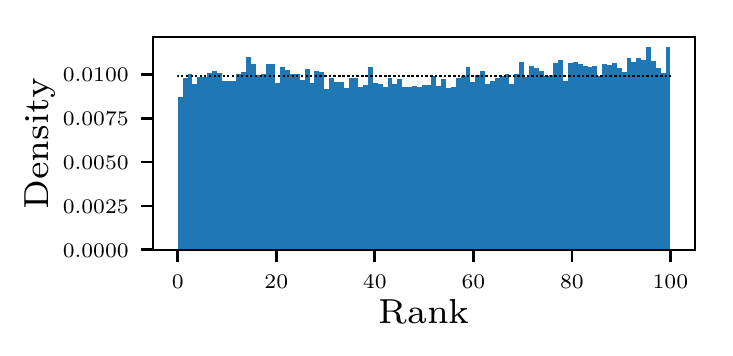}
  \caption{$r=0.068$, linear observations.}
  \label{sfig:ks-linear-letkf-rank-hist}
  \end{subfigure}
  \begin{subfigure}[b]{0.45\linewidth}
    \centering
    \includegraphics[width=\linewidth]{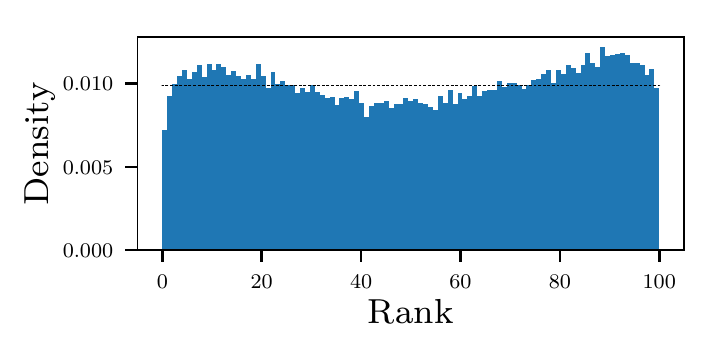}
    \caption{$r=0.064$, non-linear observations.}
    \label{sfig:ks-nonlinear-letkf-rank-hist}
  \end{subfigure}
  \caption{Rank histograms for single local \ac{etkf} runs on \ac{ks} \acp{ssm}.}
  \label{sfig:ks-letkf-rank-hist}
\end{figure}

The effect of the non-Gaussianity induced by the non-linear observation operator can also be seen by comparing rank histograms (i.e. the ranks of the true state values within the ensemble across all time and spatial indices) for single runs of the local \ac{etkf} on the linearly and non-linearly observed \ac{ks} \acp{ssm}, as shown in \cref{sfig:ks-linear-letkf-rank-hist} and \cref{sfig:ks-nonlinear-letkf-rank-hist} respectively. The localisation radius $r$ was set to the value found in the grid searches to give the lowest mean estimate \ac{rmse}. For a well calibrated ensemble the rank histograms should be close to uniform (indicated by the dashed black line on the plots). While for the linearly observed case the minor departures from uniformity can be plausibly attributed to sampling noise, the histogram for the non-linearly observed case has a clear `double-humped' non-uniform shape, with this suggesting the ensemble estimates of the filter distributions have greater kurtosis than the true filtering distributions.  

\begin{figure}[!t]
  \centering
  \includegraphics[width=\textwidth]{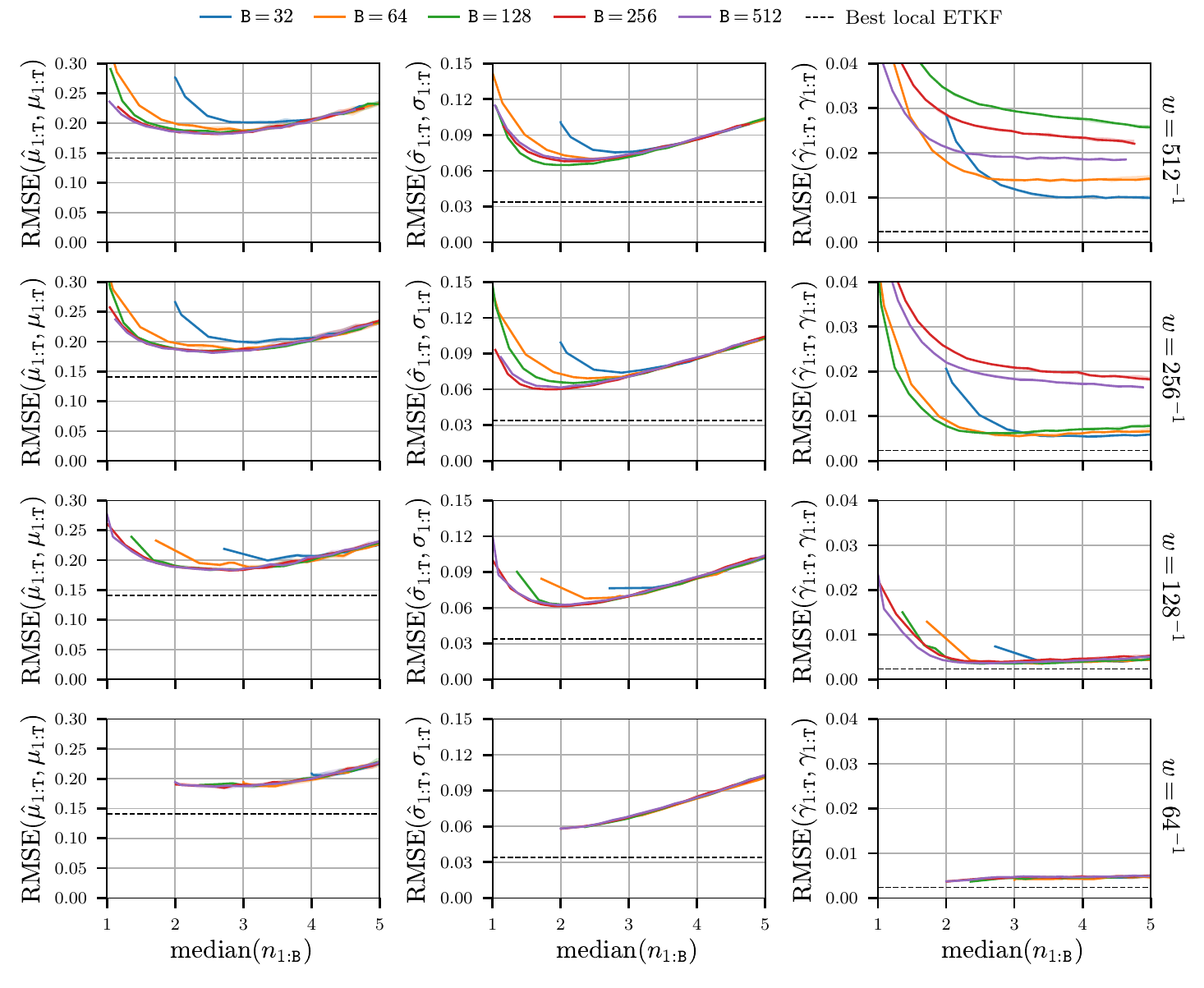}
  \caption{Comparison of accuracy of \ac{sletpf} estimates on linearly observed \ac{ks} \ac{ssm}.
  }
  \label{fig:ks-linear-sletpf-eff-obs-vs-rmse-mean-std-smoothness}
\end{figure}

\begin{figure}[t!]
  \centering
  \includegraphics[width=\textwidth]{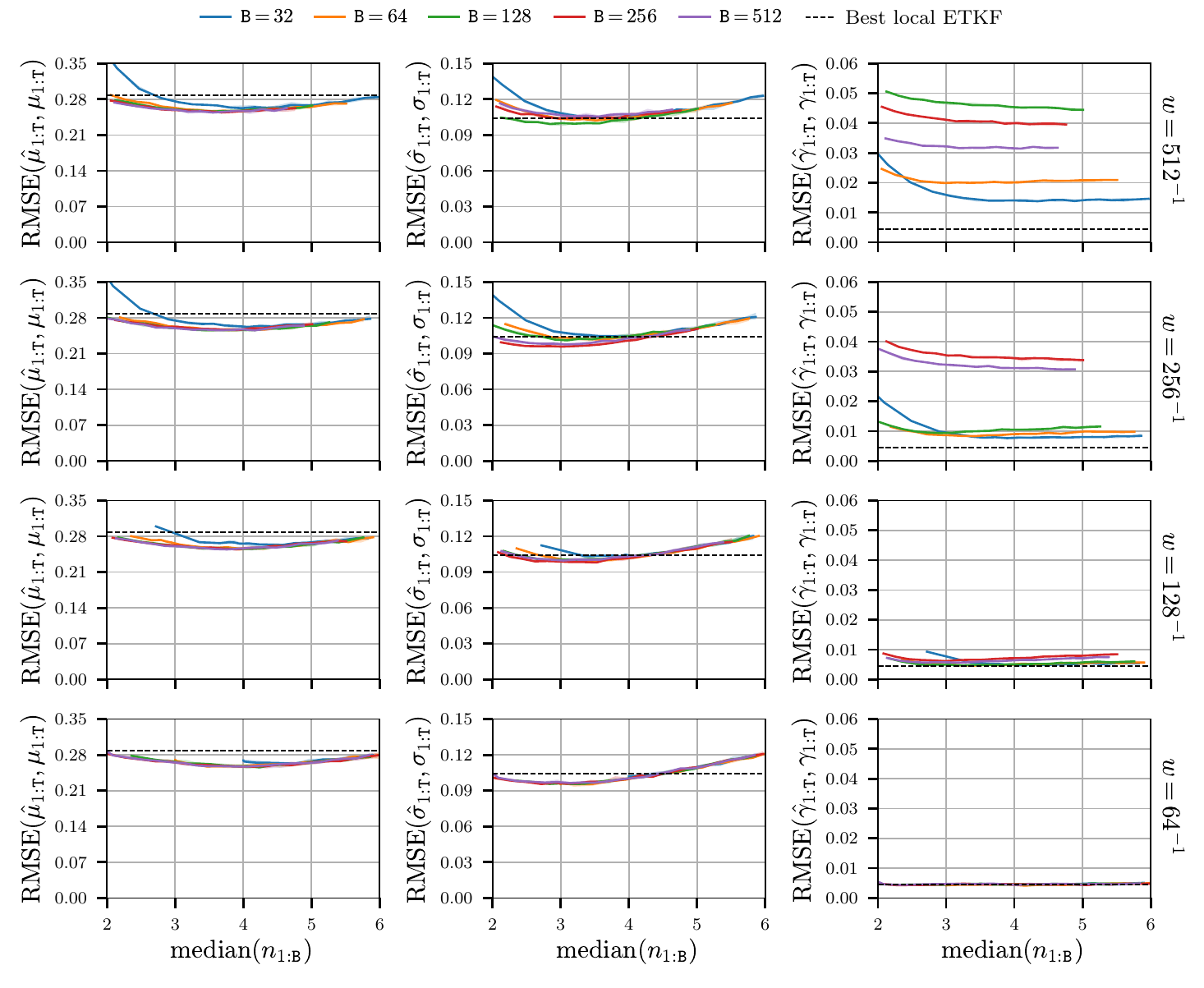}
  \caption{Comparison of accuracy of \ac{sletpf} estimates on non-linearly obs. \ac{ks} \ac{ssm}. 
  } 
  \label{fig:ks-nonlinear-sletpf-eff-obs-vs-rmse-mean-std-smoothness}
\end{figure}

For the \ac{sletpf} runs we use the same method to construct the \acp{pou} as described in the preceding section for the \ac{st} model experiments. We again considered \acp{pou} with $\cns{B} \in \lbrace 32, 64, 128, 265, 512 \rbrace$ number of patches and smoothing kernel widths of $w \in \lbrace 512^{-1}, 256^{-1}, 128^{-1}, 64^{-1} \rbrace$. For each $(\cns{B}, w)$ pair we tested all localisation radii $r \in \lbrace 0.001, 0.002, \dots 0.030 \rbrace$ where $\median(n_{\srange{\cns{B}}})$ was in the range $[1, 5]$ for the linearly observed \ac{ks} \ac{ssm} and in the range $[2, 6]$ for the non-linearly observed \ac{ks} \ac{ssm}. For each $(\cns{B}, w, r)$ parameter triple tested, we performed five independent filtering runs, with the median values recorded for the three metrics shown by the coloured curves in \cref{fig:ks-linear-sletpf-eff-obs-vs-rmse-mean-std-smoothness} for the linearly observed \ac{ssm} and in \cref{fig:ks-nonlinear-sletpf-eff-obs-vs-rmse-mean-std-smoothness} for the non-linearly observed \ac{ssm}, along with the best values achieved by local \ac{etkf} on each metric by the dashed horizontal lines. The plots in \cref{fig:ks-linear-sletpf-eff-obs-vs-rmse-mean-std-smoothness,fig:ks-nonlinear-sletpf-eff-obs-vs-rmse-mean-std-smoothness} have the same format as  \cref{fig:st-linear-sletpf-eff-obs-vs-rmse-mean-std-smoothness,fig:st-nonlinear-sletpf-eff-obs-vs-rmse-mean-std-smoothness} for the \ac{st} model experiments.

From the linearly observed \ac{ks} \ac{ssm} results in \cref{fig:ks-linear-sletpf-eff-obs-vs-rmse-mean-std-smoothness} we see that the \ac{sletpf} was outperformed across all parameter settings and metrics by the best local \ac{etkf} results. This reinforces the point that local \ac{enkf} methods are a strongly performant approach and can often be the best choice even in models with non-linear dynamics, where the Gaussianity assumptions are not valid, due to their robust performance when using small ensemble sizes. A further advantage of local \ac{enkf} methods over local \acp{pf} is that they naturally maintain smoothness properties of the state field particles as evidenced by the low smoothness coefficient errors achieved by the local \ac{etkf} across all model configurations. Local \ac{pf} type approaches such as the \ac{sletpf} algorithm proposed here should generally therefore be considered as a fallback solution for cases where local \ac{enkf} methods are known, or at least suspected, to give poor accuracy.

Considering the performance of the \ac{sletpf} on the linearly observed \ac{ks} \ac{ssm} for different $(\cns{B}, w)$ parameter settings we see similar trends as observed for the \ac{st} model experiments though with some difference in the details. The differences in performances on the mean and standard deviation \ac{rmse} metrics for \acp{pou} with different numbers of patches $\cns{B}$ for a fixed smoothing kernel width $w$ show less variation than seen in the \ac{st} model experiments. Even for the hard \acp{pou} case without smoothing ($w = 512^{-1}$, top-row of \cref{fig:ks-linear-sletpf-eff-obs-vs-rmse-mean-std-smoothness}), only the runs with a \ac{pou} with $\cns{B} = 32$ patches show a significant drop in mean and standard deviation estimate accuracies across most $\median(n_{\srange{\cns{B}}})$ values, and for the $\cns{B} = 32$ case the relative drops in accuracies are still quite minor. The most obvious effect of increasing the kernel width $w$ in this model is therefore in the improved accuracy of the smoothness coefficient estimates for larger $w$ values. This suggests that in the \ac{ks} model, although using a smoother \ac{pou} does reduce the introduction of artificial discontinuities into the state field particles, these discontinuities have less of a negative effect on filtering performance than for the \ac{st} model, perhaps due to a stronger diffusive smoothing element to the model dynamics.

The results for the non-linearly observed \ac{ks} \ac{ssm} in \cref{fig:ks-nonlinear-sletpf-eff-obs-vs-rmse-mean-std-smoothness} show similar relative performances for the different \ac{sletpf} configurations as for the linearly observed case, with a general increase in the absolute \ac{rmse} values across the board.  The corresponding increase in the \ac{rmse} values for optimal tunings of the local \ac{etkf} are however significantly larger, meaning that for this model the \ac{sletpf} approaches show a minor improvement in the accuracy of the mean estimates compared to the local \ac{etkf} across virtually all configurations and performs comparably in terms of the accuracy of the standard deviations estimates, having slightly better performance for some configurations and slightly poorer for others. Again the smoothness of the \ac{pou} used does not seem to have a strong effect on performance in terms of the mean and standard deviation estimates here, with the main change as the smoothing kernel width $w$ is increased the improved accuracy of the smoothness coefficient estimates corresponding to improved reproduction of the smoothness of the fields under the true filtering distributions.

\begin{figure}[t!]
  \centering
  \begin{subfigure}[b]{0.45\linewidth}
    \centering
  \includegraphics[width=\linewidth]{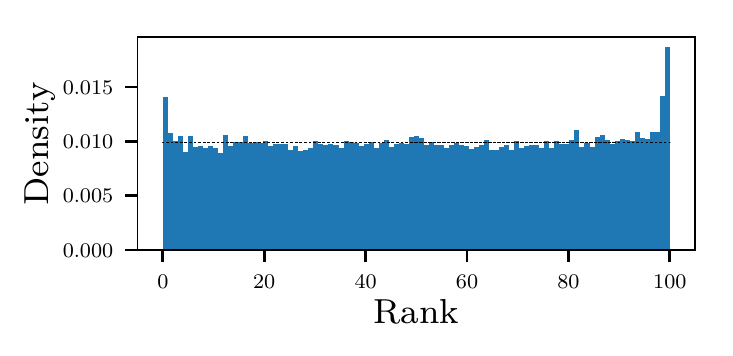}
  \caption{$\cns{B}=512, w = 512^{-1}, r = 0.040$.}
  \label{sfig:ks-nonlinear-letpf-rank-hist}
  \end{subfigure}
  \begin{subfigure}[b]{0.45\linewidth}
    \centering
    \includegraphics[width=\linewidth]{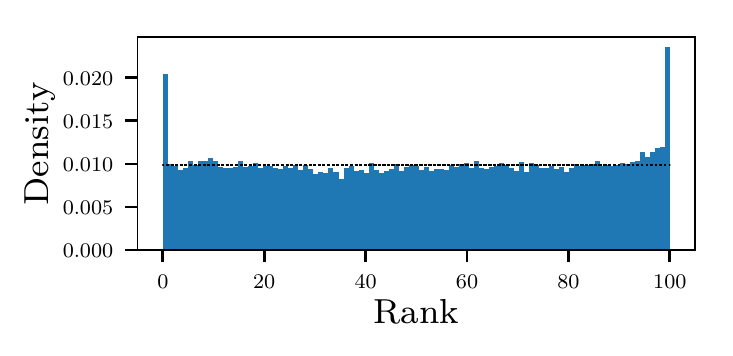}
    \caption{$\cns{B}=64, w = 128^{-1}, r = 0.022$.}
    \label{sfig:ks-nonlinear-sletpf-64-rank-hist}
  \end{subfigure}
  \caption{Rank histograms for single \ac{sletpf} runs on non-linearly observed \ac{ks} \ac{ssm}.}
  \label{fig:ks-nonlinear-sletpf-rank-hist}
\end{figure}

As for the local \ac{etkf} ensemble estimates of the \ac{ks} \ac{ssm} filtering distributions, we can also use rank histograms for the \ac{sletpf} ensembles as an alternative check of the calibration of the filtering distribution estimates. The rank histogram for an ensemble generated for the non-linearly observed \ac{ks} \ac{ssm} by a \ac{sletpf} with a \ac{pou} with $\cns{B} = 512$ patches and kernel width $w = 512^{-1}$ (i.e. corresponding to the per-node local \ac{etpf}) is shown in \cref{sfig:ks-nonlinear-letpf-rank-hist}, and for an ensemble generated for the non-linearly observed \ac{ks} \ac{ssm} by a \ac{sletpf} with a \ac{pou} with $\cns{B} = 64$ patches and kernel width $w = 128^{-1}$ in \cref{sfig:ks-nonlinear-sletpf-64-rank-hist}. In both cases the localisation radius $r$ was set to the value from the grid search giving the lowest mean estimate \ac{rmse}. Compared to the corresponding rank histogram for the local \ac{etkf} in \cref{sfig:ks-nonlinear-letkf-rank-hist}, the histograms for both \ac{sletpf} configurations are much closer to uniform. The peaks at the extreme ranks in both histograms are characteristic of the ensembles underestimating the dispersion of the filtering distribution in the tails, with this discrepancy appearing to be stronger in the \ac{sletpf} using fewer patches here.

\begin{figure}[t!]
  \centering
  \includegraphics[width=\textwidth]{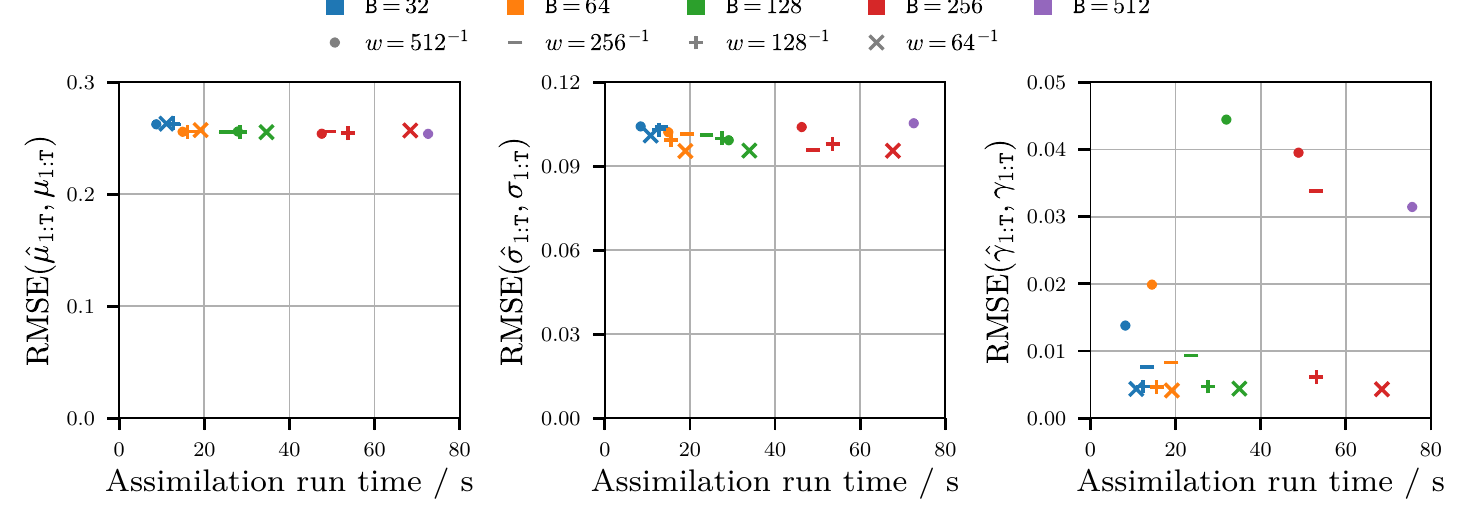}
  \caption{Accuracy versus run time for \ac{sletpf} in non-linearly observed \ac{ks} \ac{ssm}.
  } 
  \label{fig:ks-nonlinear-sletpf-runtime-vs-rmse-mean-std-smoothness}
\end{figure}

As in \cref{fig:st-nonlinear-sletpf-runtime-vs-rmse-mean-std-smoothness} for the transformed \ac{st} model runs, it is instructive to also compare the relative computational cost of the different \ac{sletpf} configurations versus their performance on the three filtering accuracy metrics. \cref{fig:ks-nonlinear-sletpf-runtime-vs-rmse-mean-std-smoothness} shows the time taken to perform the assimilation updates (horizontal axes) versus the value recorded for each of the three \ac{rmse} metrics (vertical axes), for each of the $(\cns{B}, w)$ \ac{pou} configurations.  The markers show the median values across the five runs for the localisation radius $r$ which achieved the minimum value for that particular metric for the $(\cns{B}, w)$ values in question. Due to the decreased drop-off in filtering accuracy for \acp{pou} with fewer number of patches $\cns{B}$ compared to \ac{st} models, here we see we are able to achieve even larger improvements in computational efficiency compared to the local \ac{etpf} scheme of \citet{cheng2015assimilating} (corresponding to $\cns{B} = 512$, $w=51^{-1}$, $\color{Orchid}\bullet$) while retaining the same filtering accuracy. In particular the \acp{sletpf} with $\cns{B} = 64$ patches in the \ac{pou} ($\color{orange}\bullet+-\times$) are able to achieve the same mean estimate accuracy, a slight improvement in the accuracy of the standard deviation estimates, and a substantial improvement in the accuracy of the smoothness coefficient estimates, while having an assimilation time that is around a quarter of the \ac{sletpf} which computes separate \ac{ot} transport maps for each mesh node ($\cns{B} = 512$). Further in this case the filtering accuracy is largely unaffected by the choice of smoothing kernel width $w$, other than an improvement in the smoothness coefficient estimates for larger $w$ values. At the cost of a slight increase in all three \acp{rmse}, the \acp{sletpf} with \acp{pou} with $\cns{B} = 32$ patches give a further approximate factor two decrease in assimilation time, leading to around a eight times decrease in assimilation time compared to the per-node local \ac{etpf}.
\section{Discussion}\label{sec:conclusion}

In this article we have proposed a new scheme for constructing local particle filters for state inference in \ac{spde} models of spatially-extended dynamical systems. The local \ac{etpf} \citep{cheng2015assimilating} although having the desirable property of improved robustness to non-Gaussianity in the filtering distributions compared to local \ac{enkf} approaches has two key shortcomings: (i) the state fields produced by the assimilation step fail to maintain the smoothness properties of the predictive ensemble members, potentially leading to numerical instabilities when used to filter \ac{spde} models and (ii) as an \ac{ot} problem must be solved for every node in the spatial mesh, the assimilation updates can be costly for dense meshes.

Our approach to solving both issues is to softly partition the spatial domain using a \emph{partition of unity}: a finite set of non-negative bump functions which tile the domain and sum to unity at all points. By computing an \ac{ot} map for the patch of the spatial domain associated with each bump function and then using the bump functions to smoothly interpolate these maps across the domain, we are able to smoothly combine different regions of the predictive ensemble particles. 

As well as allowing the smoothness of the spatial fields to be maintained during the assimilation step, the proposed approach reduces the $\otilde(\cns{M}\cns{P}^3)$ cost of the per-node local \ac{etpf} assimilation updates to $\otilde(\cns{B}\cns{P^3} + |\set{M}|\cns{P}^2 + \cns{M}\cns{P})$. If we increase the mesh resolution by using a larger number of nodes $\cns{M}$, while keeping the number of patches $\cns{B}$ and number of subsampled nodes $|\set{M}|$ fixed, the computational cost of the assimilation update only need to scale at rate $\otilde(\cns{M}\cns{P})$ with $\cns{M}$, which could be considered as the lower bound for an update to $\cns{P}$ particles of $\otilde(\cns{M})$ dimension.

We demonstrated in the numerical experiments that the resulting scheme is able to produce, at often significantly reduced computational cost, ensemble estimates of the filtering distributions for state space models with equivalent accuracy and improved smoothness compared to the local \ac{etpf} of \citet{cheng2015assimilating}. Although the experiments were restricted to models on one-dimensional spatial domains, in most applications of interest the spatial domain will be two or three-dimensional. Our proposed scheme naturally carries over to this setting and as the mesh sizes in such models will tend to be significantly higher, the potential computational savings are even larger. Further, while we concentrated here on filtering in spatial models which are observed at point locations, our scheme could be extended to models with spatially distributed observations by partitioning the spatial domain according to the geometry of the observation processes.

The localisation approach to overcoming weight degeneracy when applying \acp{pf} to spatial models considered here could also be combined with other methods for improving \ac{pf} performance in high-dimensional \acp{ssm}. In particular \emph{tempering} approaches split the usual single  prediction and assimilation update per observation time into multiple updates which target a sequence of distributions bridging between the filtering distributions at adjacent observation times \citep[][]{frei2013bridging,johansen2015blocks,beskos2017stable,svensson2018learning,herbst2019tempered}. Tempering could be paired with our framework to further improve its robustness to high-dimensional and strongly informative observations, with the use of multiple assimilation updates per observation time when tempering making the reduced computational cost and improved smoothness preservation of our approach particularly important.

\newpage

\appendix

\numberwithin{figure}{section}
\numberwithin{table}{section}

\section{Ensemble transform Kalman filter}
\label{app:ensemble-transform-kalman-filter}

In this Appendix we describe the details of the \ac{etkf} assimilation update \citep{bishop2001adaptive} and show how it can be expressed in the form of the \ac{letf} framework discussed in \cref{subsec:linear-ensemble-transform-filters}. We first introduce predictive and filtering \emph{ensemble matrices} respectively defined as
\begin{equation} \label{eq:ensemble-matrices}
  \pred{\rmtx{X}}_{t} = \lsb
    \pred{\rvct{x}}\pss{1}_{t} ~
    \pred{\rvct{x}}\pss{2}_{t} ~
    \cdots ~
    \pred{\rvct{x}}\pss{\cns{P}}_{t}
  \rsb\tr
  \quad\textrm{and}\quad
  \rmtx{X}_{t} = \lsb
    \rvct{x}\pss{1}_{t} ~
    \rvct{x}\pss{2}_{t} ~
    \cdots ~
    \rvct{x}\pss{\cns{P}}_{t}
  \rsb\tr.
\end{equation}
Using the following linear operators
\begin{equation}\label{eq:delta-and-epsilon-defs}
  \vct{\varepsilon}= \recip{\cns{P}}\onevct[\cns{P}]\tr,
  \qquad
  \mtx{\Delta} = \recip{\sqrt{\cns{P}-1}} ( \idmtx[\cns{P}] - \onevct[\cns{P}]\vct{\varepsilon}),
\end{equation}
the predictive and filtering ensemble means can then be compactly expressed
\begin{equation}\label{eq:empirical-means}
  \pred{\rvct{m}}_t\tr = \vct{\varepsilon}\pred{\rmtx{X}}_t,
  \quad\textrm{and}\quad
  \rvct{m}_t\tr = \vct{\varepsilon}\rmtx{X}_t,,
\end{equation}
and similarly the predictive and filtering ensemble covariances can be written
\begin{equation}\label{eq:empirical-covariances}
  \pred{\rmtx{C}}_t = (\mtx{\Delta}\pred{\rmtx{X}}_t)\tr(\mtx{\Delta}\pred{\rmtx{X}}_t)
  \quad\textrm{and}\quad
  \rmtx{C}_t = (\mtx{\Delta}\rmtx{X}_t)\tr(\mtx{\Delta}\rmtx{X}_t).
\end{equation}
Assuming initially linear-Gaussian observations as in \cref{eq:linear-gaussian-observation} then by substituting the expressions for the empirical covariances \cref{eq:empirical-covariances} into the Kalman filter covariance assimilation update in \cref{eq:kf-assimilation-update-covariance} and applying the identity $\idmtx[\cns{P}] - \mtx{\Delta}\pred{\rmtx{Y}}_t (\mtx{R}_t + \pred{\rmtx{Y}}_t\tr\mtx{\Delta}^2\pred{\rmtx{Y}}_t)^{-1}\pred{\rmtx{Y}}_t\tr\mtx{\Delta} = (\idmtx[\cns{P}] + \mtx{\Delta}\pred{\rmtx{Y}}_t\mtx{R}_t^{-1}\pred{\rmtx{Y}}_t\tr\mtx{\Delta})^{-1}$ with $\pred{\rmtx{Y}}_t = \pred{\rmtx{X}}_t\mtx{H}_t\tr$ we have
\begin{equation}\label{eq:etkf-covariance-update}
  (\mtx{\Delta}\rmtx{X}_t)\tr(\mtx{\Delta}\rmtx{X}_t) =
  (\mtx{\Delta}\pred{\rmtx{X}}_t)\tr
  \left(
  \idmtx[\cns{P}]
  +
  \mtx{\Delta}
  \pred{\rmtx{Y}}_t
  \mtx{R}^{-1}_t
  \pred{\rmtx{Y}}_t\tr
  \mtx{\Delta}
  \right)^{-1}
  (\mtx{\Delta}\pred{\rmtx{X}}_t).
\end{equation}
Definining $\rmtx{S}_t$ as the symmetric matrix square-root of the central term in the right-hand-side of \cref{eq:etkf-covariance-update}, i.e.
\begin{equation}\label{eq:etkf-square-root-definition}
  \rmtx{S}^2_t =
  \rmtx{S}_t\rmtx{S}_t = 
  \left(
  \idmtx[\cns{P}]
  +
  \mtx{\Delta}\pred{\rmtx{Y}}_t\mtx{R}^{-1}_t\pred{\rmtx{Y}}_t\tr\mtx{\Delta}
  \right)^{-1}
\end{equation}
then we can compute a family of solutions of \cref{eq:etkf-covariance-update} for the filtering ensemble projection $\mtx{\Delta}\rmtx{X}_t$ in terms of the predictive ensemble projection $\mtx{\Delta}\pred{\rmtx{X}}_t$ as
\begin{equation}
  \mtx{\Delta}\rmtx{X}_t = \mtx{Q}\rmtx{S}_t \mtx{\Delta}\pred{\rmtx{X}}_t
\end{equation}
where $\mtx{Q}$ is an arbitary $\cns{P}\times\cns{P}$ orthogonal matrix. For the \ac{etkf} generally $\mtx{Q} = \idmtx[\cns{P}]$ is chosen, corresponding to directly transforming by the symmetric square-root.

Now considering the Kalman assimilation update for the mean in \cref{eq:kf-assimilation-update-mean}, subsituting the expressions for the ensemble empirical means and covariances in \cref{eq:empirical-means,eq:empirical-covariances} and using the definition of the square-root matrix $\rmtx{S}_t$ in \cref{eq:etkf-square-root-definition} we have that
\begin{equation}
  \vct{\varepsilon}\rmtx{X}_t =
  \vct{\varepsilon}\pred{\rmtx{X}}_t + 
  (
    \vct{y}_t\tr - 
    \vct{\varepsilon}\pred{\rmtx{Y}}_t
  )
  \mtx{R}_t^{-1}\pred{\rmtx{Y}}_t\tr\mtx{\Delta}
  \rmtx{S}_t^2
  \mtx{\Delta}\pred{\rmtx{X}}_t.
\end{equation}
From the definition of $\mtx{\Delta}$ in \cref{eq:delta-and-epsilon-defs} we have that $\idmtx[\cns{P}] = \onevct[\cns{P}]\vct{\varepsilon} + \sqrt{\cns{P}-1}\mtx{\Delta}$ and so
\begin{align}
  \rmtx{X}_t &= 
  \onevct[\cns{P}]\vct{\varepsilon}\rmtx{X}_t + \sqrt{\cns{P}-1}\mtx{\Delta}\rmtx{X}_t\\
  &=\label{eq:etkf-assimilation-update}
  \left(
  \onevct[\cns{P}] \vct{\varepsilon} +
  \onevct[\cns{P}] (
    \vct{y}_t\tr - 
    \vct{\varepsilon}\pred{\rmtx{Y}}_t
  )
  \mtx{R}_t^{-1}
  \pred{\rmtx{Y}}_t\tr
  \mtx{\Delta} \rmtx{S}_t^2 \mtx{\Delta} +
  \sqrt{\cns{P}-1}  \rmtx{S}_t \mtx{\Delta}
  \right)
  \pred{\rmtx{X}}_t,
\end{align}
with the matrix term in parentheses defining the coefficients $\rvar{a}_t\pss{\srange{\cns{P}},\srange{\cns{P}}}$ of an \ac{letf} assimilation update as in \cref{eq:letf-assimilation-update}.

In the above it was assumed the observation model is linear-Gaussian. In the case of a more general observation model of the form
\begin{equation}\label{eq:non-linear-additive-gaussian-observation-model}
  \op{G}_t(\vct{x}, \vct{v}) = \op{H}_t(\vct{x}) + \vct{v},
  \quad
  \rvct{v}_t \sim \gau(\vct{0}, \mtx{R}_t),
\end{equation}
where now $\op{H}_t$ is a potentially non-linear operator, then by observing that all occurences of $\mtx{H}_t$ in \cref{eq:etkf-assimilation-update,eq:etkf-square-root-definition} are via $\pred{\rmtx{Y}}_t = \pred{\rmtx{X}}_t\mtx{H}_t\tr$, for non-linear $\op{H}_t$ we can instead define the \emph{predictive observation ensemble matrix} $\pred{\rmtx{Y}}_t$ as
\begin{equation}
  \pred{\rmtx{Y}}_{t} = \lsb
    \op{H}_t(\pred{\rvct{x}}\pss{1}_{t}) ~
    \op{H}_t(\pred{\rvct{x}}\pss{2}_{t}) ~
    \cdots ~
    \op{H}_t(\pred{\rvct{x}}\pss{\cns{P}}_{t})
  \rsb\tr.
\end{equation}
The \ac{etkf} formulation of a square-root \ac{enkf} has the advantage of only requiring computing cubic-cost matrix operations for matrices of size $\cns{P}\times\cns{P}$ (due to the conditional independence assumptions $\mtx{R}_t$ is block diagonal and so the cost of computing $\mtx{R}_t^{-1}$ is at worst $\mathcal{O}(\cns{L}\cns{K}^3)$ with in general $\cns{K} \ll \cns{P}$).

For all \ac{enkf} methods, the assimilation updates are only consistent with the analytic assimilation update in \cref{eq:assimilation-update} as $\cns{P} \to \infty$ for linear-Gaussian models. In models where the state update and observation operators are only weakly nonlinear, the filtering distribution at each time index $\pi_{t}$ can remain `close' to Gaussian and the \ac{enkf} updates will often give reasonable estimates of the filtering distribution \citep{evensen2009data}. For models with highly non-Gaussian filtering distributions \ac{enkf} methods will typically perform poorly however.

A local version of the \ac{etkf} algorithm was proposed in \citet{hunt2007efficient}. In the global \ac{etkf} assimilation update summarised in \cref{eq:etkf-assimilation-update} the linear transform coefficients depend on the current predictive state ensemble values $\pred{\rvct{x}}_{t}\pss{\srange{\cns{P}}}$ only via a $\cns{P}\times\cns{K}\cns{L}$ observation ensemble matrix $\pred{\rmtx{Y}}_t$. The local \ac{etkf} algorithm scales the dependence of the update coefficients at each mesh node on the columns of $\pred{\rmtx{Y}}_t$ via a \emph{localisation function} $\ell_r : [0, \infty) \to [0, 1]$ satisfying the conditions in \cref{eq:localisation-function-conditions} for some localisation radius $r > 0$, such that observations at a distance more than $r$ from the mesh node are ignored in the corresponding local assimilation update.  

For each of the $\cns{M}$ mesh nodes a \emph{localisation kernel} is then defined by applying $\ell_r$ to the distances between the mesh nodes and the observation locations
\begin{equation}
  \vct{k}_m\tr = [
    \ell_r(d(s_m, s^{\textrm{o}}_1))^{\frac{1}{2}} \, \onevct[\cns{K}]\tr,\,
    \cdots,\,
    \ell_r(d(s_m, s^{\textrm{o}}_{\cns{L}}))^{\frac{1}{2}} \,\onevct[\cns{K}]\tr
  ]
  \quad \forall m\in\range{\cns{M}}.
\end{equation}
We can then define local effective observation noise precision matrices $\tilde{\mtx{R}}^{-1}_{t,\srange{\cns{M}}}$
\begin{equation}\label{eq:local-etkf-obs-prec-matrix}
  \tilde{\mtx{R}}_{t,m}^{-1} = 
  \mtx{R}_{t}^{-1} \odot (\vct{k}_m \vct{k}_m\tr)
  \quad \forall m \in\range{\cns{M}}
\end{equation}
where $\odot$ indicate the elementwise or Hadamard product between equal sized tensors. The local \ac{etkf} assimilation update is then
\begin{equation}\label{eq:local-etkf-assimilation-update}
  \rmtx{X}_{t,m} =
  \left(
  \onevct[\cns{P}] \vct{\varepsilon} +
  \onevct[\cns{P}] (
    \vct{y}_{t}\tr - 
    \vct{\varepsilon}\pred{\rmtx{Y}}_{t}
  )
  \tilde{\mtx{R}}_{t,m}^{-1}
  \pred{\rmtx{Y}}_{t}\tr
  \mtx{\Delta} \tilde{\rmtx{S}}_{t,m}^2 \mtx{\Delta} +
  \sqrt{\cns{P}-1}  \tilde{\rmtx{S}}_{t,m} \mtx{\Delta}
  \right)
  \pred{\rmtx{X}}_{t,m},
\end{equation}
where the local square root matrix $\tilde{\rmtx{S}}_{t,m}$ is defined
\begin{equation}\label{eq:local-etkf-square-root-definition}
  \tilde{\rmtx{S}}^2_{t,m} =
  \tilde{\rmtx{S}}_{t,m}\tilde{\rmtx{S}}_{t,m} = 
  \left(
  \idmtx[\cns{P}]
  +
  \mtx{\Delta}\pred{\rmtx{Y}}_{t}
  \tilde{\mtx{R}}^{-1}_{t,m}
  \pred{\rmtx{Y}}_{t,m}\tr\mtx{\Delta}
  \right)^{-1}.
\end{equation}
This local assimilation update is equivalent to replacing each observation ensemble matrix term $\pred{\rmtx{Y}}_t$ and observation vector term $\vct{y}_t$ in the global assimilation update in \cref{eq:etkf-assimilation-update} with $\pred{\rmtx{Y}}_t \odot (\onevct[\cns{P}] \vct{k}_m\tr)$ and $\vct{y}_t \odot \vct{k}_m$ respectively. As $\vct{k}_m$ has zero entries for all indices corresponding to observation locations more than $r$ in distance from $s_m$, in practice when implementing the local \ac{etkf} assimilation update the computations can be performed with only the non-zero submatrices of $\pred{\rmtx{Y}}_t \odot (\onevct[\cns{P}] \vct{k}_m\tr)$ and $\vct{y}_t \odot \vct{k}_m$ and corresponding submatrix of $\mtx{R}_{t}^{-1}$.

As separate assimilation updates need to be computed for each mesh node the computational cost of the local \ac{etkf} scales linearly with the number of mesh nodes $\cns{M}$. The computation for each mesh node is of order $\mathcal{O}(\cns{P}^3)$ due to requirement to calculate a matrix decomposition of the $\cns{P}\times\cns{P}$ matrix inside the parentheses on the right hand side of \cref{eq:local-etkf-square-root-definition}. On a sequential architecture the overall computation time will therefore have a $\mathcal{O}(\cns{M}\cns{P}^3)$ scaling. As each of the local assimilation updates can be independently computed in parallel, with a large number of parallel compute nodes the assimilation update can still be computed efficiently for models with large mesh sizes $\cns{M}$ however as shown in the numerical experiments in \citet{hunt2007efficient}.
\newpage
\section{Alternative particle filter proposals}
\label{app:alt-particle-filter-proposals}

Rather than propagating according to the forward dynamics of the generative model, it is possible to instead propose new particle values from different conditional distributions (which may depend on future observed values) and adjust the expression for the importance weights in \cref{eq:bootstrap-particle-weights} accordingly. Typically the resulting expression for the importance weights is given in terms of the transition density of the state updates, however as noted previously this density will often be intractable to compute. Alternative state proposals can however instead be formulated by changing the distribution the state noise variables are drawn from. If each state noise vector $\rvct{u}_{t}\pss{p}$ is sampled from a distribution with a known density $d\pss{p}_{t}$ with respect to $\mu_{t}$ and the predictive ensemble particles computed as in \cref{eq:empirical-prediction-update}, then unnormalised importance weights for the propagated particles can be computed as
\begin{equation}\label{eq:generalised-importance-weights}
  \tilde{\rvar{w}}_{t}\pss{p} =
    g_t\lpa \vct{y}_t \gvn \op{F}_{t}(\rvct{x}\pss{p}_{t-1}, \rvct{u}\pss{p}_{t})\rpa
    d\pss{p}_{t}(\rvct{u}\pss{p}_{t})^{-1}
  \quad
  \forall p \in \range{\cns{P}}.
\end{equation}
The corresponding normalised weights can then be used in the empirical filtering distribution approximation in \cref{eq:bootstrap-particle-weights} and resampling update in \cref{eq:pf-assimilation-update}. If we restrict the state noise proposal density $d\pss{p}_{t}$ to be dependent on only the previous particle $\vct{x}\pss{p}_{t-1}$ and current observation $\vct{y}_{t}$ in order to maintain the online nature of the algorithm, then the proposal distributions which minimise the variance of the importance weights have densities with respect to $\mu_{t}$
\begin{equation}\label{eq:optimal-proposal-density-noise-variables}
  d\pss{p}_{t}(\vct{u}) =
  \frac
    {g_t\lpa \vct{y}_t \gvn \op{F}_{t}(\rvct{x}\pss{p}_{t-1}, \vct{u})\rpa}
    {\int_{\set{U}}
       g_t\lpa\vct{y}_t\gvn\op{F}_{t}(\rvct{x}\pss{p}_{t-1}, \vct{u}')\rpa
     \,\mu_t(\dr\vct{u}')}
     \quad
     \forall p \in \range{\cns{P}}.
\end{equation}
In this case the unnormalised weights in \cref{eq:generalised-importance-weights} are independent of the state noise variables $\rvct{u}_{t}\pss{\srange{\cns{P}}}$. Although this `optimal' proposal is more typically expressed as a conditional distribution on $\pred{\rvct{x}}^{p}_{t}$ given $\rvct{x}^{p}_{t-1}$ this alternative formulation is equivalent. In general it will not be possible to generate samples from the optimal proposal, however it may be possible to for example find a tractable approximation to use as a proxy.

In cases where the optimal proposal is tractable or can be well approximated, the resulting \ac{pf} algorithm can significantly outperform the basic bootstrap \ac{pf} in terms of the ensemble size required for a given accuracy in the filtering distribution estimates.
\newpage
\section{Visualisation of smooth local LETF assimilation update}
\label{app:smooth-local-letf-assimilation-vis}

\begin{figure}[!h]
  \includegraphics[width=0.975\textwidth]{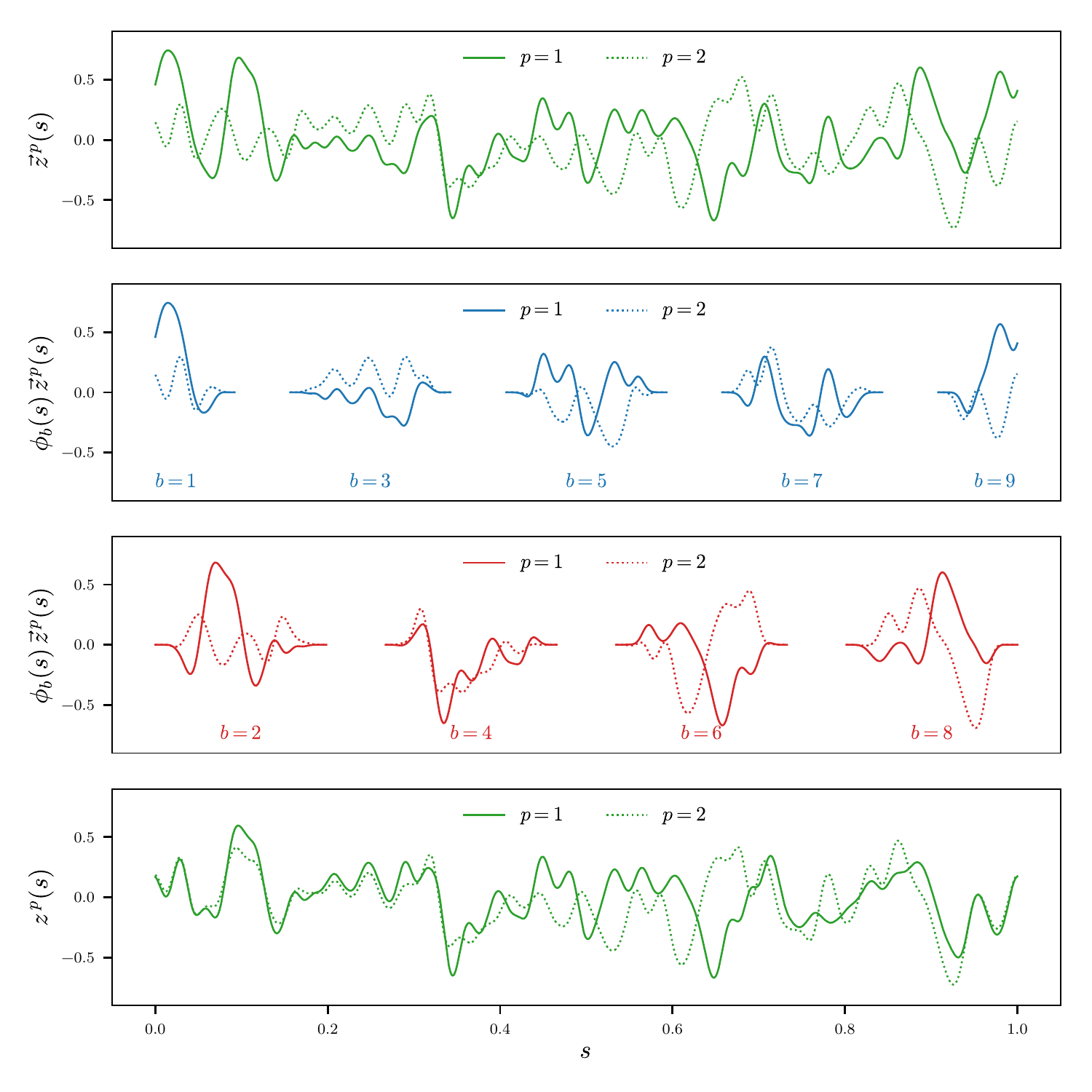}
  \caption{Example of applying smooth local \ac{letf} assimilation update in \cref{eq:sletf-assimilation-update} on a one-dimensional spatial domain $\set{S} = [0,1]$ using the \ac{pou} from \cref{fig:pou-example} with $\cns{P}=2$ particles.} 
  \label{fig:smooth-letf-1d-example}
\end{figure}

Consider a spatial domain which is the same unit interval $\set{S} = [0,1]$ as used in \cref{fig:pou-example} and a \ac{pou} chosen as the smooth bump functions $\phi_{\srange{9}}$ shown there. The top panel in  \cref{fig:smooth-letf-1d-example} shows two smooth predictive distribution particle realisations $\pred{z}\pss{\srange{2}}$. The central two\footnote{The separation of products with odd and even indexed bump functions on to separate panels in \cref{fig:smooth-letf-1d-example} is simply for visual clarity.} panels show the products $\phi_b(s) \pred{z}\pss{\,p}(s)~\forall b\in\range{9},p\in\range{2}$, which can also seen to be smooth functions of the spatial coordinate $s$ and compactly supported on the patches $\hat{\set{S}}_{\srange{9}}$. The bottom panel shows the filtering distribution particle realisations $z\pss{\srange{2}}$ computed using the assimilation update in \cref{eq:sletf-assimilation-update} for a randomly generated set of coefficients $\hat{\rvar{a}}^{\srange{2},\srange{2}}_{\srange{9}}$ satisfying the conditions in \cref{eq:sletf-stochastic-matrix-conditions}, with these post-assimilation fields maintaining the smoothness of the predictive fields.

\newpage
\section{Partitioning the spatial domain}
\label{app:partitioning-the-spatial-domain}

\begin{figure}[!h]

  \centering
  \begin{subfigure}[t]{0.42\linewidth}
  \includegraphics[width=\linewidth]{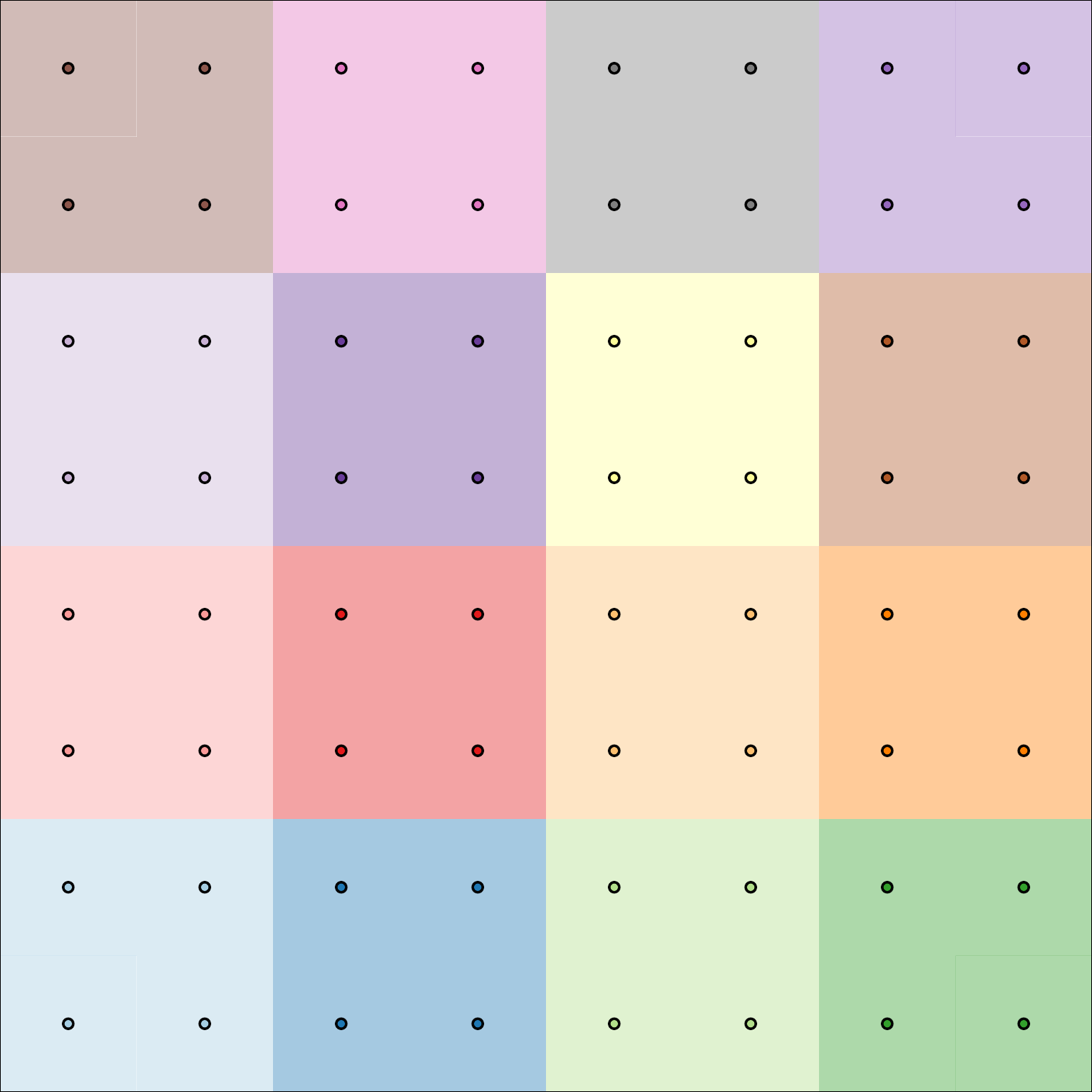}
  \caption{Rectilinear observation locations.}
  \label{sfig:regular-obs-partition}
  \end{subfigure}
  ~~~
  \centering
  \begin{subfigure}[t]{0.42\linewidth}
  \includegraphics[width=\linewidth]{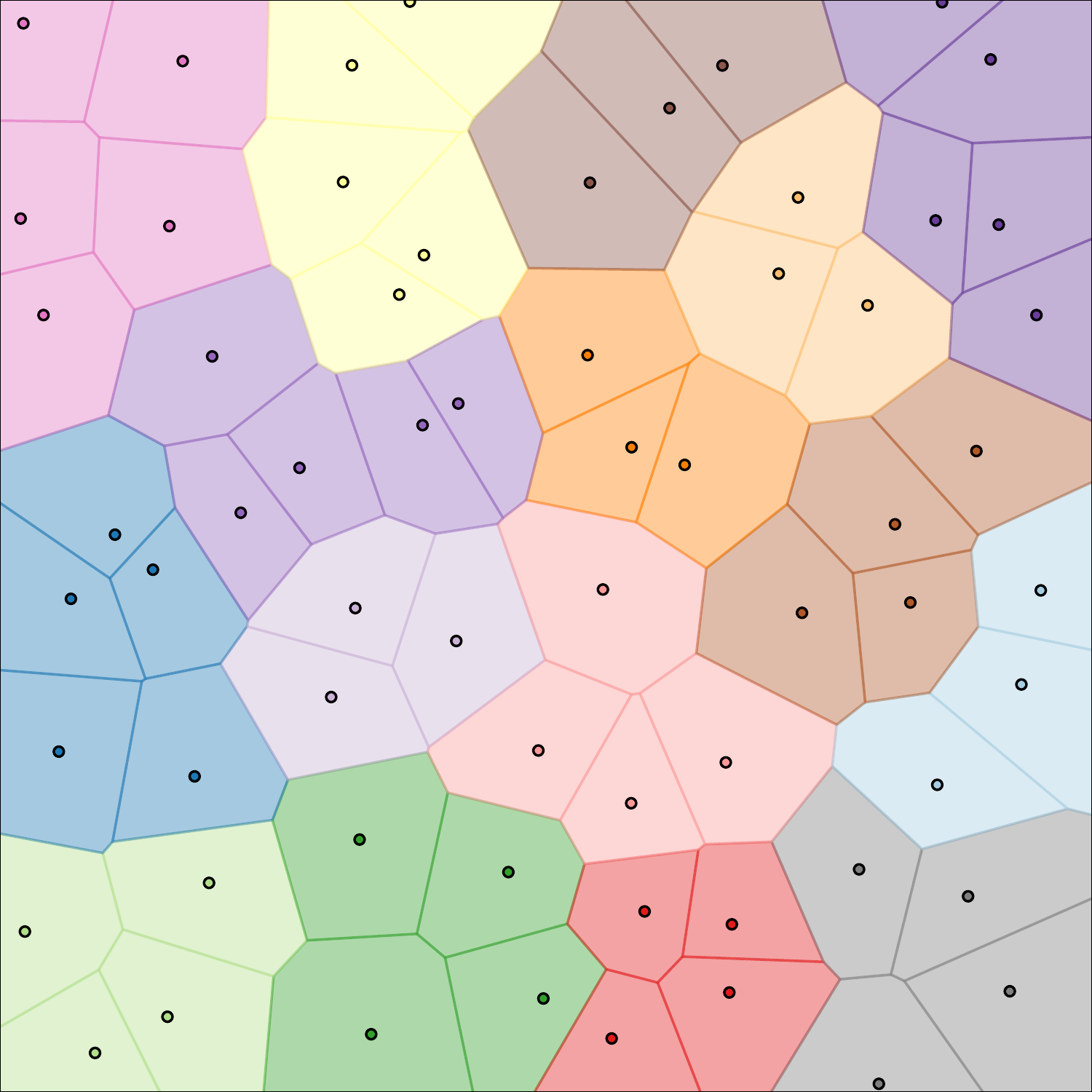}
  \caption{Irregular observation locations.}
  \label{sfig:irregular-obs-partition}
  \end{subfigure}

  \caption{Examples of partitioning a space based on observation locations for a two-dimensional spatial domain. Panel \subref{sfig:regular-obs-partition} shows a partition (indicated by coloured regions) for observations located on a equispaced rectilinear grid (shown by circular markers). Panel \subref{sfig:irregular-obs-partition} shows a partition for an irregularly located set of observations, with the observation locations initially clustered (indicated by colours of markers) before partitioning based on the Voronoi cells associated with each cluster of observation locations (cells shown by bordered polygonal regions).} 
  \label{fig:partition-space-2d-example}
\end{figure}

In order to control the number of observations used to compute each local weight in the \ac{sletpf} scheme, we recommend choosing the partition of the spatial domain used to define the \ac{pou} such that each patch contains roughly the same number of observations. For observations located on a rectilinear grid this can easily be achieved by partitioning the space in to rectilinear blocks aligned with the observation grid and each containing the same number of observations (an example is shown in \cref{sfig:regular-obs-partition}). For irregularly spaced observations, one option is to first group the observation locations in to similarly sized clusters using for example a $k$-means algorithm. The spatial domain can then be partitioned using a Voronoi diagram generated from the observation locations, with all the cells corresponding to observations in a single cluster then merged to form a single contiguous region. This leads to a partition of the spatial domain into a set of regions which each contain a roughly number of observations and such that the numbers of additional observations close to the region boundaries are minimised. A example of applying this scheme to a set of irregularly located observation points is shown in \cref{sfig:irregular-obs-partition}. In both the rectilinear and irregular spacing cases, a soft \ac{pou} can then be generated from the resulting partition by convolving with a mollifier function as described in \cref{subsec:partition-of-unity}.
\newpage
\section{Transformed state-space models}
\label{app:transformed-ssms}

One of our primary motivations for considering \ac{pf}-based methods was the claim that they are more robust to non-Gaussianity in the filtering distributions compared to \ac{enkf} methods. While this can shown to be the case in the large ensemble limit for non-localised \ac{pf} algorithms (including the \ac{etpf}) compared to \ac{enkf} methods, it does not necessarily follow that, when using small ensemble sizes, a local \ac{etpf} would be expected to outperform a local \ac{enkf} in models with non-Gaussian filtering distributions. Further even if there is a benefit to using the local \ac{etpf} compared to the local \ac{enkf}, this does not necessarily carry over to our proposed smooth and scalable local \ac{etpf} scheme.

Therefore to assess the affect on the relative performance of the local ensemble filters methods being considered of non-Gaussianity in the filtering distributions while controlling as far as possible other factors which might affect performance, we use a simple scheme to map a tractable linear-Gaussian \ac{ssm} to a transformed \ac{ssm} with non-Gaussian filtering distributions. In particular let $\op{T} : \set{X} \to \set{X}$ be a diffeomorphism on the state space, with $\op{T}^{-1}$ denoting its inverse, which we assume we can also compute. If we define $\rvct{x}'_t = \op{T}(\rvct{x}_t)~\forall t \in \range{\cns{T}}$ then the conditional distribution on $\rvct{x}'_t$ given observations $\rvct{y}_{\srange{t}}=\vct{y}_{\srange{t}}$ will be $\op{T}_\sharp\pi_t$ for any time index $t \in \range{\cns{T}}$, i.e. the \emph{push-forward} of the filtering distribution $\pi_t$ under the map $\op{T}$. If $\op{T}$ is non-linear then if $\pi_t$ is Gaussian $\op{T}_\sharp\pi_t$ will in general be non-Gaussian.

Importantly for our purposes we can construct a \ac{ssm} acting directly on the transformed states $\rvct{x}'_{\srange{\cns{T}}}$. In particular for a \emph{base} \ac{ssm} with state update and observation operators $\op{F}_{\srange{\cns{T}}}$ and $\op{G}_{\srange{\cns{T}}}$, we can define a $\op{T}$-\emph{transformed} \ac{ssm} with state update and observation operators $\op{F}'_{\srange{\cns{T}}}$ and $\op{G}'_{\srange{\cns{T}}}$ given by
\begin{alignat}{2}
  \label{eq:generate-initial-state-transformed}
  \rvct{x}'_{1} 
  &= \op{F}'_1(\rvct{u}_1) 
  = \op{T} \circ \op{F}_{1}(\rvct{u}_{1}),
  &&\quad \rvct{u}_{1} \sim \mu_{1},
  \\
  \label{eq:generate-next-state-transformed}
  \rvct{x}'_{t} 
  &= \op{F}'_{t}(\rvct{x}'_{t-1}, \rvct{u}_{t}) 
  = \op{T} \circ \op{F}_{t}(\op{T}^{-1}(\rvct{x}'_{t-1}), \rvct{u}_{t}),
  &&\quad \rvct{u}_{t} \sim \mu_{t}
  \quad \forall t \in \range[2]{\cns{T}},
  \\
  \label{eq:generate-observation-transformed}
  \rvct{y}_{t} &= \op{G}'_t(\rvct{x}'_t,\rvct{v}_t) 
  = \op{G}_t(\op{T}^{-1}(\rvct{x}'_{t}), \rvct{v}_{t}),
  &&\quad \rvct{v}_{t} \sim \nu_{t}
  \quad \forall t \in \range{\cns{T}}
\end{alignat}
and with observation densities $g'_{\srange{\cns{T}}}$ defined by
\begin{equation}
  g'_t(\vct{y}_t\gvn\vct{x}'_t) = 
  g_t(\vct{y}_t\gvn\op{T}^{-1}(\vct{x}'_t))
  \quad \forall t \in\range{\cns{T}}.
\end{equation}
We can therefore run ensemble filter algorithms on the $\op{T}$-transformed \ac{ssm} to directly compute ensemble estimates of the transformed filtering distributions $\pi'_{\srange{\cns{T}}}$ with by construction $\pi'_t = \op{T}_{\sharp}\pi_t~\forall t \in\range{\cns{T}}$. If the base \ac{ssm} is linear-Gaussian and so a \ac{kf} can be used to exactly compute the Gaussian filtering distributions $\pi_{\srange{\cns{T}}}$, we can compute accurate unbiased Monte Carlo estimates of expectations under the transformed filtering distributions  $\pi'_{\srange{\cns{T}}}$ as we can generate $\cns{N}$ independent samples from each $\pi'_t$ by generating $\cns{N}$ independent samples from the Gaussian filtering distribution $\pi_t$ and pushing each of the samples through the map $\op{T}$.

This scheme therefore provides a method for constructing a non-Gaussian \ac{ssm} for which we can easily compute accurate Monte Carlo estimates of the true filtering distribution means $\mu_{\srange{\cns{T}}}$, standard deviations $\sigma_{\srange{\cns{T}}}$ and smoothness coefficients $\gamma_{\srange{\cns{T}}}$ as defined in \cref{eq:filtering-dist-mean-and-std-true,eq:filtering-dist-smoothness-coefficient-true} and so evaluate the \ac{rmse} accuracy metrics described in the preceding section for ensemble estimates of the filtering distributions. By using a large number of independent samples $\cns{N}$ in the Monte Carlo estimates we can ensure the $\mathcal{O}(\cns{N}^{-\frac{1}{2}})$ Monte Carlo error is negligible compared to the error in the ensemble estimates.
\newpage
\section{Model details}

\subsection{Stochastic turbulence model}

\label{app:stochastic-turbulence-model-details}

\begin{table}[ht!]
  \begin{tabular}{ll}
    Number of mesh nodes & $\cns{M} = 512$ \\
    Number of observation times & $\cns{T} = 200$ \\
    Number of observation locations & $\cns{L} = 64$ \\
    Time step & $\delta = 2.5$\\
    Diffusion coefficient & $\theta_1 = 4\times10^{-5}$\\
    Advection coefficient & $\theta_2 = 0.1$\\
    Damping coefficient & $\theta_3 = 0.1$\\
    Transformation scale factor & $\theta_4 = 5$\\
    State noise kernel length scale & $\vartheta = 4\times10^{-3}$\\
    State noise kernel amplitude & $\alpha = 0.1$\\
    Observation noise standard deviation & $\varsigma = 0.5$\\
  \end{tabular}
  \caption{Stochastic turbulence model parameter settings}
  \label{tab:st-model-params}
\end{table}

\begin{figure}[ht!]
  \centering
  \begin{subfigure}[b]{\linewidth}
    \centering
    \includegraphics[width=0.9\textwidth]{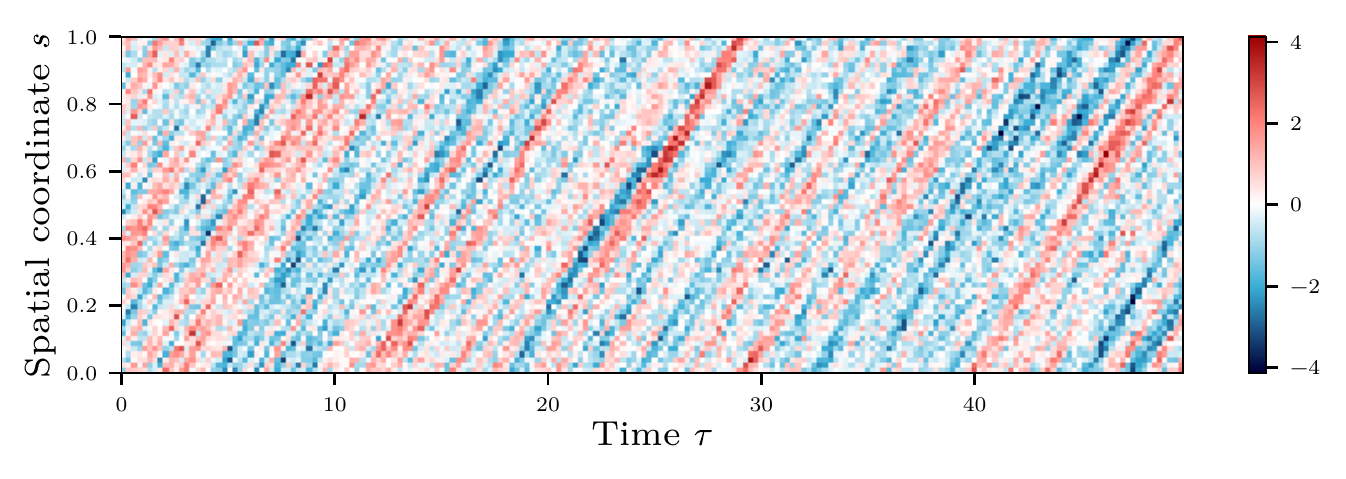}
    \caption{Noisy observation sequence $\vct{y}_{\srange{\cns{T}}}$.}
    \label{sfig:st-sim-obs}
  \end{subfigure}
  \begin{subfigure}[b]{\linewidth}
    \centering
    \includegraphics[width=0.9\textwidth]{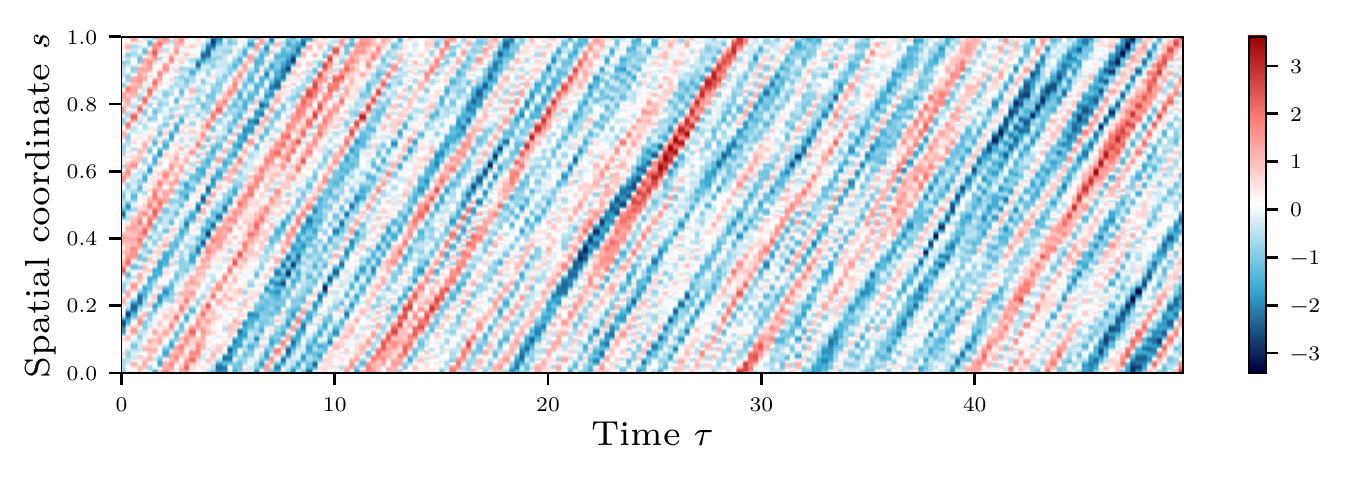}
    \caption{State sequence $\vct{z}_{\srange{\cns{T}}}$ for linear-Gaussian \ac{ssm}.}
    \label{sfig:st-linear-sim-state}
  \end{subfigure}
  \begin{subfigure}[b]{\linewidth}
    \centering
    \includegraphics[width=0.9\textwidth]{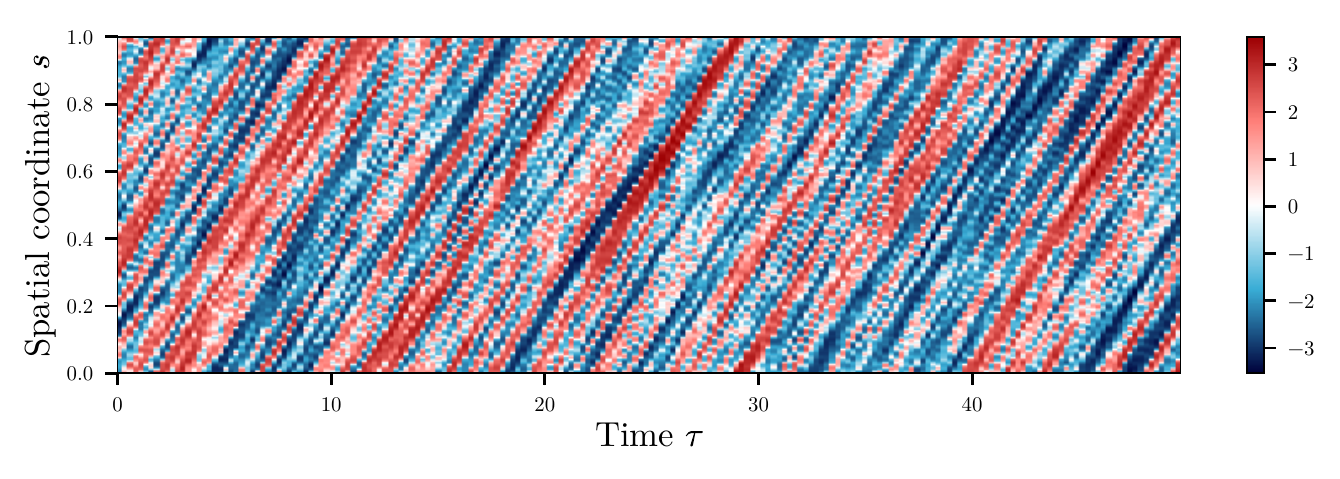}
    \caption{State sequence $\vct{z}'_{\srange{\cns{T}}}$ for transformed \ac{ssm}.}
    \label{sfig:st-nonlinear-sim-state}
  \end{subfigure}
  \caption{Simulated sequences used in experiments for \ac{st} \acp{ssm}.} 
  \label{fig:st-simulated-state-and-obs}
\end{figure}

We define a regular mesh of nodes $s_{\srange{\cns{M}}}$ and basis functions $\beta_{\srange{\cns{M}}}$
\begin{equation}\label{eq:spectral-nodal-basis-function}
  s_m = \frac{m-1}{\cns{M}}
  ~~\textrm{and}~~
  \beta_m(s) = 
  \frac{\sinc(2\uppi\cns{M}(s-s_m))\cos(\uppi(s-s_m))}{\sinc(\uppi(s-s_m))}
  ~~\forall m \in\range{\cns{M}}
\end{equation}
with the space-time varying processes $\rvar{\zeta}$ and $\rvar{\eta}$ and kernel function $\kappa$ then being defined respectively in terms of the finite set of time-varying processes $\rvar{\chi}_{\srange{\cns{M}}}$ and $\rvar{\upsilon}_{\srange{\cns{M}}}$ and coefficients $\lambda_{\srange{\cns{M}}}$ as
\begin{align}\label{eq:spectral-nodal-process-expansions}
  \rvar{\zeta}(s,\tau) &= 
  \sumrange{m}{1}{\cns{M}} \rvar{\chi}_m(\tau) \,\beta_m(s),
  \\
  \rvar{\eta}(s,\tau) &= 
  \sumrange{m}{1}{\cns{M}} \rvar{\upsilon}_m(\tau) \,\beta_m(s),
  \\
  \textrm{and}\quad
  \kappa(s) &= 
  \sumrange{m}{1}{\cns{M}} \lambda_m \,\beta_m(s).
\end{align}
The basis functions $\beta_{\srange{\cns{M}}}$ and nodes $s_{\srange{\cns{M}}}$ satisfy \cref{eq:spatial-basis-function-conditions} such that $\rvar{\chi}_m(\tau)$, $\rvar{\upsilon}_m(\tau)$ and $\lambda_m$ correspond to the values of respectively $\rvar{\zeta}(s_m, \tau)$, $\rvar{\eta}(s_m,\tau)$ and $\kappa(s_m)$ for any mesh node $s_m$. We define $\tilde{\rvar{\chi}}_{\srange[0]{\cns{K}}}(\tau) = \textsc{dft}(\rvar{\chi}_{\srange{\cns{M}}}(\tau))$, $\tilde{\rvar{\upsilon}}_{\srange[0]{\cns{K}}}(\tau) = \textsc{dft}(\rvar{\upsilon}_{\srange{\cns{M}}}(\tau))$ and $\tilde{\lambda}_{\srange[0]{\cns{K}}} = \textsc{dft}(\lambda_{\srange{\cns{M}}})$ with $\cns{K} = \floor*{\frac{\cns{M}}{2}}$ and $\textsc{dft}$ indicating the discrete Fourier transform, with the Fourier coefficient $\tilde{x}_k$ for a real sequence $x_{\srange{\cns{M}}}$ being computed as
\begin{equation}\label{eq:discrete-fourier-transform}
  \tilde{x}_k = 
  \textsc{dft}_k(x_{\srange{\cns{M}}}) = 
  \recip{\cns{M}}\sumrange{m}{1}{\cns{M}}
    x_{m} \exp\left(-\frac{i2\uppi k m}{\cns{M}}\right)
  \in
  \begin{cases}
  \reals & \textrm{if }k \in \lbrace 0, \textstyle\frac{\cns{M}}{2} \rbrace, \\
  \complexs & \textrm{if }k \in \range[1]{\ceil*{\textstyle\frac{\cns{M}}{2}}-1}.
  \end{cases}
\end{equation}
Then we have the following equivalent spectral expansions for $\rvar{\zeta}$, $\rvar{\eta}$ and $\kappa$
\begin{align}
  \label{eq:spectral-fourier-process-expansion-state}
  \rvar{\zeta}(s,\tau) &= 
  \sumrange{k}{-\cns{K}}{\cns{K}} \alpha_k \tilde{\rvar{\chi}}_{k}(\tau) \exp(i\omega_k s),\\
  \label{eq:spectral-fourier-process-expansion-noise}
  \rvar{\eta}(s,\tau) &= 
  \sumrange{k}{-\cns{K}}{\cns{K}} \alpha_k \tilde{\rvar{\upsilon}}_{k}(\tau) \exp(i\omega_k s),\\
  \label{eq:spectral-fourier-process-expansion-kernel}
  \rvar{\kappa}(s) &= 
  \sumrange{k}{-\cns{K}}{\cns{K}} \alpha_k \tilde{\lambda}_{k} \exp(i\omega_k s),
\end{align}
with the convention that negative indices to the Fourier coefficients indicate complex conjugation, e.g. $\tilde{\lambda}_{-k} = \tilde{\lambda}_k^*$, and $\alpha_{\srange[-\cns{K}]{\cns{K}}}$ and $\omega_{\srange[-\cns{K}]{\cns{K}}}$ are defined as
\begin{equation}\label{eq:spectral-fourier-process-expansion-coefficients}
  \alpha_k = \begin{cases}
  \frac{1}{2} & \textrm{if } k = 0,\\
  1 & \textrm{if } |k| \in \range[1]{\textstyle\ceil*{\frac{\cns{M}}{2}}-1},\\
  \frac{1}{4} & \textrm{if } |k| = \frac{\cns{M}}{2},
  \end{cases}
  \quad\textrm{and}\quad
  \omega_k = 2\uppi k ~~\forall k \in\range[-\cns{K}]{\cns{K}}.
\end{equation}
Using \cref{eq:spectral-fourier-process-expansion-state} we then have that spatial derivatives of $\rvar{\zeta}$ can be computed as
\begin{equation}\label{eq:spectral-fourier-process-expansion-state-derivatives}
  \partial^n_s \rvar{\zeta}(s,\tau) = 
  \sumrange{k}{-\cns{K}}{\cns{K}} \alpha_k (i\omega_k)^n\bar{\rvar{\chi}}_{k}(\tau) \exp(i\omega_k s)
  \quad \forall n\in\naturals.
\end{equation}
Substituting the expansions in \cref{eq:spectral-fourier-process-expansion-noise,eq:spectral-fourier-process-expansion-state,eq:spectral-fourier-process-expansion-state-derivatives} for the processes and spatial derivatives into \cref{eq:spde-stochastic-turbulence} and using the convolution theorem gives
\begin{equation}
  \sumrange{k}{-\cns{K}}{\cns{K}} \alpha_k \left( 
    \dr\tilde{\rvar{\chi}}_k - 
    (-\theta_1\omega_k^2 + i\theta_2\omega_k - \theta_3)\tilde{\rvar{\chi}}_k\,\dr\tau -
    \tilde{\lambda}_k \,\dr\tilde{\rvar{\upsilon}}_k
  \right)\exp(i\omega_k s) = 0.
\end{equation}
Integrating both sides over $\set{S}$ against a suitable orthogonal set of test functions
\begin{equation}\label{eq:st-test-functions}
  h_j(s) = 
  \exp(-i\omega_js) ~ \forall j\in\range[0]{\left(\textstyle\ceil*{\frac{\cns{M}}{2}}-1\right)} 
  ~\textrm{and}~
  h_{\frac{\cns{M}}{2}}(s) = \cos(\cns{M}\uppi s)~\textrm{if }\cns{M}\textrm{ is even},
\end{equation}
we arrive at the following system of \acp{sde}
\begin{align}\label{eq:st-fourier-sdes}
  \dr\tilde{\rvar{\chi}}_k(\tau) &=
  (-\theta_1\omega_k^2 + i\theta_2\omega_k-\theta_3)\tilde{\rvar{\chi}}_k(\tau)\,\dr\tau +
  \tilde{\lambda}_k\,\dr\tilde{\rvar{\upsilon}}_k(\tau)
  ~~\forall k \in\range[0]{\left(\textstyle\ceil*{\frac{\cns{M}}{2}}-1\right)},\\
  \label{eq:st-fourier-sde-nyquist}
  \dr\tilde{\rvar{\chi}}_{\frac{\cns{M}}{2}}(\tau) &=
  (-\theta_1\omega_k^2 -\theta_3)\tilde{\rvar{\chi}}_{\frac{\cns{M}}{2}}(\tau)\,\dr\tau +
  \tilde{\lambda}_{\frac{\cns{M}}{2}}\,\dr\tilde{\rvar{\upsilon}}_{\frac{\cns{M}}{2}}(\tau)
  ~~\textrm{if }\cns{M}\textrm{ is even}.
\end{align}
Assuming that the noise Fourier coefficients $\tilde{\rvar{\upsilon}}_{\srange[0]{\cns{K}}}$ are independent Wiener processes, real-valued for the zero- and Nyquist-frequency coefficients ($\tilde{\rvar{\upsilon}}_0$ and $\tilde{\rvar{\upsilon}}_{\frac{\cns{M}}{2}}$) and complex-valued for the remaining coefficients, then the transition distributions for this system have analytic solutions
\begin{equation}\label{eq:st-fourier-transition-dist}
\begin{gathered}
  \tilde{\rvar{\chi}}_k(\tau) \gvn \tilde{\rvar{\chi}}_k(0)
  \sim
  \gau\left( 
    \exp\left( \xi_k\tau\right) \tilde{\rvar{\chi}}_k(0),\,
    \frac{\tilde{\lambda}_k^2}{2\psi_k}\left(1 - \exp(-2\psi_k\tau)\right)
  \right),
  \\
  \textrm{with }
  \psi_k = \theta_1\omega_k^2 + \theta_3
  \textrm{ and }
  \xi_k = \begin{cases} 
    i\theta_2\omega_k - \psi_k & \textrm{if } k \neq \frac{\cns{M}}{2}\\
    -\psi_k & \textrm{if } k = \frac{\cns{M}}{2}
  \end{cases},
\end{gathered}
~~ k\in\range[0]{\cns{K}}.
\end{equation}
where we have overloaded the notation for a Gaussian distribution $\gau$ to extend to complex-valued variables with the convention that for a complex-valued random variable $\rvar{z} \in \complexs$, complex mean parameter $\mu \in \complexs$ and real variance $\sigma^2 \in \reals_{> 0}$, that
\begin{equation}
\begin{aligned}
  &\rvar{z} \sim \gau(\mu,\sigma^2)
  \implies\\
  &\qquad
  \Re(\rvar{z}) \sim \gau\left(\Re(\mu),\frac{\sigma^2}{2}\right),~
  \Im(\rvar{z}) \sim \gau\left(\Im(\mu),\frac{\sigma^2}{2}\right)
  ~\textrm{and}~
  \Re(\rvar{z}) \perp \Im(\rvar{z}).
\end{aligned}
\end{equation}
The Fourier coefficients $\tilde{\rvar{\chi}}_{\srange[0]{\cns{K}}}$ then also have Gaussian stationary distributions
\begin{equation}\label{eq:st-fourier-stationary-dist}
  \tilde{\rvar{\chi}}_k(\infty) \sim
  \gau\left( 0,\, \frac{\tilde{\lambda}_k^2}{2\psi_k}\right)
  \quad\forall k\in\range[0]{\cns{K}}.
\end{equation}
We assume the system is observed at $\cns{T}$ time points with $\tau_t = (t-1)\delta ~~\forall t\in\range{\cns{T}}$ and that the Fourier coefficients of the initial state at time $\tau_1 = 0$ are generated from the stationary distributions in \cref{eq:st-fourier-stationary-dist}. Identifying
\begin{equation}
  \rvar{z}_t(s) = \rvar{\zeta}(s, \tau_t) 
  \quad\textrm{and}\quad
  \rvct{x}_{t,\srange{\cns{M}}} = \rvct{\chi}_{\srange{\cns{M}}}(\tau_t)
  \quad\forall t \in \range{\cns{T}}
\end{equation}
we have that the state update operators can be written
\begin{align}
  \label{eq:st-init-state}
  \rvct{x}_{1,\srange{\cns{M}}} &= \textsc{dft}^{-1}\left(\vct{a}_{\srange[0]{\cns{K}}}\odot \rvct{u}_{1,\srange[0]{\cns{K}}} \right),
  \\
  \label{eq:st-state-update}
  \rvct{x}_{t,\srange{\cns{M}}} &= 
  \textsc{dft}^{-1}\left(  
    \vct{b}_{\srange[0]{\cns{K}}} \odot \textsc{dft}(\rvct{x}_{t-1,\srange{\cns{M}}})
    +
    \vct{c}_{\srange[0]{\cns{K}}} \odot \rvct{u}_{t,\srange[0]{\cns{K}}}
  \right)
  \quad \forall t\in\range[2]{\cns{T}},
\end{align}
where $\vct{a}_{\srange[0]{\cns{K}}}$, $\vct{b}_{\srange[0]{\cns{K}}}$ and $\vct{c}_{\srange[0]{\cns{K}}}$ are length $\cns{K} + 1$ vectors with
\begin{equation}
  \label{eq:st-fourier-update-coefficients}
  a_k = \frac{\tilde{\lambda}_k}{\sqrt{2\psi_k}},~
  b_k = \exp(\xi_k\delta),~
  c_k = a_k \sqrt{1 - \exp(-2\psi_k\delta)}
  \quad \forall k \in \range[0]{\cns{K}},
\end{equation}
and the state noise variables $\rvar{u}_{\srange{\cns{T},\srange[0]{\cns{K}}}}$ are real-valued for the zero- and Nyquist-frequency components and complex otherwise and have Gaussian distributions
\begin{equation}\label{eq:st-state-noise-distributions}
  \rvar{u}_{t,k} \in 
  \begin{cases}
    \reals & \textrm{if }k \in \lbrace 0, \textstyle\frac{\cns{M}}{2} \rbrace,\\
    \complexs & \textrm{if }k \in \range[1]{\ceil*{\textstyle\frac{\cns{M}}{2}}-1},
  \end{cases}
  ~
  \rvar{u}_{t,k} \sim \gau(0, 1)
  \quad \forall t \in \range{\cns{T}}, k\in\range[0]{\cns{K}}.
\end{equation}
The system is observed at $\cns{L}$ equispaced mesh nodes with $s^{\textrm{o}}_l = s_{\frac{\cns{M}}{\cns{L}}(l - \frac{1}{2})} ~\forall l\in\range{\cns{L}}$ and a simple linear-Gaussian observation model assumed
\begin{equation}\label{eq:st-obs-model}
  \rvar{y}_{t,l} = 
  \rvar{z}_t(s^{\textrm{o}}_l) + \rvar{v}_{t,l} =
  \rvar{x}_{\frac{\cns{M}}{\cns{L}}(l - \frac{1}{2})} + \rvar{v}_{t,l},~
  \rvar{v}_{t,l} \sim \gau(0, \varsigma^2)
  \quad \forall t \in \range{\cns{T}}, l \in \range{\cns{L}}.
\end{equation}
The state noise kernel Fourier coefficients $\tilde{\lambda}_{\srange[0]{\cns{K}}}$ are chosen to represent a squared-exponential kernel with length-scale parameter $\vartheta$ and amplitude parameter $\alpha$
\begin{equation}\label{eq:st-noise-kernel-fourier-coeffs}
  \tilde{\lambda}_k = \alpha \exp(-\omega_k^2 \vartheta^2 )
  \quad \forall k \in\range[0]{\cns{K}}.
\end{equation}

\subsection{Kuramoto--Sivashinksy model}

\label{app:kuramoto-sivashinsky-model-details}

\begin{table}[t!]
  \begin{tabular}{ll}
    Number of mesh nodes & $\cns{M} = 512$ \\
    Number of observation times & $\cns{T} = 200$ \\
    Number of observation locations & $\cns{L} = 64$ \\
    Number of integrator steps between observations & $\cns{S} = 10$ \\
    Integrator time step & $\delta = 0.25$\\
    Length scale parameter & $\theta_1 = 32\uppi$\\
    Damping coefficient & $\theta_2 = \frac{1}{6}$\\
    State noise kernel length scale & $\theta_3 = \theta_1^{-1}$\\
    State noise kernel amplitude & $\theta_4 = \theta_1^{-\frac{1}{2}}$\\
    Observation noise standard deviation & $\varsigma = 0.5$\\
  \end{tabular}
  \caption{Kuramoto-Sivashinksy model parameter settings}
  \label{tab:ks-model-params}
\end{table}

\begin{figure}[t!]
  \centering
  \begin{subfigure}[b]{\linewidth}
    \centering
    \includegraphics[width=0.9\textwidth]{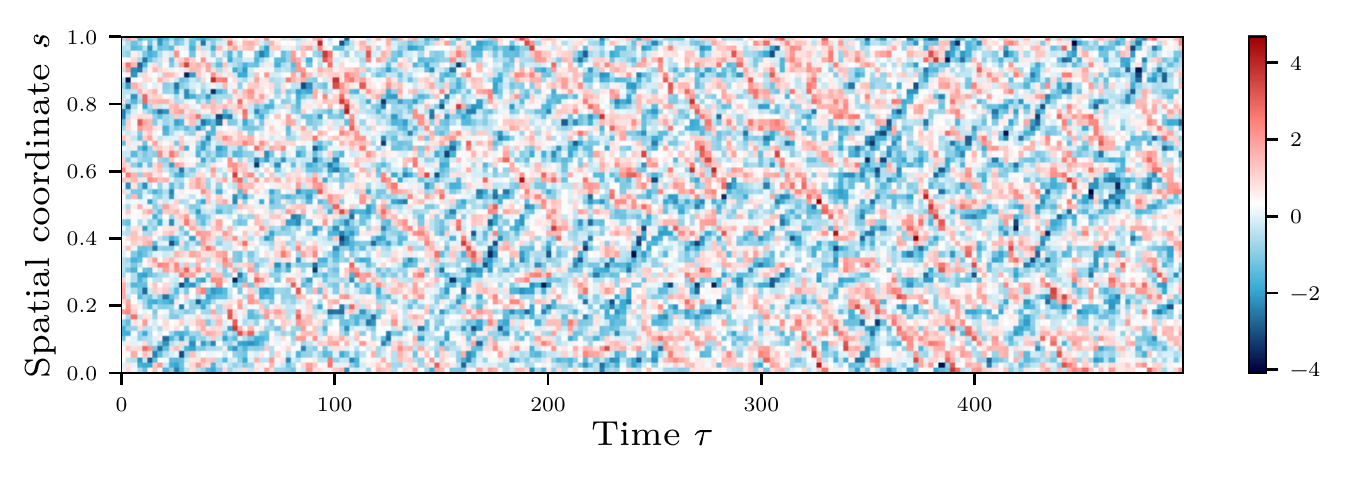}
    \caption{Noisy observation sequence $\vct{y}_{\srange{\cns{T}}}$ with linear observation operator.}
    \label{sfig:ks-linear-sim-obs}
  \end{subfigure}
  \begin{subfigure}[b]{\linewidth}
    \centering
    \includegraphics[width=0.9\textwidth]{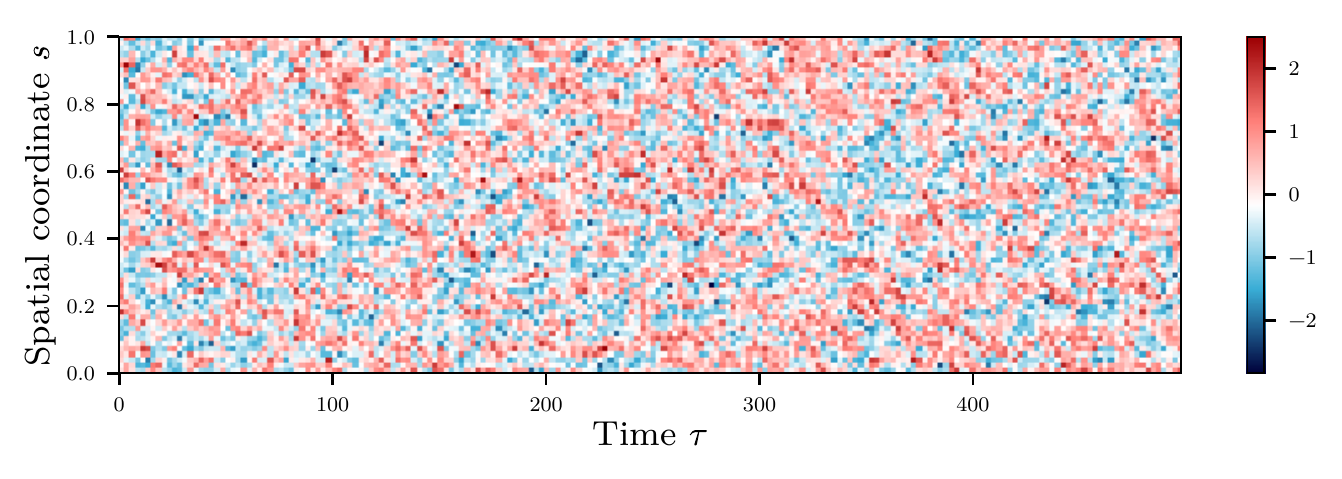}
    \caption{Noisy observation sequence $\vct{y}_{\srange{\cns{T}}}$ with nonlinear observation operator.}
    \label{sfig:ks-nonlinear-sim-obs}
  \end{subfigure}
  \begin{subfigure}[b]{\linewidth}
    \centering
    \includegraphics[width=0.9\textwidth]{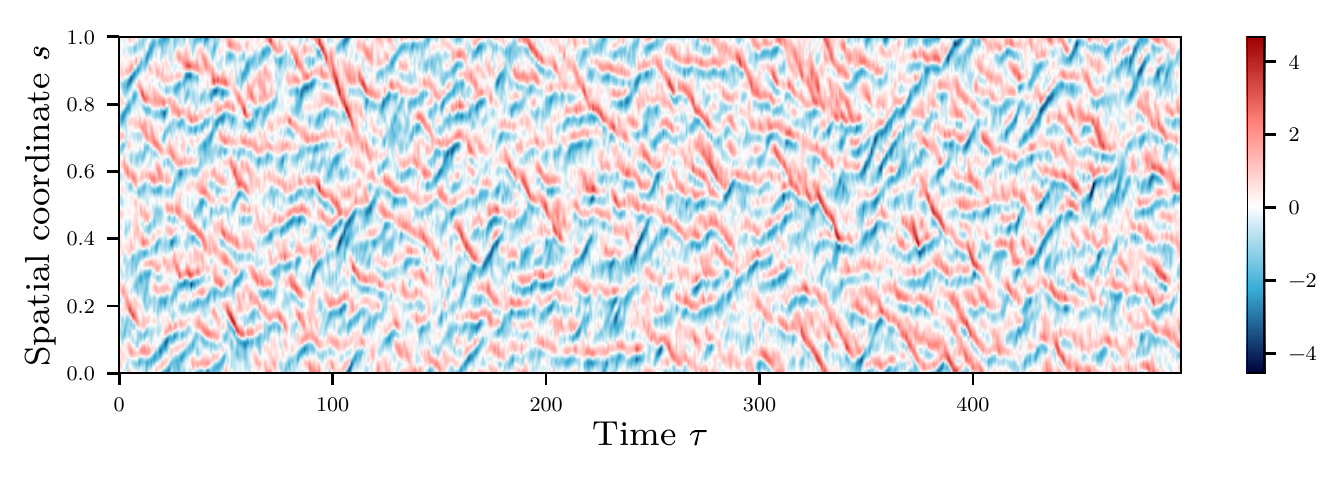}
    \caption{True state sequence $\vct{z}_{\srange{\cns{T}}}$ used to generate observations.}
    \label{sfig:ks-sim-state}
  \end{subfigure}
  \caption{Simulated sequences used in experiments with stochastic \ac{ks} \acp{ssm}.} 
  \label{fig:ks-simulated-state-and-obs}
\end{figure}

We use the same spectral approach in as in the \ac{st} model to define the basis function expansions of the processes $\rvar{\zeta}$ and $\rvar{\eta}$ and kernel $\kappa$ in terms of coefficients $\xi_{\srange{\cns{M}}}$, $\upsilon_{\srange{\cns{M}}}$ and $\lambda_{\srange{\cns{M}}}$ (see \cref{eq:spectral-fourier-process-expansion-state,eq:spectral-fourier-process-expansion-noise,eq:spectral-fourier-process-expansion-noise}). The non-linear $\zeta^2$ term in the drift component of the \ac{spde} cannot be exactly expressed as a linear combination of the basis function $\beta_{\srange{\cns{M}}}$, and so we cannot directly form a system of \acp{sde} to solve as in the \ac{st} model. We make the approximation that
\begin{equation}
  \zeta(s,\tau)^2 = 
  \sumrange{m}{1}{\cns{M}}
  \sumrange{n}{1}{\cns{M}}
  \chi_m(\tau)\chi_n(\tau) \beta_m(s)\beta_n(s)
  \approx
  \sumrange{m}{1}{\cns{M}}
  \chi_m(\tau)^2 \beta_m(s).
\end{equation}
At the mesh nodes $s_{\srange{\cns{M}}}$ this gives the correct values but gives a different interpolation at points between the nodes; for dense meshes however the error introduced is small. Using this approximation the following system of \acp{sde} can be derived in the Fourier coefficients $\tilde{\xi}_{\srange[0]{\cns{K}}}$, $\tilde{\upsilon}_{\srange[0]{\cns{K}}}$ and $\tilde{\lambda}_{\srange[0]{\cns{K}}}$
\begin{equation}\label{eq:ks-fourier-sde}
  \dr\tilde{\rvar{\chi}}_k(\tau) =
  \left(
  \left(\frac{\omega_k^2}{\theta_1^2} - \frac{\omega_k^4}{\theta_1^4}-\theta_2\right)\tilde{\rvar{\chi}}_k(\tau) + 
    N_k(\tilde{\rvar{\chi}}_{\srange[0]{\cns{K}}})
  \right)\dr\tau +
  \tilde{\lambda}_k\,\dr\tilde{\rvar{\upsilon}}_k(\tau)
  ~~~\forall k \in \range[0]{\cns{K}}
\end{equation}
with the noise Fourier coefficients $\tilde{\rvar{\upsilon}}_{\srange{\cns{K}}}$ again assumed to be (complex-valued) Wiener processes and the non-linear $N_k$ terms in the drift defined by
\begin{equation}
  \op{N}_k(\tilde{\rvar{\chi}}_{\srange[0]{\cns{K}}}) = \begin{cases}
    \frac{i\omega_k}{2\theta_1}\textsc{dft}_k(\textsc{dft}^{-1}(\tilde{\rvar{\chi}}_{\srange[0]{\cns{K}}}(\tau))^2) &
    \textrm{if}~ k \in\range[0]{\left(\textstyle\ceil*{\frac{\cns{M}}{2}}-1\right)},
    \\
    0 & \textrm{if} ~ k = \frac{\cns{M}}{2}.
  \end{cases}
\end{equation}
The state noise kernel Fourier coefficients $\tilde{\lambda}_{\srange[0]{\cns{K}}}$ are as in the \ac{st} model chosen to represent a squared-exponential kernel as defined in \cref{eq:st-noise-kernel-fourier-coeffs}.

Due to the non-linear terms, the system of \acp{sde} in \cref{eq:ks-fourier-sde} does not have an analytic solution. Therefore we numerically integrate the system using a heuristic combination of a exponential-time differencing fourth-order Runge-Kutta scheme \citep{cox2002exponential} to time step forward according to the drift term and a Euler-Maruyama discretisation to account for the diffusion term. To reduce the time discretisation error we use $\textrm{S}$ integrator steps with time step $\delta$ between each of the $\cns{T}$ observation times $\tau_t = (t-1)\cns{S}\delta ~\forall t\in\range{\cns{T}}$. The state transition operator $\op{F}_{t}$ then correspond to the map from a previous state vector $\rvct{x}_{t-1}$ and state noise variable $\rvct{u}_t$ (consisting of the concatenation of $\cns{S}$ simulated Wiener process increments) to the state vector $\rvct{x}_t$ by peforming $S$ integrator steps. The state transition operators are non-linear and the density of the corresponding state transition distribution does not have a closed form solution. 

For the observation operators we considered two cases - a linear-Gaussian observation model and a non-linear observation operator. Although due to the non-linear state transition operators the filtering distributions are non-Gaussian irrespective of the observation operator used, in practice we found the local \ac{enkf} was able to generate accurate ensemble estimates of the filtering distributions when using a simple linear-Gaussian observation model, suggesting the filtering distributions remain close to Gaussian despite the non-linear state dynamics. As our focus is on inference in \acp{ssm} for which existing local \ac{enkf} approaches perform poorly in, we also considered an alternative model configuration in which a non-linear function of the model state is noisily observed.

In both the linear and non-linear cases system is assume to be observed at $\cns{L}$ equispaced mesh nodes with $s^{\textrm{o}}_l = s_{\frac{\cns{M}}{\cns{L}}(l - \frac{1}{2})} ~\forall l\in\range{\cns{L}}$. For the linear case the observation model is assumed to be equivalent to that assumed for the \ac{st} model,
\begin{equation}\label{eq:ks-obs-model-linear}
  \rvar{y}_{t,l} = 
  \rvar{z}_t(s^{\textrm{o}}_l) + \rvar{v}_{t,l} =
  \rvar{x}_{\frac{\cns{M}}{\cns{L}}(l - \frac{1}{2})} + \rvar{v}_{t,l},~
  \rvar{v}_{t,l} \sim \gau(0, \varsigma^2)
  \quad \forall t \in \range{\cns{T}}, l \in \range{\cns{L}}.
\end{equation}
The non-linear case is directly analogous other than the state values being observed via a hyperbolic tangent non-linearity:
\begin{equation}\label{eq:ks-obs-model-nonlinear}
  \rvar{y}_{t,l} = 
  \tanh(\rvar{x}_{\frac{\cns{M}}{\cns{L}}(l - \frac{1}{2})}) + \rvar{v}_{t,l},~
  \rvar{v}_{t,l} \sim \gau(0, \varsigma^2)
  \quad \forall t \in \range{\cns{T}}, l \in \range{\cns{L}}.
\end{equation}
Although seemingly minor change in the model, as illustrated in the experimental results, introducing this non-linearity was sufficient to significantly degrade the filtering performance of the local \ac{etkf}.
\newpage
\section{Full grid search results for local ETKF}
\label{app:full-letkf-grid-search-results}

\begin{figure}[ht!]
  \centering
  \begin{subfigure}[b]{\linewidth}
    \centering
    \includegraphics[width=0.95\textwidth]{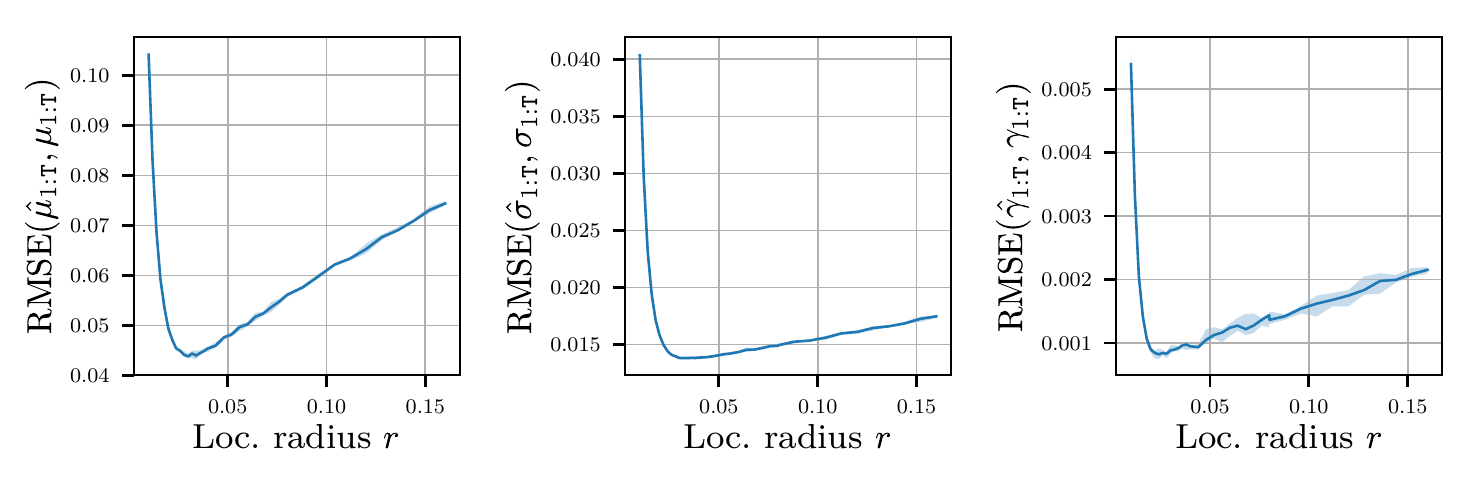}
    \caption{Linear-Gaussian \ac{st} \ac{ssm}.}
  \end{subfigure}
  \begin{subfigure}[b]{\linewidth}
    \centering
    \includegraphics[width=0.95\textwidth]{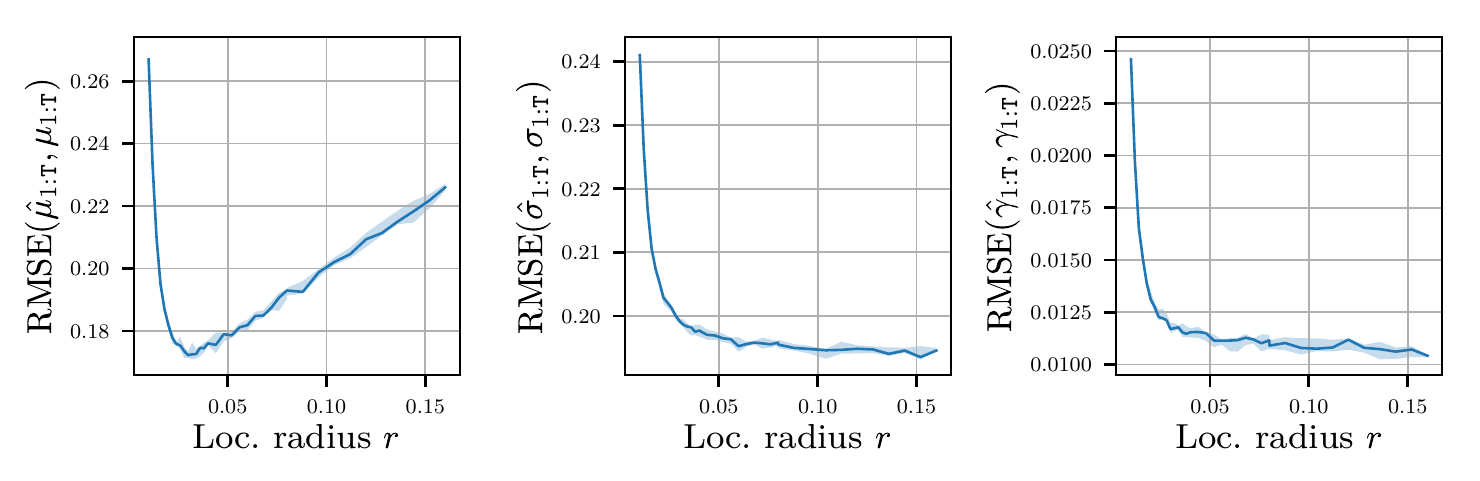}
    \caption{Transformed \ac{st} \ac{ssm}.}
  \end{subfigure}
  \begin{subfigure}[b]{\linewidth}
    \centering
    \includegraphics[width=0.95\textwidth]{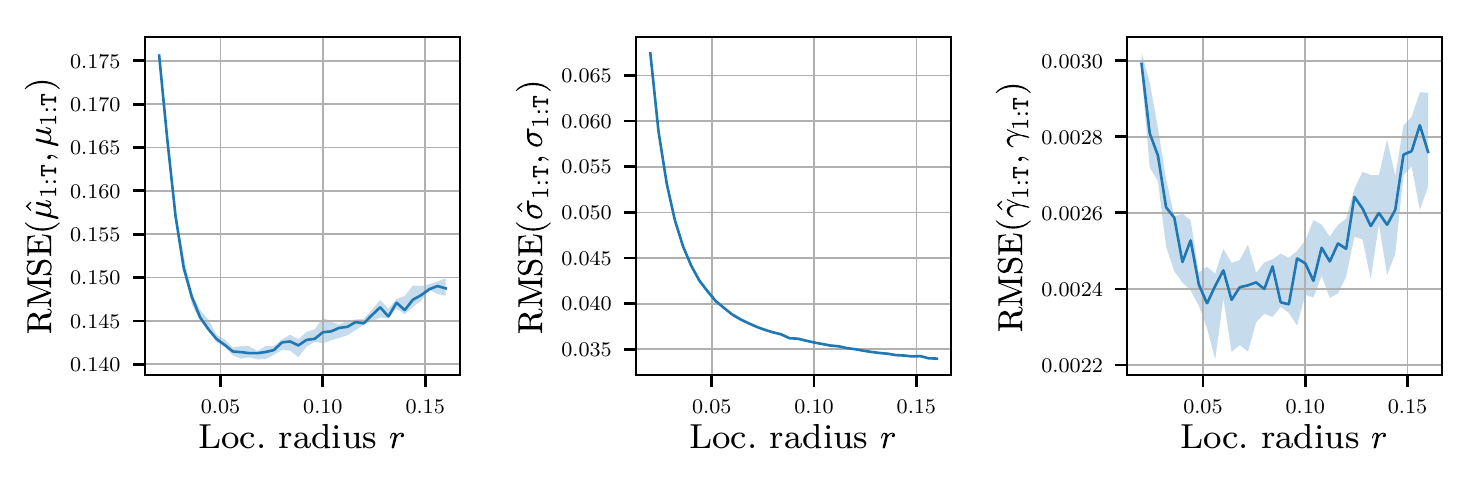}
    \caption{Linearly observed \ac{ks} \ac{ssm}.}
  \end{subfigure}
  \begin{subfigure}[b]{\linewidth}
    \centering
    \includegraphics[width=0.95\textwidth]{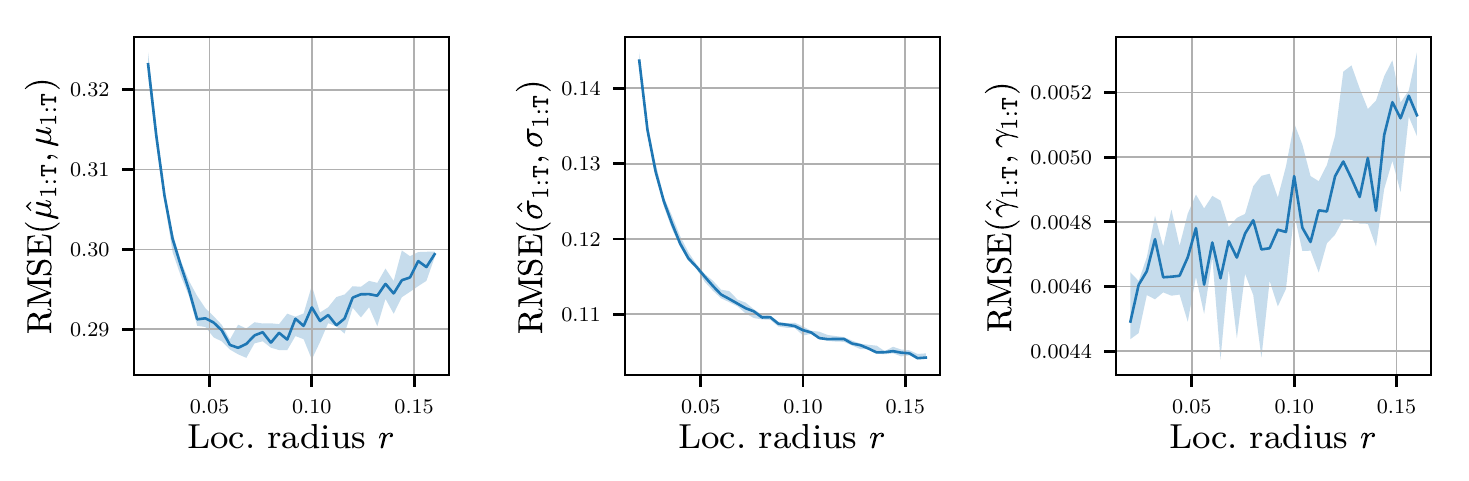}
    \caption{Non-linearly observed \ac{ks} \ac{ssm}.}
  \end{subfigure}
  \caption{Values of metrics for all localisation radii $r$ for local \ac{etkf} on four \acp{ssm} considered in experiments. In all cases the curve shows the median value across five independent runs and the filled region the minimum to maximum range.} 
  \label{fig:full-letkf-grid-search-results}
\end{figure}
\newpage
\section{Details of MCMC runs for KS models}
\label{app:ks-mcmc-details}

A non-centred parametrisation was used for the Hamiltonian Monte Carlo chains for the two \ac{ks} \acp{ssm} \citep{papaspiliopoulos2007general}, with the target smoothing distribution formulated in terms of the $\cns{M}\cns{T}\cns{S} \approx 10^6$ dimensional set of state noise variables $\rvct{u}_{\srange{\cns{T}}}$ which are independently and identically distributed standard normal variables under the prior, with the observation sequence $\vct{y}_{\srange{\cns{T}}}$ then having a Gaussian conditional distribution given $\rvct{u}_{\srange{\cns{T}}}$. The step-size for the integrator of the Hamiltonian dynamics was manually tuned once for each \ac{ssm} using short pilot chains with a fixed number of integrator steps to achieve an average acceptance probability in the range $[0.6, 0.9]$ \citep{betancourt2014optimizing}, with in both \acp{ssm} a step size $2.5\times 10^{-3}$ found to give an acceptance rate is the target range. The integrator used was a variant of the standard leapfrog / St\"{o}rmer-Verlet integrator which uses an alternative splitting of the Hamiltonian to leverage an exact analytic solution for the Hamiltonian dynamics under the quadratic potential energy component due to the Gaussian prior \citep{shahbaba2014split}. The number of integrator steps used to generate the Hamiltonian dynamics trajectory in each chain transition was dynamically set on each iteration using a variant of the \emph{No-U-Turn sampler} scheme \citep{hoffman2014no,betancourt2017conceptual}, with the chains for both \acp{ssm} performing approximately $2\times 10^3$ steps per transition on average. For each \ac{ssm} the total wall clock time to run the five chains in parallel on a Intel Xeon E5-2620 v4 8-core CPU was around one week.

\begin{figure}
  \centering
  \includegraphics[width=\textwidth]{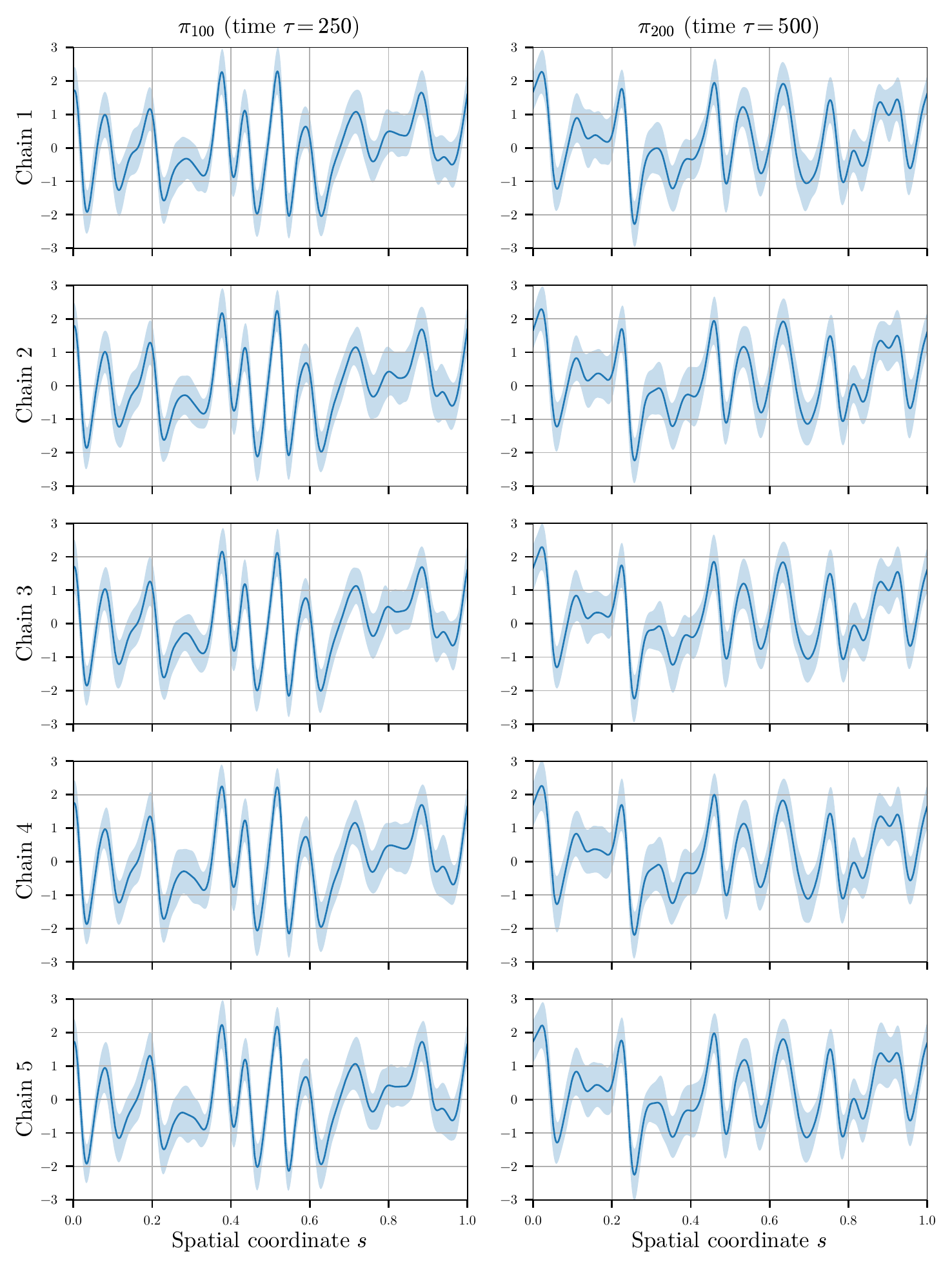}
  \caption{Comparison of estimates of first and second moments of filtering distributions $\pi_{100}$ and $\pi_{200}$ for linearly observed \ac{ks} \ac{ssm} using final 100 samples from each of 5 chains (curves show the estimated mean and the filled region the mean $\pm$ two standard deviations).}
  \label{fig:ks-linear-chains-comparison}
\end{figure}

\begin{figure}
  \centering
  \includegraphics[width=\textwidth]{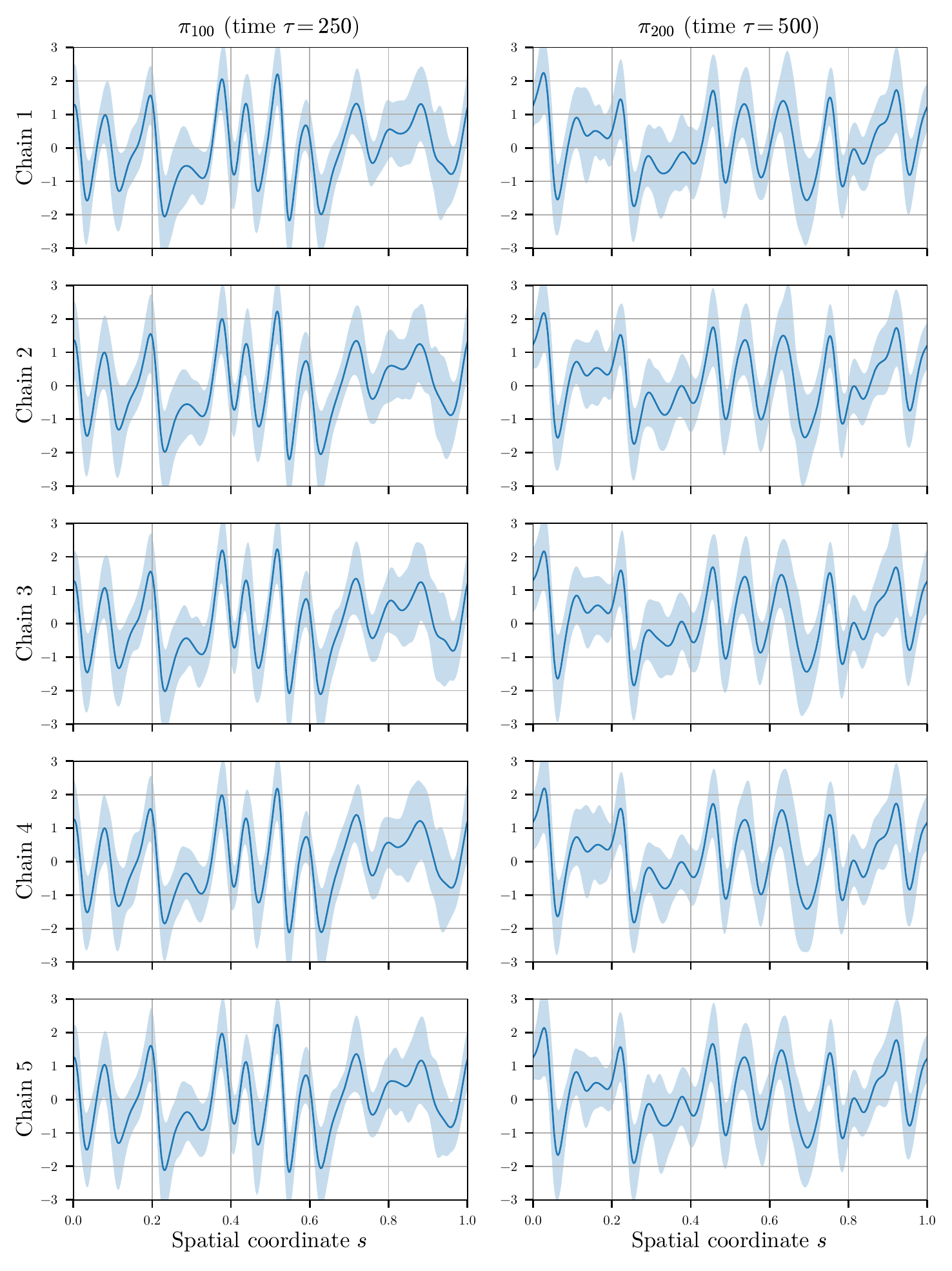}
  \caption{Comparison of estimates of first and second moments of filtering distributions $\pi_{100}$ and $\pi_{200}$ for non-linearly observed \ac{ks} \ac{ssm} using final 100 samples from each of 5 chains (curves show the estimated mean and the filled region the mean $\pm$ two standard deviations).}
  \label{fig:ks-nonlinear-chains-comparison}
\end{figure}

All chains were initialised from the true state noise sequence $\vct{u}_{\srange{\cns{T}}}$ used to generate the observations, which corresponds to a single exact sample from the target distribution $\prob(\rvct{u}_{\srange{\cns{T}}} \in \dr\vct{u} \gvn \rvct{y}_{\srange{\cns{T}}} = \vct{y}_{\srange{\cns{T}}})$ as the $(\vct{u}_{\srange{\cns{T}}}, \vct{y}_{\srange{\cns{T}}})$ pair was originally generated from the corresponding joint distribution $\prob(\rvct{u}_{\srange{\cns{T}}} \in \dr\vct{u}, \rvct{y}_{\srange{\cns{T}}} = \dr\vct{y})$. Although typically it would be preferable for the robustness of convergence diagnostics based on comparisons between chains to initialise each of the chains independently from an over-dispersed distribution compared to the target such as the prior, here we found the step-size required to robustly achieve an average acceptance probability in the range $[0.6, 0.9]$ for chains initialised from the prior to be much smaller than for chains initialised from the `true' noise sequence $\vct{u}_{\srange{\cns{T}}}$, likely due to the differing geometry of the target distribution in the tails (where initialisations from the prior are likely to fall) and typical set, which $\vct{u}_{\srange{\cns{T}}}$ as an exact sample from the target should be within. Given the long chain run times even when using the larger step size, a pragmatic choice was therefore made to use a common initialisation. This initialisation scheme and relatively small number of samples in each chain means there is a risk that the chains therefore only explored a subset of the target distributions' typical sets. As partial evidence against this being the case, visual checks of the estimates of the first and second moments of a subset of the filtering distributions $\pi_{\srange{\cns{T}}}$ using the final 100 samples from each of the chains suggest that the estimates from the different chains are consistent with each other (see examples in \cref{fig:ks-linear-chains-comparison,fig:ks-nonlinear-chains-comparison}).

\newpage
\bibliography{refs}
\bibliographystyle{imsart-nameyear}

\end{document}